\documentclass[aps,prd,preprint,superscriptaddress]{revtex4}
\usepackage{graphicx}
\usepackage{amsmath,amssymb}
\usepackage{bm}% bold math
\usepackage[caption=false]{subfig}
\usepackage{color}
\usepackage{setspace}
\usepackage{soul}
\usepackage{times}

\def\p{\partial}

\def\=:{=\hspace{-.7em}\raisebox{1.1ex}{.}\hspace{.1em}\raisebox{-0.2ex}{.} }

\newcommand {\beq}{\begin{eqnarray}}
\newcommand {\eeq}{\end{eqnarray}}
\newcommand {\non}{\nonumber\\}

\newcommand {\1}[1]{\frac{1}{#1}}

\newcommand {\del}{\partial}

\newcommand {\tr}{{\rm tr}\,}

\begin{document}


\title{Fractional instantons and bions in 
the $O(N)$ model\\ with twisted boundary conditions}



\author{Muneto Nitta}
\affiliation{Department of Physics, and Research and Education Center for Natural 
Sciences, Keio University, Hiyoshi 4-1-1, Yokohama, Kanagawa 223-8521, Japan
}
%\thanks{}



\date{\today}
\begin{abstract}
Recently, multiple fractional instanton configurations 
with zero instanton charge, called bions, 
have been revealed to play important roles 
in quantum field theories on compactified spacetime. 
In two dimensions, fractional instantons and bions 
have been extensively studied 
in the ${\mathbb C}P^{N-1}$ model
and the Grassmann sigma model 
on ${\mathbb R}^1 \times S^1$ 
with the ${\mathbb Z}_N$ symmetric twisted boundary condition.
Fractional instantons in these models 
are domain walls with a localized $U(1)$ modulus twisted half 
along their world volume.
In this paper,
we classify fractional instantons and bions 
in the $O(N)$ nonlinear sigma model 
on ${\mathbb R}^{N-2} \times S^1$ with 
more general twisted boundary conditions 
in which arbitrary number of fields change sign.
We find that fractional instantons have more general composite structures, 
that is, 
a global vortex with an Ising spin 
(or a half-lump vortex),
a half sine-Gordon kink on a domain wall,
or a half lump on a ``space-filling brane" in the $O(3)$ model 
(${\mathbb C}P^1$ model)
on ${\mathbb R}^{1} \times S^1$, 
and 
a global monopole with an Ising spin 
(or a half-Skyrmion monopole),  
a half sine-Gordon kink on a global vortex, 
a half lump on a domain wall,
or a half Skyrmion on a ``space-filling brane"
in the $O(4)$ model (principal chiral model or Skyrme model)
on ${\mathbb R}^{2} \times S^1$.
We also construct bion configurations in these models. 

\end{abstract}

%\keywords{}

\maketitle

\section{Introduction}

Instantons have been known for long time to play significant roles in 
non-perturbative dynamics of quantum field theories 
such as supersymmetric QCD. 
Recently, 
multiple fractional instanton configurations 
with zero instanton charge, called bions, 
have been revealed to play important roles 
in quantum field theories on compactified spacetime 
\cite{Unsal:2007vu, Unsal:2007jx, Shifman:2008ja, 
Poppitz:2009uq, Poppitz:2012sw,Argyres:2012vv, Argyres:2012ka, 
Dunne:2012ae, Dunne:2012zk, Dabrowski:2013kba, Dunne:2013ada, 
Cherman:2013yfa, Basar:2013eka, Dunne:2014bca, Cherman:2014ofa,
Bolognesi:2013tya, Misumi:2014jua, Misumi:2014bsa, Misumi:2014raa,
Shermer:2014wxa,Anber:2014sda}.
The prime example which has been studied extensively is QCD 
with adjoint fermions (adj.) on ${\mathbb R}^3 \times S^1$. 
Bions can be classified into two classes, 
magnetic (charged) bions 
carrying a magnetic charge,
and neutral bions carrying no magnetic charge.
Magnetic bions are conjectured to lead semiclassical confinement 
in QCD (adj.) on ${\mathbb R}^{3} \times S^{1}$ 
\cite{Hosotani:1983xw, Hosotani:1988bm, Myers:2007vc, 
Myers:2009df, Cossu:2009sq, Meisinger:2009ne, Nishimura:2009me, 
Anber:2011de, Ogilvie:2012is, Kashiwa:2013rmg, Cossu:2013ora}. 
On the other hand, 
neutral bions 
are identified as the infrared renormalons in field theory 
\cite{Argyres:2012vv, Argyres:2012ka, Dunne:2012ae, Dunne:2012zk, 
Dabrowski:2013kba, Dunne:2013ada, Cherman:2013yfa, Basar:2013eka, 
Dunne:2014bca, Cherman:2014ofa, 'tHooft:1977am, 
Fateev:1994ai, Fateev:1994dp},
and play an essential role in unambiguous and self-consistent 
semiclassical definition of quantum field theories in  
a process known as the resurgence; 
Imaginary ambiguities called renormalon ambiguities 
 arising in non-Borel-summable  perturbative series 
 exactly cancel out with those arising in neutral 
bion's amplitude  
 in the small compactification-scale regime of 
QCD (adj.) on ${\mathbb R}^{3} \times S^{1}$. 
It indicates that the full semi-classical expansion, 
referred as a resurgent expansion \cite{Ec1}, 
that includes both perturbative and non-perturbative sectors,  
leads to unambiguous and self-consistent 
definition of field theories.
In quantum mechanics this is known as
the Bogomol'nyi-Zinn-Justin prescription
\cite{Bogomolny:1980ur, ZinnJustin:1981dx, ZinnJustin:2004ib}. 

On the other hand,
two dimensional nonlinear sigma model 
enjoys a lot of common features with four-dimensional Yang-Mills theory
\cite{Polyakov} 
such as asymptotic freedom, dynamical mass generation, 
and instantons \cite{Polyakov:1975yp,Din:1980jg}.
We can further expect a similar correspondence between 
fractional instantons and bions 
in nonlinear sigma models on ${\mathbb R}^1 \times S^1$ 
and those in Yang-Mills theory on ${\mathbb R}^3 \times S^1$. 
Fractional instantons in the ${\mathbb C}P^{N-1}$ model 
\cite{Eto:2004rz} 
(see also Refs.~\cite{Bruckmann:2007zh})
and the Grassmann sigma model \cite{Eto:2006mz} 
were constructed 
on ${\mathbb R}^1 \times S^1$
with twisted boundary conditions 
by using the moduli matrix technique 
\cite{Isozumi:2004jc,Isozumi:2004vg,Eto:2005sw,Eto:2005yh}
(see Ref.~\cite{Eto:2006pg} as a review) 
and D-brane configurations \cite{Eto:2004vy,Eto:2006mz}.
Bions and the resurgence 
have been extensively studied in the ${\mathbb C}P^{N-1}$ model
\cite{Dunne:2012ae, Dunne:2012zk, Dabrowski:2013kba,
Bolognesi:2013tya, Misumi:2014jua,Shermer:2014wxa}
and the Grassmann sigma model \cite{Misumi:2014bsa} 
on ${\mathbb R}^1 \times S^1$. 
In particular in Refs.~\cite{Dunne:2012ae, Dunne:2012zk}, 
bion configurations in the ${\mathbb C}P^{N-1}$ model were studied 
based on the dilute instanton description 
with taking account of interactions between well-separated fractional 
instantons and anti-instantons, 
to show explicitly that the imaginary ambiguity in the 
amplitude of neutral bions has the same magnitude with an 
opposite sign as the leading ambiguity 
arising from the non-Borel-summable series expanded around 
the perturbative vacuum.
The ambiguities at higher orders are canceled by amplitudes 
of bion molecules  
and the full trans-series expansion around 
the perturbative and non-perturbative vacua results in 
unambiguous semiclassical definition of field theories.
Furthermore, neutral bion ansatz  
beyond exact solutions were found in 
the ${\mathbb C}P^{N-1}$ model \cite{Misumi:2014jua} 
and the Grassmann model \cite{Misumi:2014bsa} 
in terms of the moduli matrix
and was found to be consistent with the results 
from the well-separated instanton gas calculus 
\cite{Dunne:2012ae, Dunne:2012zk} 
from all ranges of separations. 
Bions and resurgence were also studied for 
principal chiral models \cite{Cherman:2013yfa,Cherman:2014ofa} 
and quantum mechanics \cite{Dunne:2013ada, Basar:2013eka, 
Dunne:2014bca}. 

In order to understand more precise structures of 
fractional instantons and bions in generic field theories,
it is worth to remind that fractional instantons in 
the ${\mathbb C}P^{N-1}$ 
and Grassmann models on 
 ${\mathbb R}^1 \times S^1$
with the ${\mathbb Z}_{N}$ twisted boundary conditions 
have a composite soliton structure 
\cite{Eto:2004rz,Eto:2006mz}. 
When the coordinate $x^2$ is a compact direction, 
fractional instantons are domain walls extending to the $x^2$ direction (perpendicular to the $x^1$ direction) 
whose world volume a $U(1)$ modulus is localized on
and twisted half along.  
Fractional instantons can be therefore regarded as 
half sine-Gordon kinks on a domain wall. 
Since a domain wall carries unit instanton (lump) charge when 
the $U(1)$ modulus is twisted once (full sine-Gordon kink) 
\cite{Nitta:2012xq,Nitta:2013cn,Kobayashi:2013ju,Jennings:2013aea}, 
the above configuration carries half instanton charge 
\cite{Auzzi:2006ju}.
In this paper, 
we refer the above domain wall and sine-Gordon kink 
as a host soliton and daughter soliton, respectively. 
The simplest among 
${\mathbb C}P^{N-1}$ model 
and the Grassmann model is
the ${\mathbb C}P^1$ model, 
which is equivalent to 
the $O(3)$ sigma model 
described by a unit three-vector of scalar fields
${\bf n}= \{n_A(x)\}$ ($A=1,2,3$) with ${\bf n}^2=1$.
The ${\mathbb Z}_2$ symmetric boundary condition
reduces to 
$(n_1,n_2,n_3)(x+R) = 
(-n_1,-n_2,+n_3)(x)$ 
in this notation.

%%%%%%%%%%%%%%%%%%%%%%%%%
\begin{table}[h]
\begin{tabular}{|c|c|c|c|c|c|} \hline
bulk  & boundary    &  fixed manifold & host soliton &  modui ${\cal M}$ of &  daughter soliton \\
space & condition & ${\cal N}$ & $\pi_n({\cal N})$, codim & host soliton  & $\pi_m({\cal M})$, codim\\
 \hline\hline 
%%%%%
%%%%
${\mathbb R}^1 \times S^1$ & $(-,+,+)$ & $S^1$ & vortex & 2 points & Ising spin \\
& & $n_2^2 + n_3^2 =1$ & $\pi_1$, 2 & $n_1 = \pm 1$ 
& $\pi_0$, 0 \\ \hline 
%%%%%
${\mathbb R}^1 \times S^1$ & $(-,-,+)$ & 2 points & domain wall & $S^1$ & SG kink \\
& & $n_3 = \pm 1$ & $\pi_0$, 1 & $n_1^2 + n_2^2 =1$ 
& $\pi_1$, 1 \\ \hline
%%%%%
${\mathbb R}^1 \times S^1$ & $(-,-,-)$ & non & space-filling & $S^2$ & lump \\
& &  $\{0\}$ & ``$\pi_{-1}$", 0 &  $n_1^2 + n_2^2 + n_3^2=1$ 
& $\pi_2$, 2 \\ \hline
\hline
%%%%%%%%%%%%%%%%%%%%%%%%%%%%%%%%%
%%%%
${\mathbb R}^2 \times S^1$ & $(-,+,+,+)$ & $S^2$ & monopole & 2 points & 
Ising spin \\
& & $n_2^2 + n_3^2 + n_4^2 =1$ & $\pi_2$, 3 & $n_1 = \pm 1$ 
& $\pi_0$, 0 \\ \hline 
%%%%
${\mathbb R}^2 \times S^1$ & $(-,-,+,+)$ & $S^1$ & vortex & $S^1$ & SG kink \\
& & $n_3^2 + n_4^2 =1$ & $\pi_1$, 2 & $n_1^2 + n_2^2 =1$ 
& $\pi_1$, 1 \\ \hline
%%%%
${\mathbb R}^2 \times S^1$ & $(-,-,-,+)$ & 2 points & domain wall & $S^2$ & lump \\
& & $n_4 = \pm 1$ & $\pi_0$, 1 & $n_1^2 + n_2^2 + n_3^2=1$ 
& $\pi_2$, 2 \\ \hline
%%%%%
${\mathbb R}^2 \times S^1$ & $(-,-,-,-)$ & non & space-filling & $S^3$  & Skyrmion \\
& & $\{0\}$ & ``$\pi_{-1}$", 0 &   $n_1^2 + n_2^2 + n_3^2  + n_4^2 =1$ 
& $\pi_3$, 3 \\ \hline
\end{tabular}
\caption{
Fractional instantons in 
the $O(3)$ model on ${\mathbb R}^1 \times S^1$ and
the $O(4)$ model on ${\mathbb R}^2 \times S^1$ 
with twisted boundary conditions. 
SG denotes sine-Gordon.
Host solitons are classified by 
$\pi_n({\cal N})$, 
where ${\cal N}$ is a fixed manifold. 
Daughter solitons are classified 
by $\pi_m({\cal M})$, where 
$M$ is a moduli space of a host soliton.
Daughter solitons are all half quantized 
carrying a half topological charge. 
There are the relations among 
the dimensionality of the homotopy groups, 
$n+m +1 = 2$ for the $O(3)$ model and
$n+m +1 = 3$ for the $O(4)$ model. 
Equivalently, the sum of codimensions of a host soliton 
and of a daughter soliton 
is 2 and 3 for the  
$O(3)$ model on ${\mathbb R}^1 \times S^1$ and
the $O(4)$ model on ${\mathbb R}^2 \times S^1$, 
respectively.  
\label{table:summary}}
\end{table}
%%%%%%%%%%%%%%%%%%%%%
\begin{figure}
\begin{center}
\begin{tabular}{c|cccc}
&
$(+1,+\1{2},+\1{2})$ & 
$(-1,-\1{2},+\1{2})$ &
$(-1,+\1{2},-\1{2})$ &  
$(+1,-\1{2},-\1{2})$\\ \hline
$(-,+,+)$
&
\includegraphics[width=0.225\linewidth,keepaspectratio]{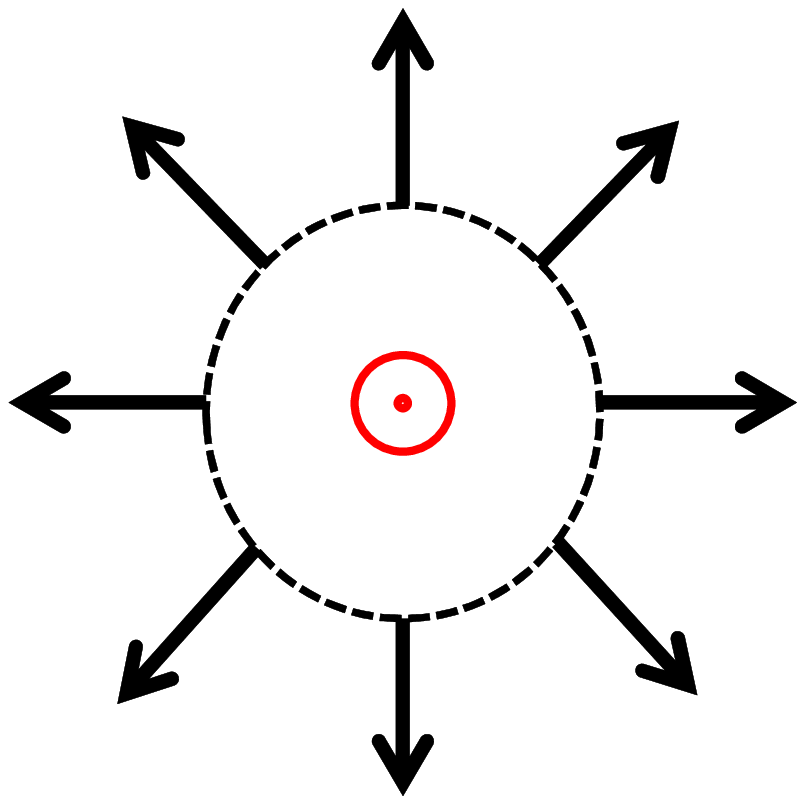} &
\includegraphics[width=0.105\linewidth,keepaspectratio]{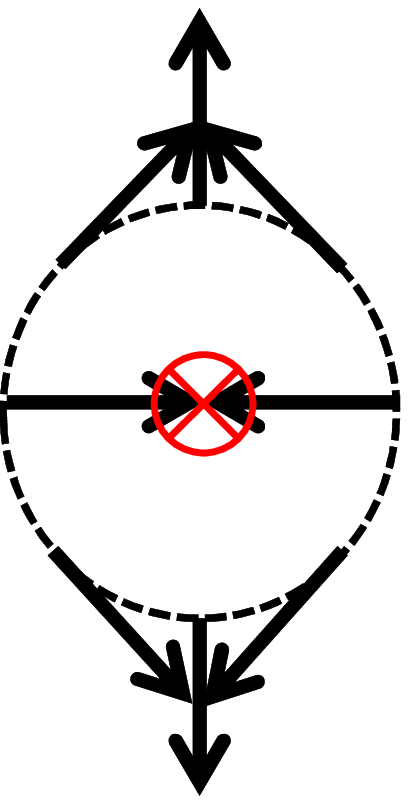} &
\includegraphics[width=0.105\linewidth,keepaspectratio]{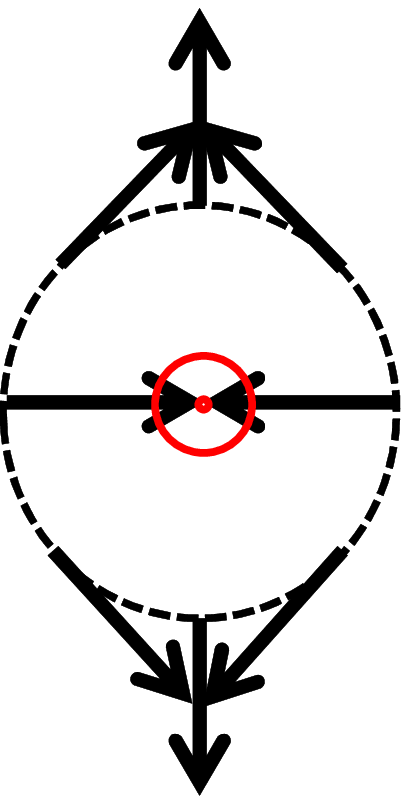} &
\includegraphics[width=0.225\linewidth,keepaspectratio]{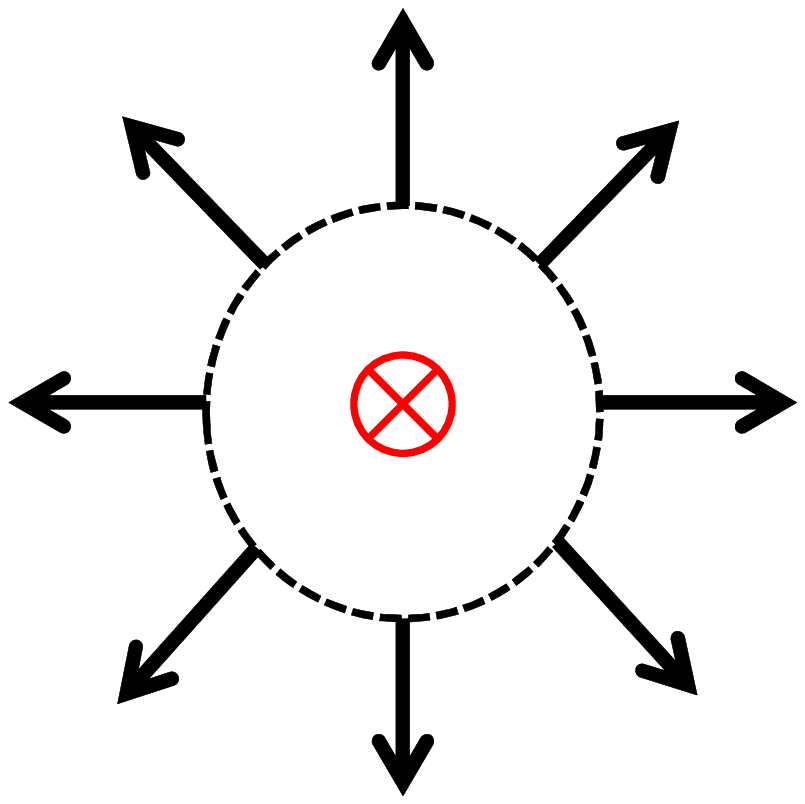} \\
&
(1a) & (1b) & (1c) & (1d)\\ \hline
$(-,-,+)$ &
\includegraphics[width=0.19\linewidth,keepaspectratio]{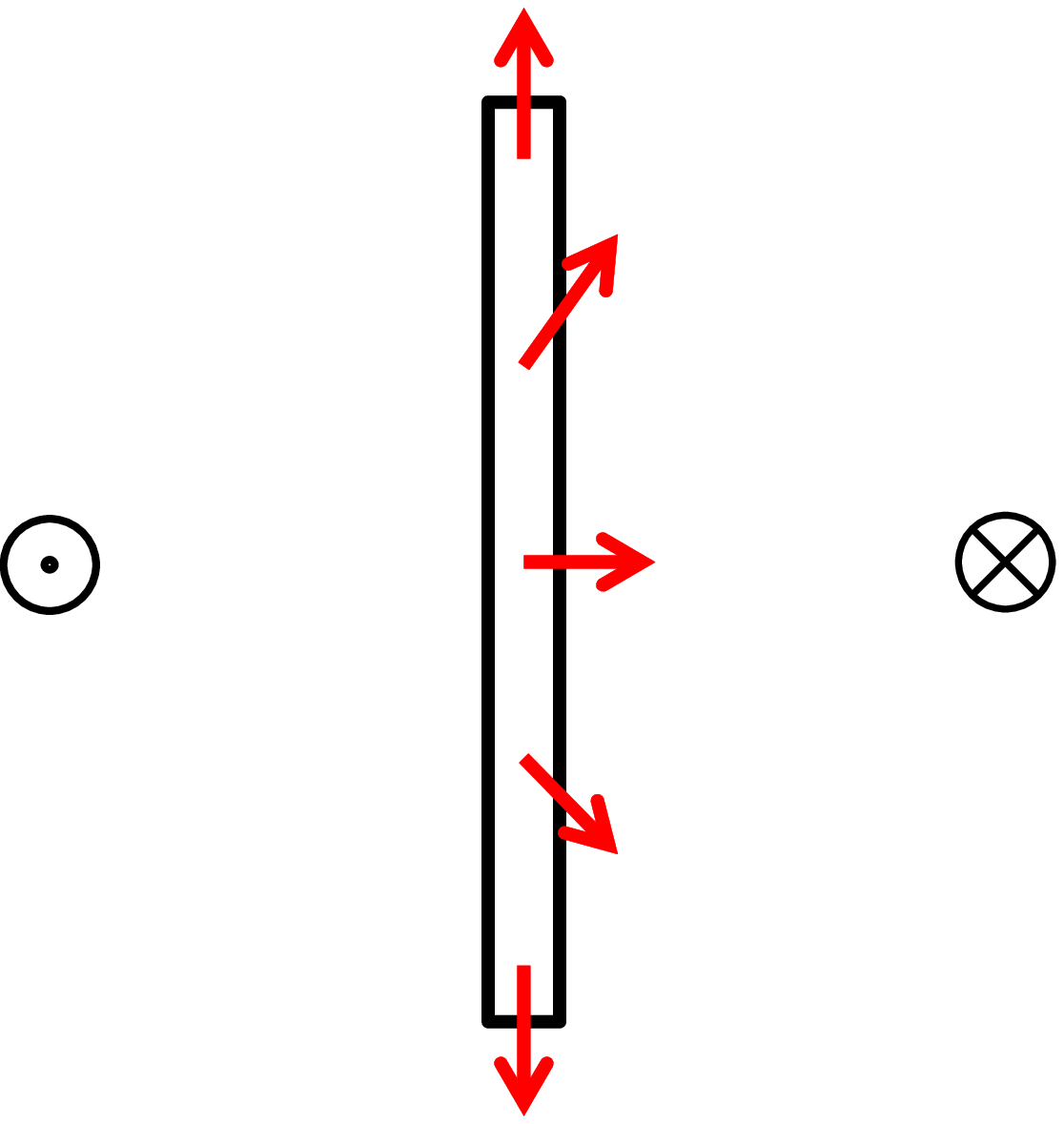} &
\includegraphics[width=0.19\linewidth,keepaspectratio]{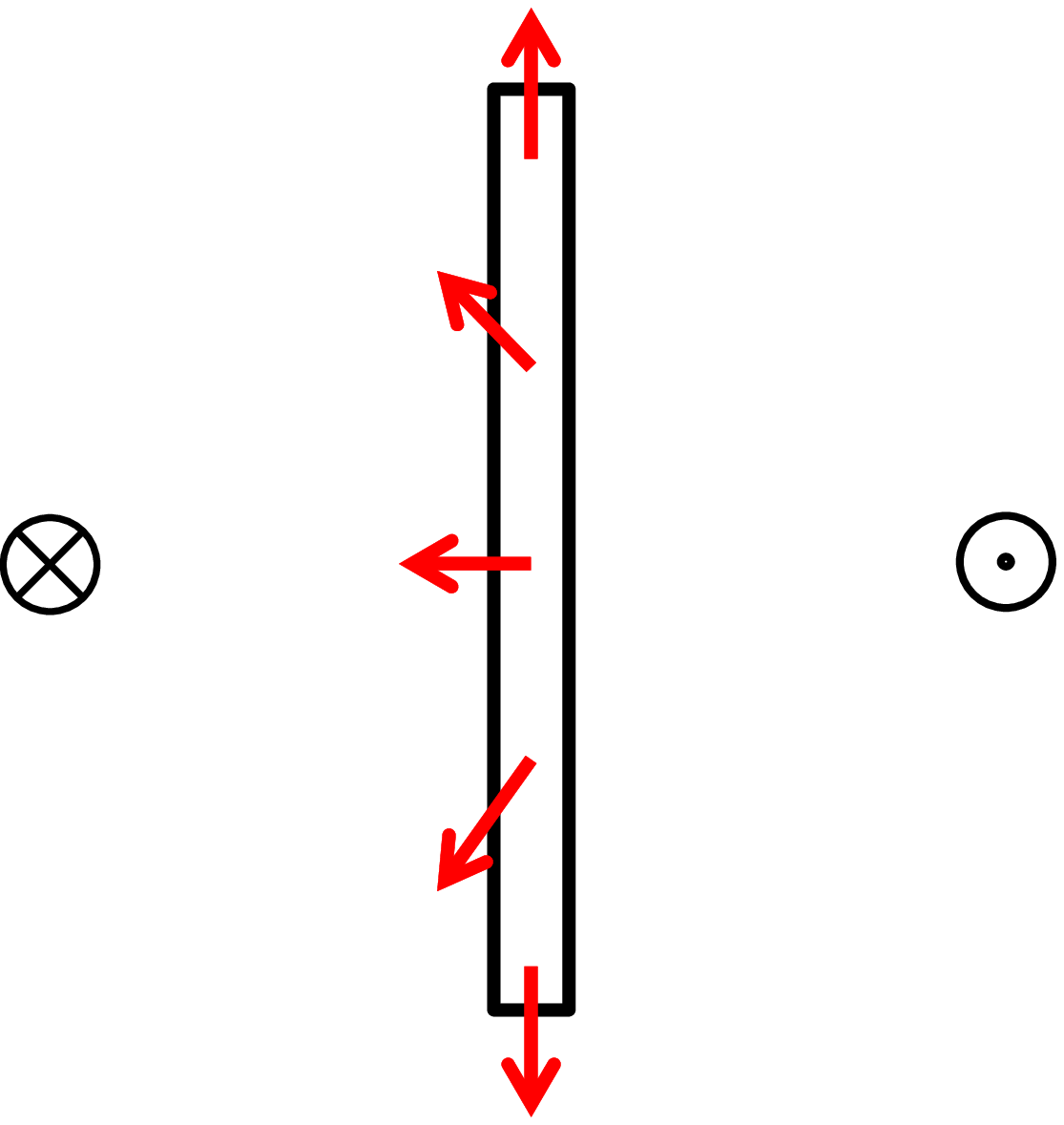} &
\includegraphics[width=0.19\linewidth,keepaspectratio]{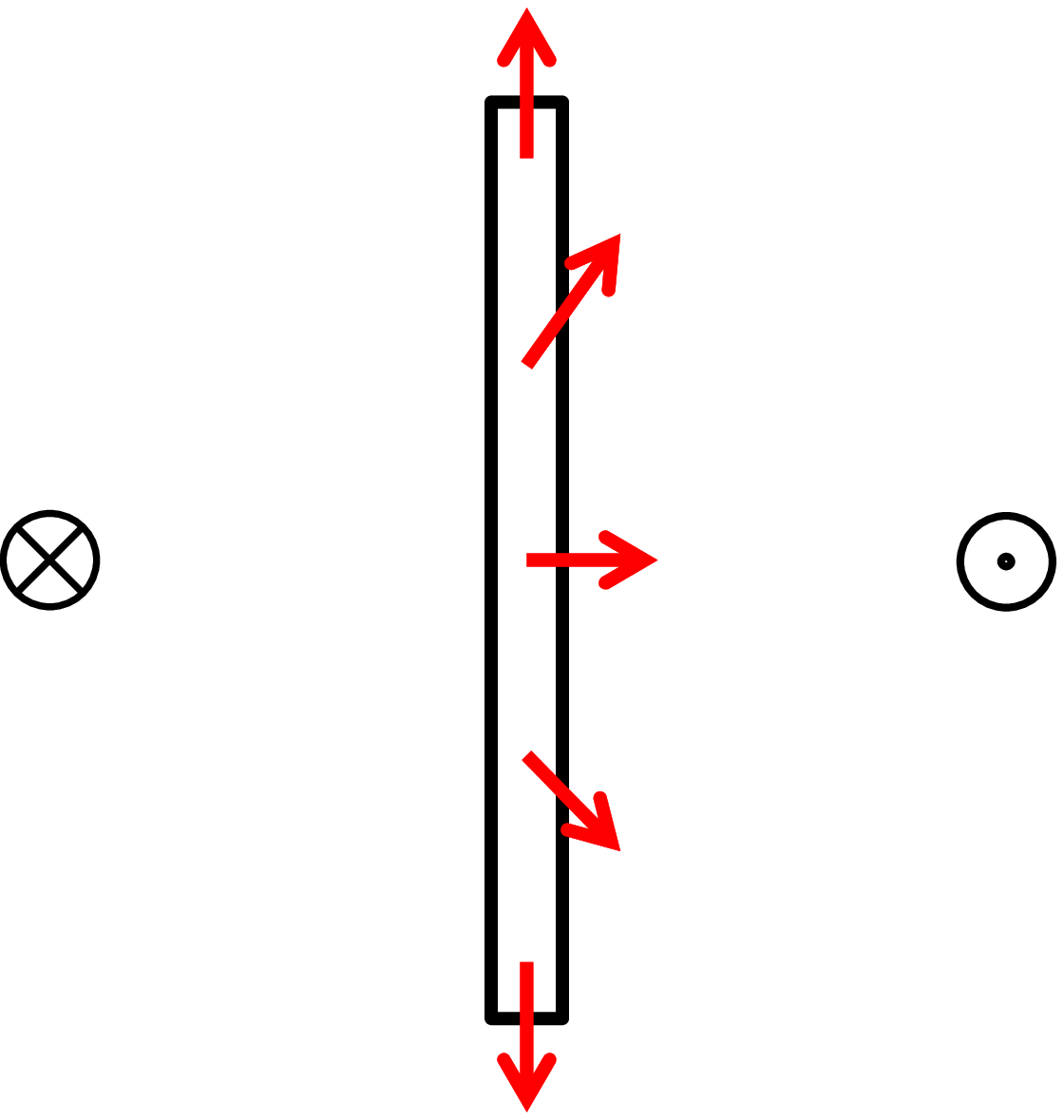} &
\includegraphics[width=0.19\linewidth,keepaspectratio]{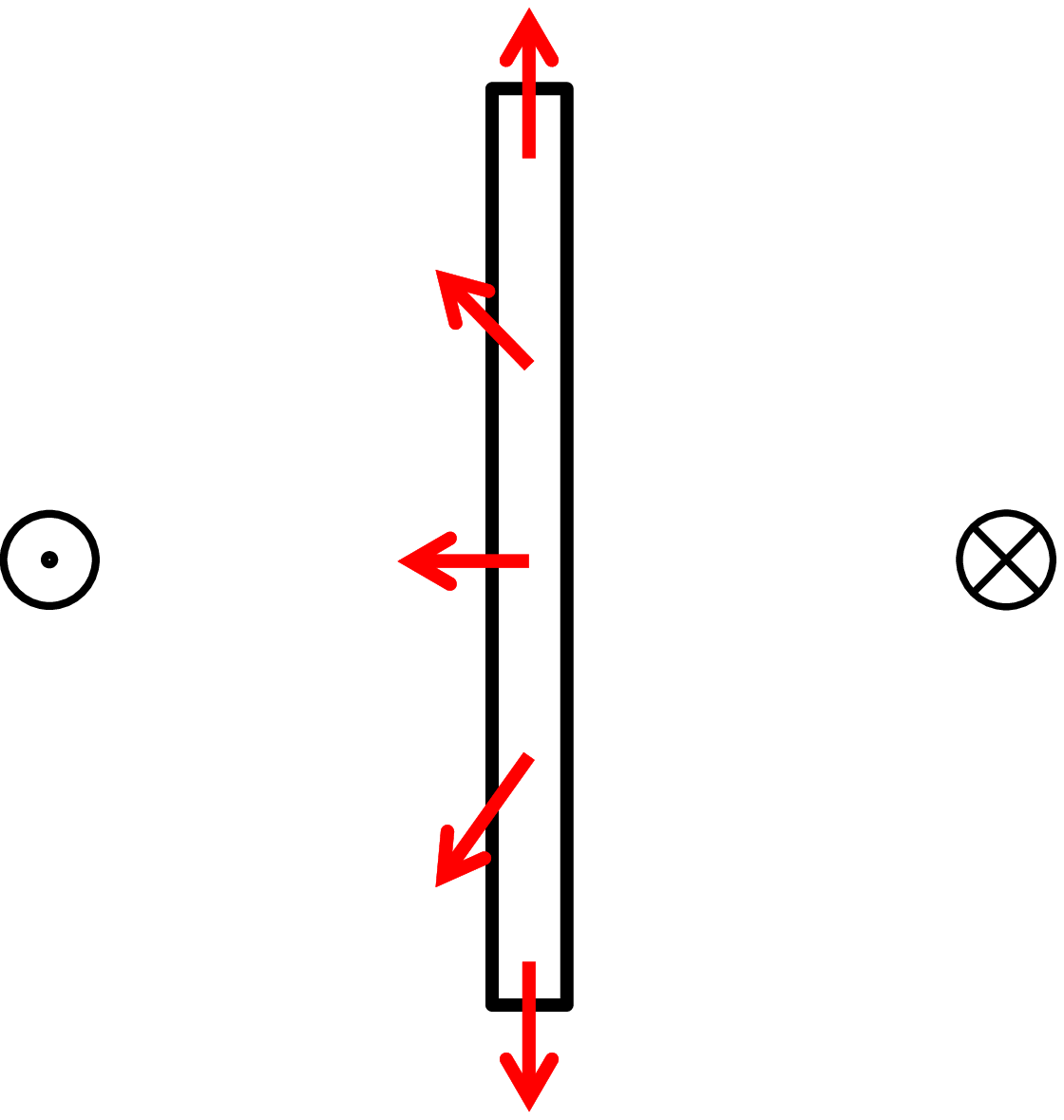} \\
&
(2a) & (2b) & (2c) & (2d)\\ \hline
$(-,-,-)$&
\includegraphics[width=0.225\linewidth,keepaspectratio]{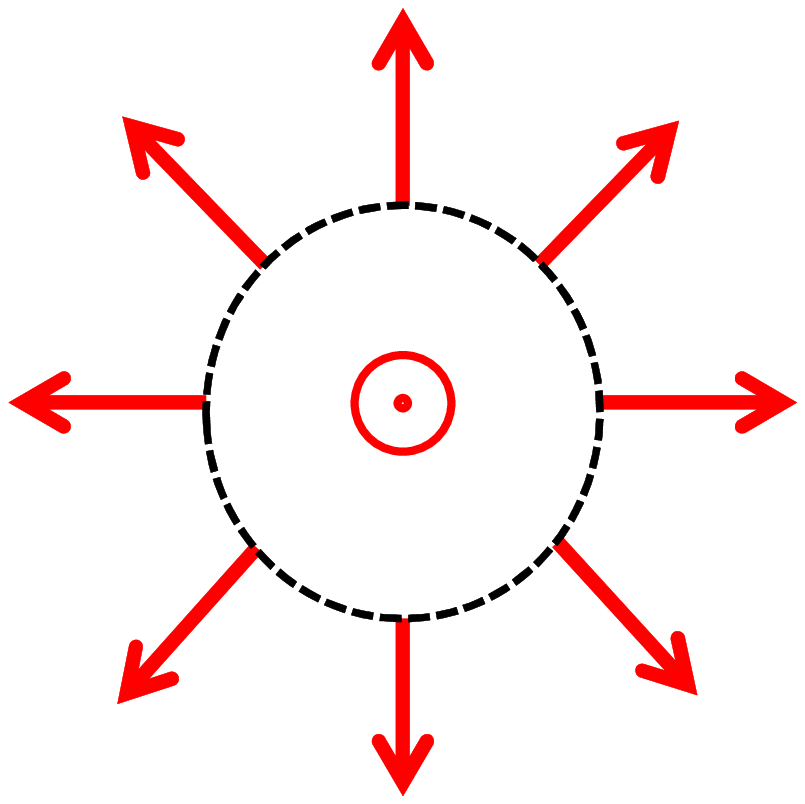} &
\includegraphics[width=0.105\linewidth,keepaspectratio]{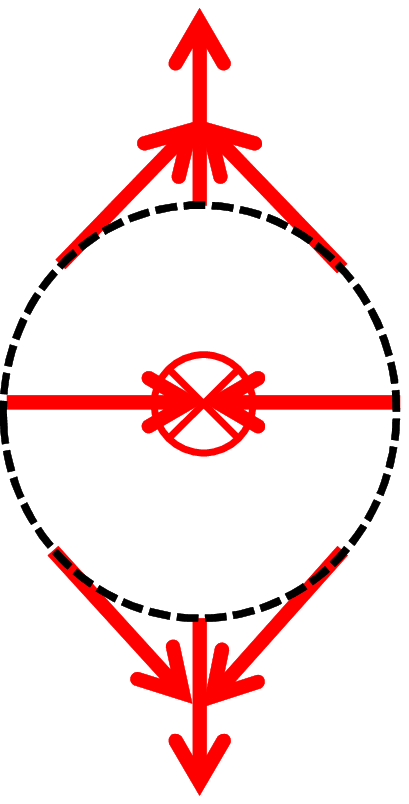} &
\includegraphics[width=0.105\linewidth,keepaspectratio]{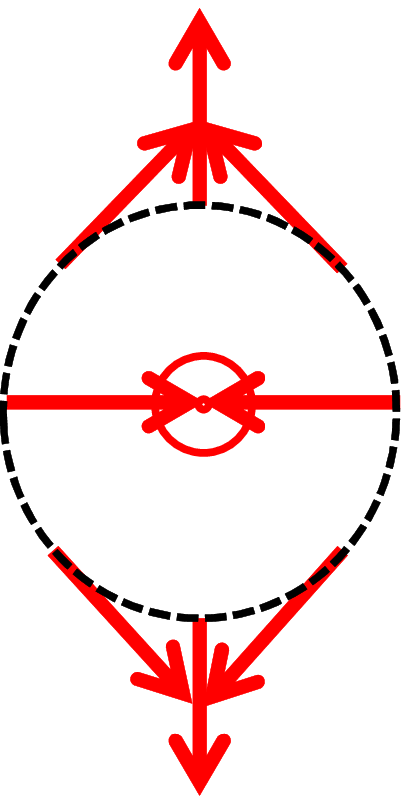} &
\includegraphics[width=0.225\linewidth,keepaspectratio]{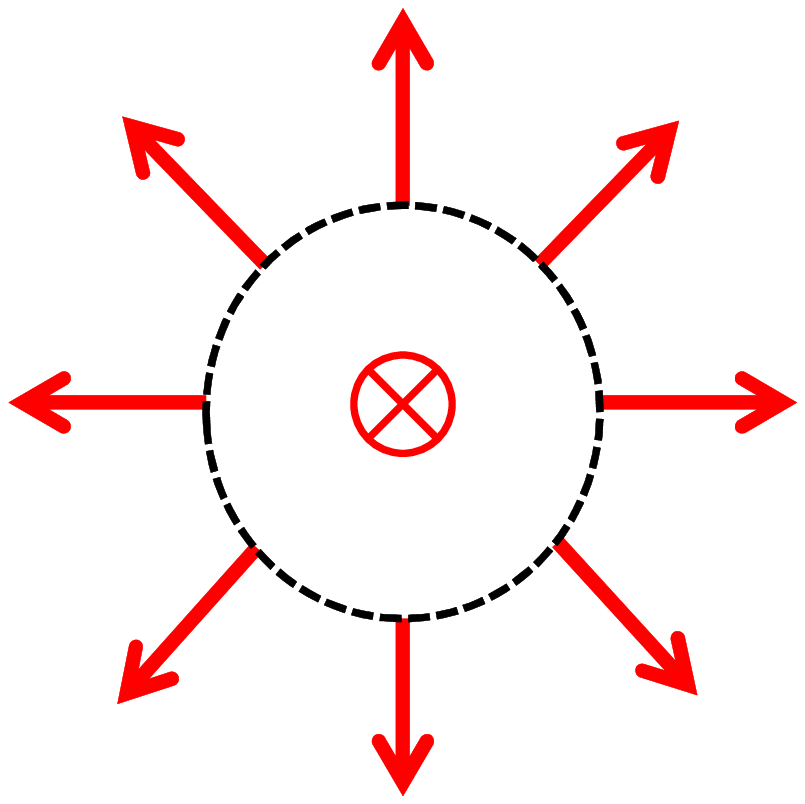} \\
&
(3a) & (3b) & (3c) & (3d)
\end{tabular}
\end{center}
\caption{
Fractional instantons in the $O(3)$ model 
with the twisted boundary conditions 
(1) $(-,+,+)$, (2) $(-,-,+)$ and (3) $(-,-,-)$. 
Black and red arrows denote the moduli space 
${\cal N}$ of vacua and the moduli space ${\cal M}$ of a host soliton, 
respectively, 
as we explain in more detail in later sections.  
The first lines indicate 
the topological charges (homotopy groups) 
characterizing (a host soliton, a daughter soliton, the total instanton charge) 
are 
$(\pi_{1},\pi_{0},\pi_2)$ for (1a)--(1d),
$(\pi_{0},\pi_{1},\pi_2)$ for (2a)--(2d), and
$(\pi_{-1},\pi_{2},\pi_2)$ for (3a)--(3d),
where $\pi_{-1}$ is merely formal. 
For each boundary condition, fractional (anti-)instantons can make 
following composite structures:   
(a)+(b) instanton, (c)+(d) anti-instanton, 
(a)+(c), (b)+(d) bions.
\label{fig:O(3)}
}
\end{figure}
%%%%%%%%%%%%%%%%%%%%%%
%%%%%%%%%%%%%%%%%%%%%%

%%%%%%%%%%%%%%%%%%%%%
\begin{figure}
\begin{center}
\begin{tabular}{c|cccc}
&
$(+1,+\1{2},+\1{2})$ & 
$(-1,-\1{2},+\1{2})$ &
$(-1,+\1{2},-\1{2})$ &  
$(+1,-\1{2},-\1{2})$\\ \hline
$(-,+,+,+)$ &
\includegraphics[width=0.24\linewidth,keepaspectratio]{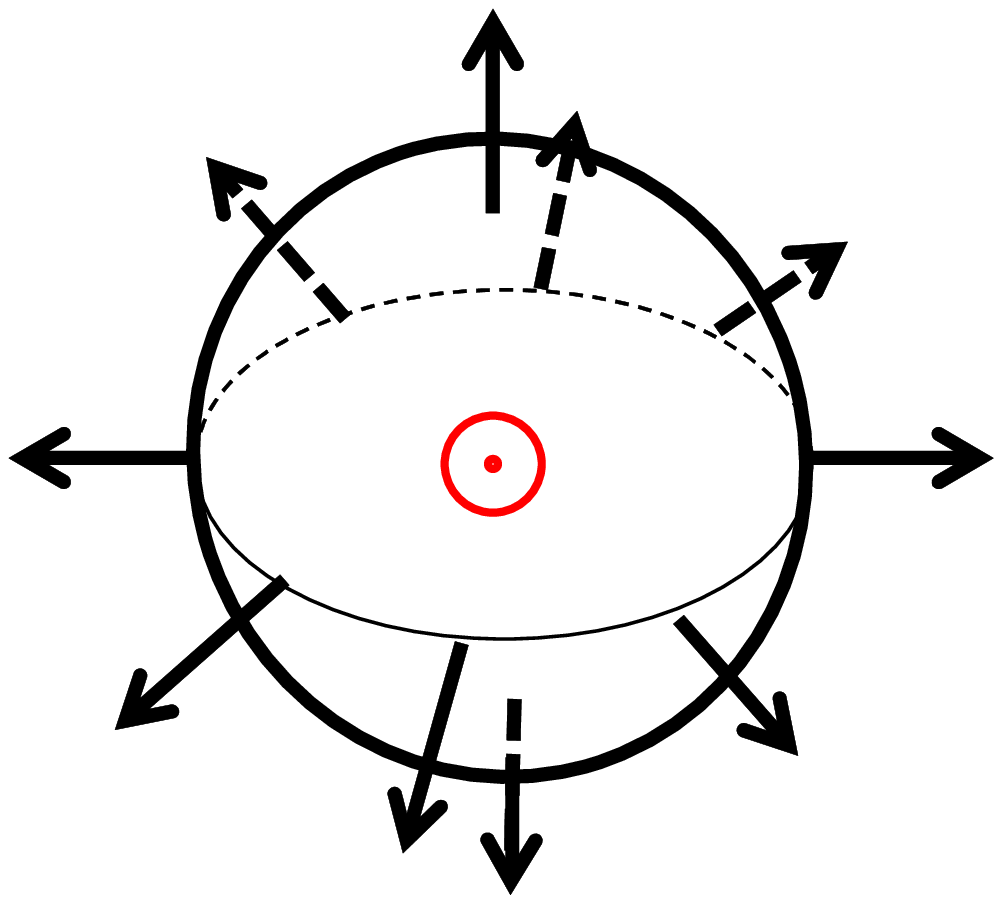} &
\includegraphics[width=0.14\linewidth,keepaspectratio]{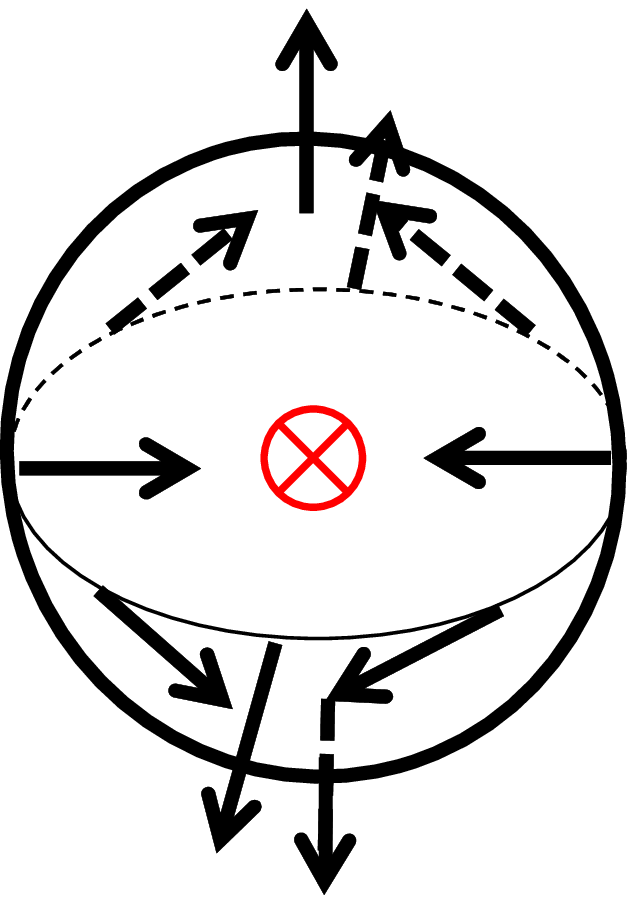} &
\includegraphics[width=0.14\linewidth,keepaspectratio]{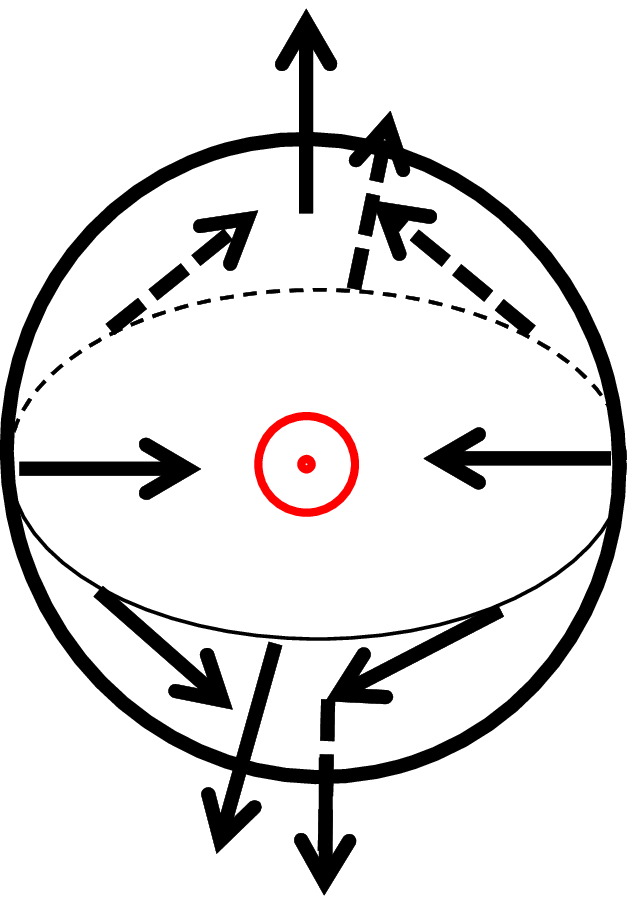} &
\includegraphics[width=0.24\linewidth,keepaspectratio]{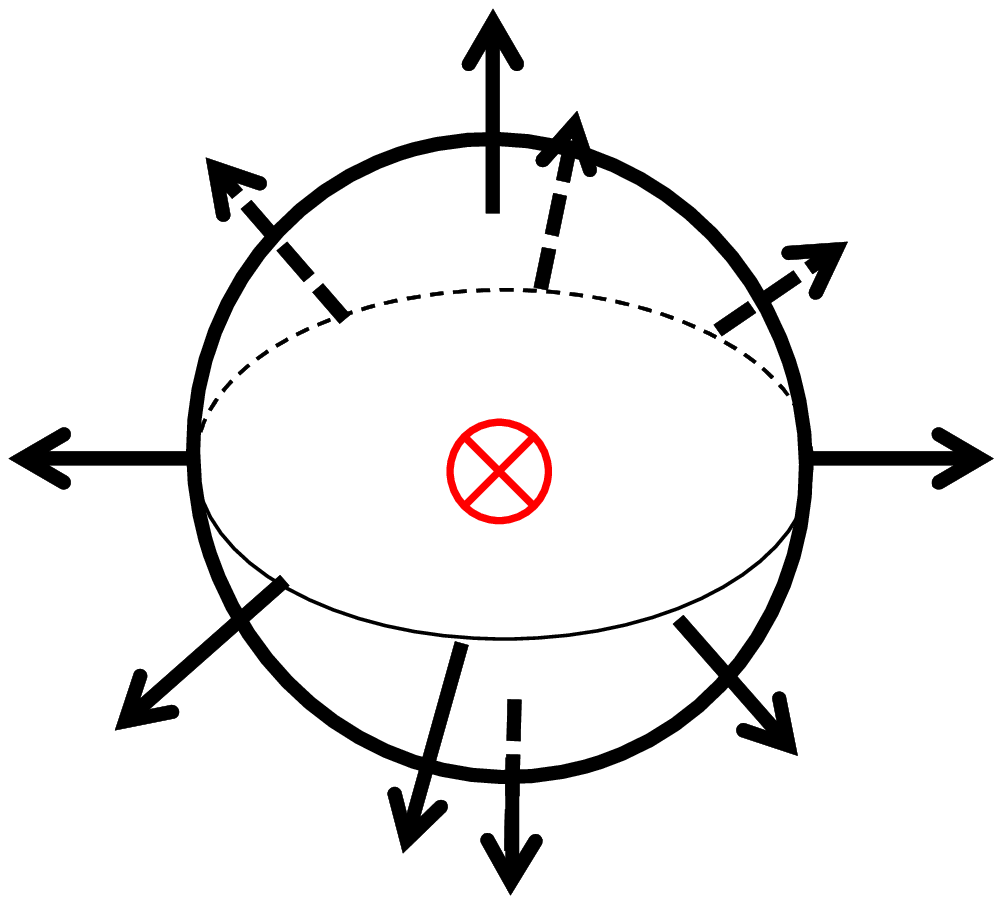} \\
&(1a) & (1b) & (1c) & (1d)\\ \hline
$(-,-,+,+)$&
\includegraphics[width=0.19\linewidth,keepaspectratio]{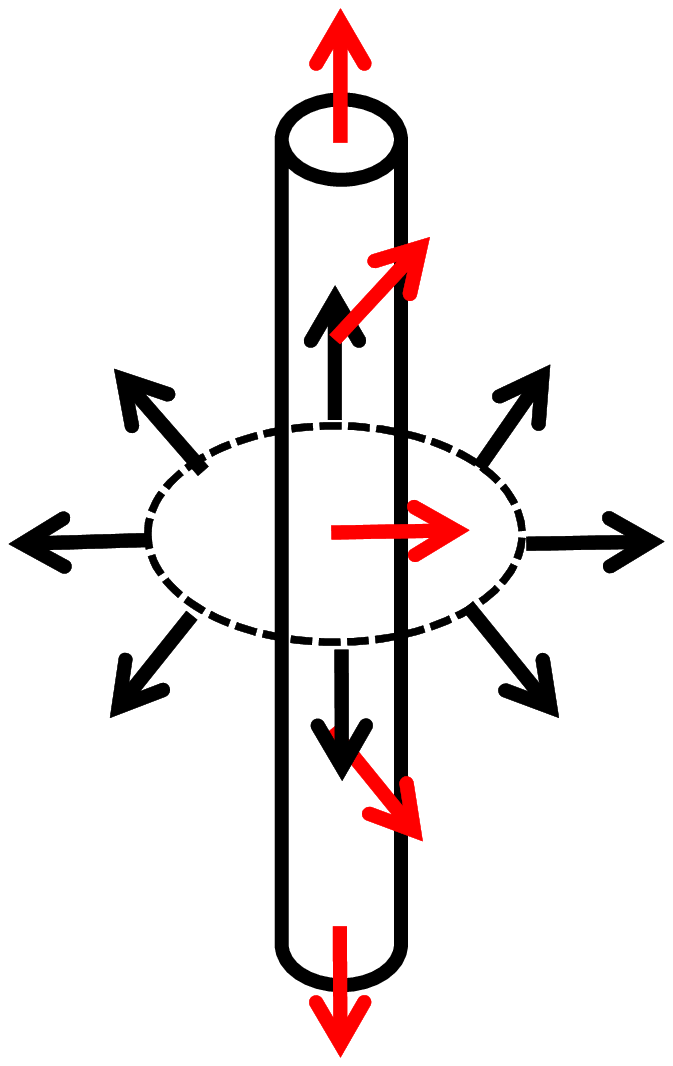} &
\includegraphics[width=0.094\linewidth,keepaspectratio]{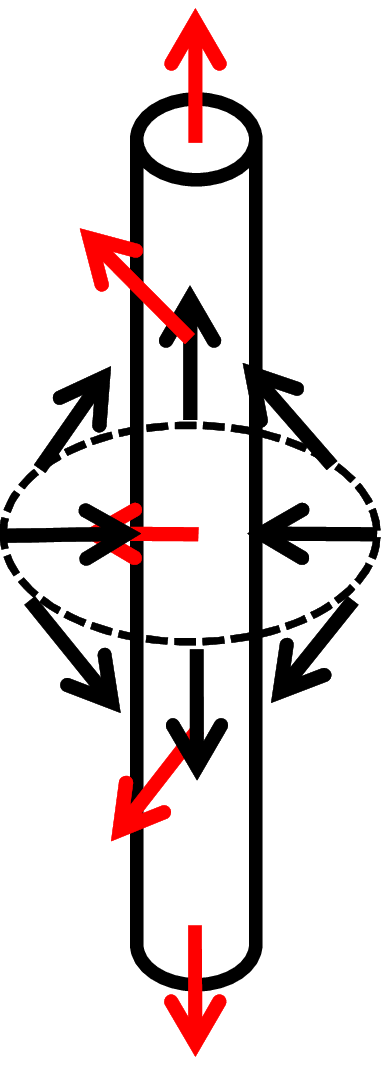} &
\includegraphics[width=0.094\linewidth,keepaspectratio]{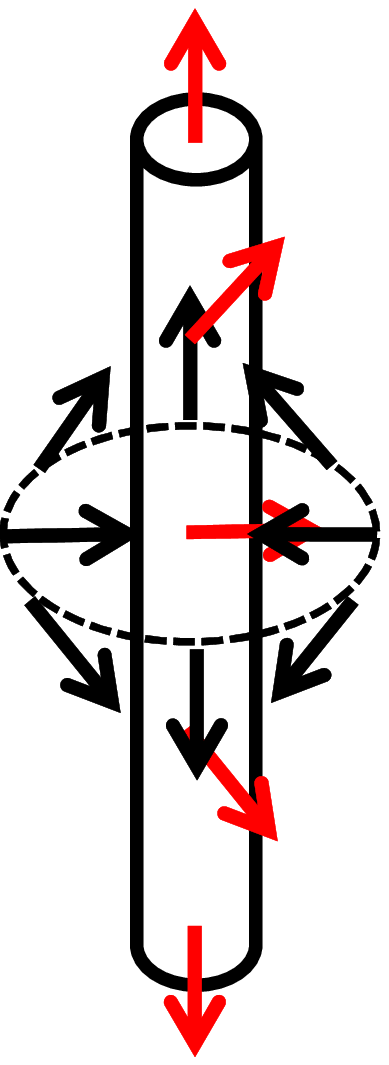} &
\includegraphics[width=0.19\linewidth,keepaspectratio]{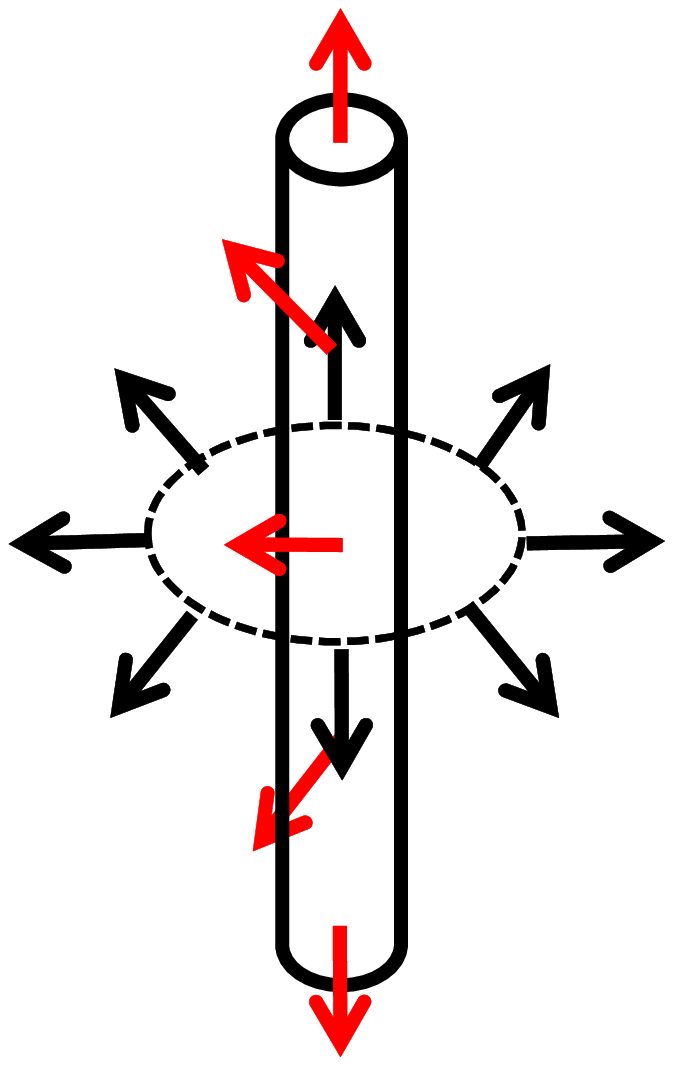} \\
&(2a) & (2b) & (2c) & (2d)\\ \hline
$(-,-,-,+)$&
\includegraphics[width=0.19\linewidth,keepaspectratio]{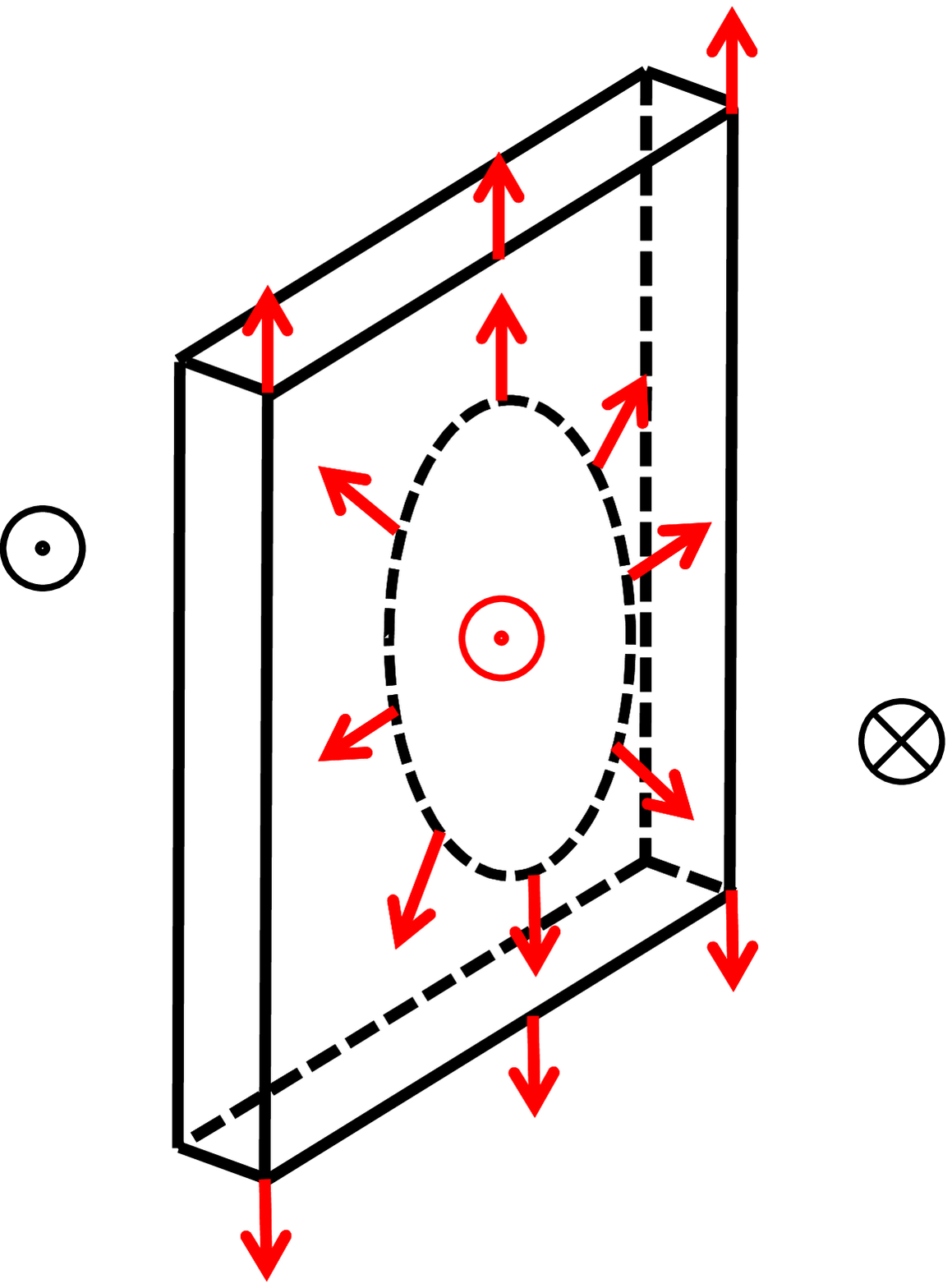} &
\includegraphics[width=0.19\linewidth,keepaspectratio]{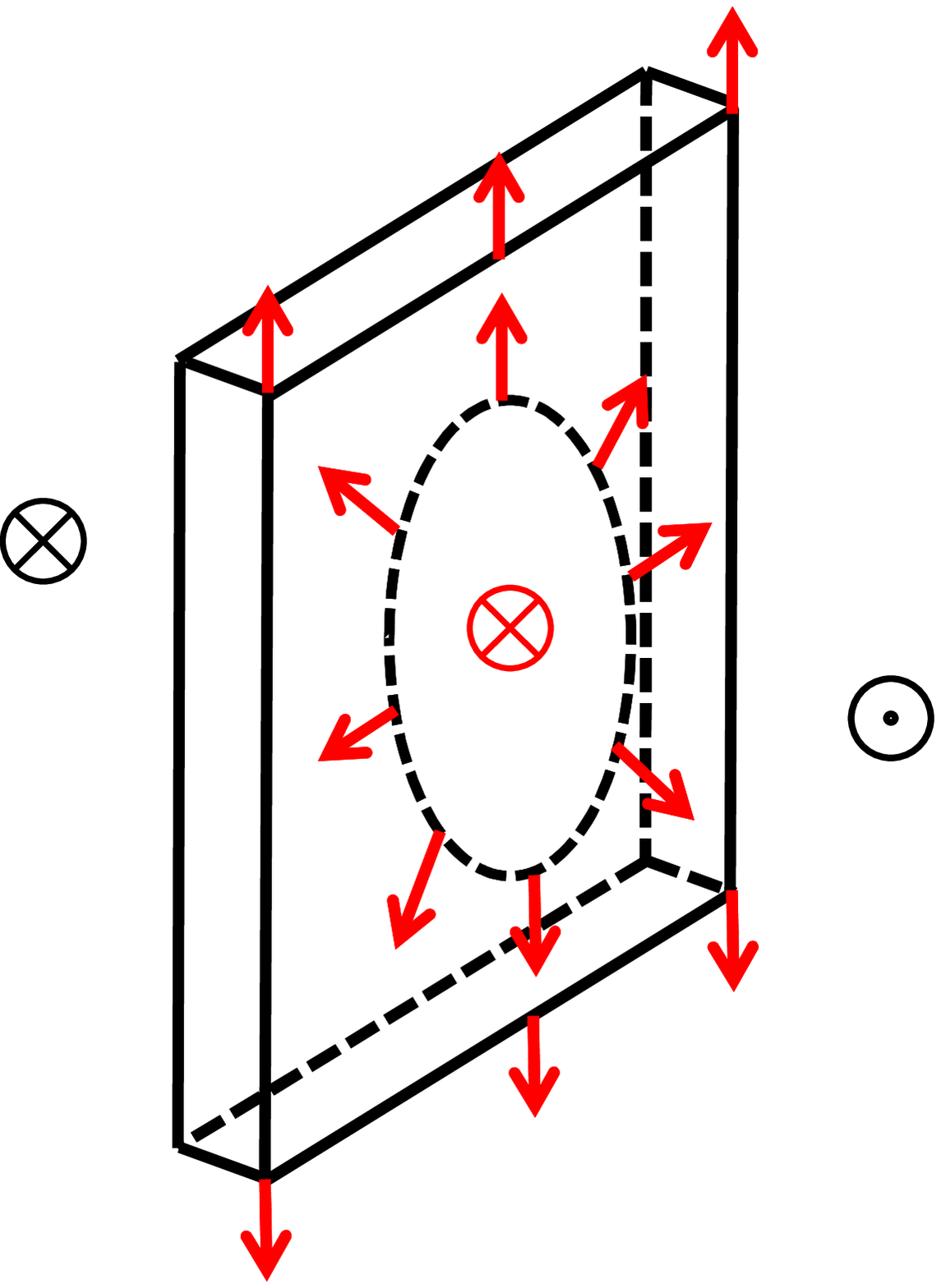} &
\includegraphics[width=0.19\linewidth,keepaspectratio]{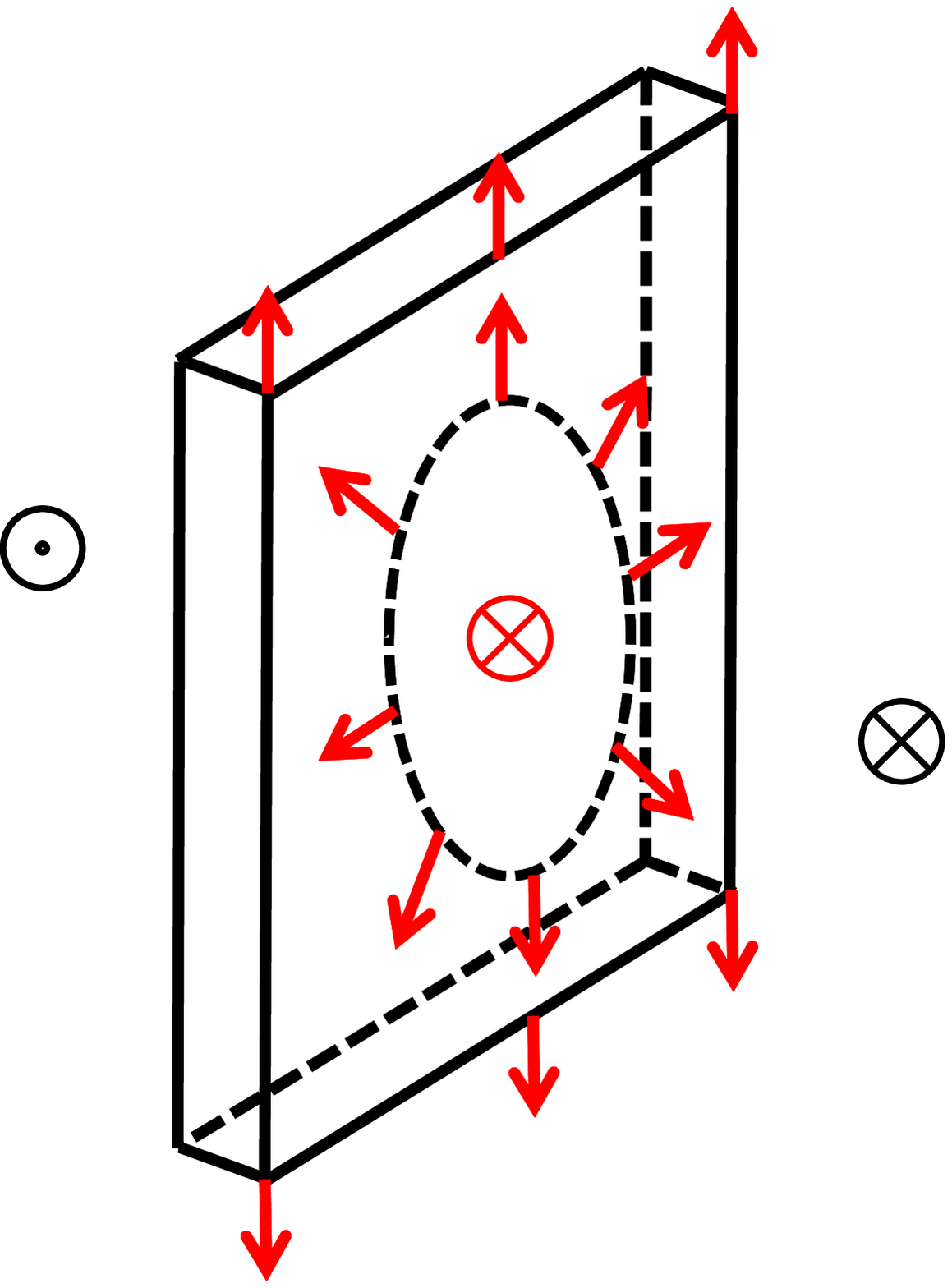} &
\includegraphics[width=0.19\linewidth,keepaspectratio]{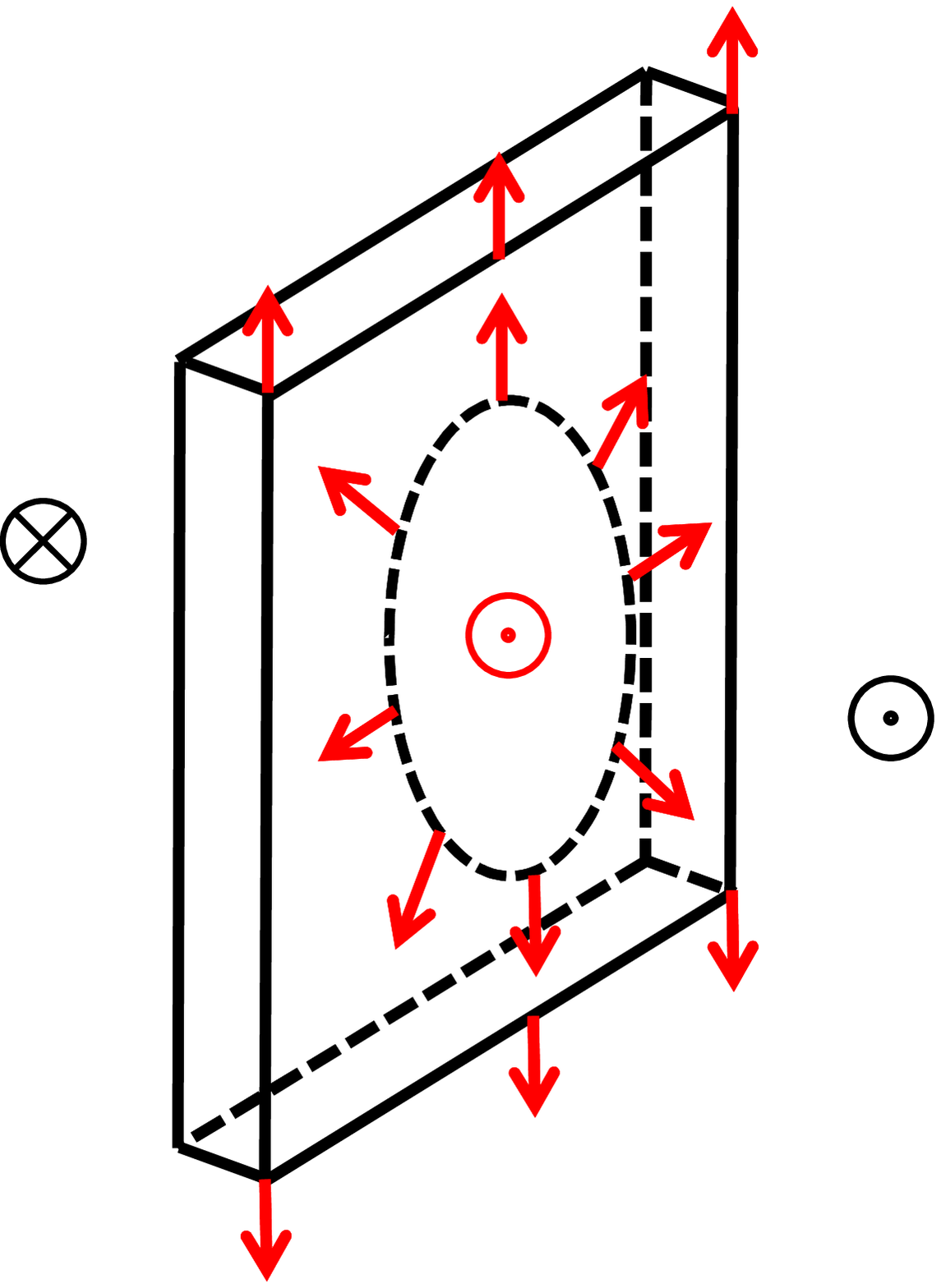} \\
&(3a) & (3b) & (3c) & (3d)\\ \hline
$(-,-,-,-)$&
\includegraphics[width=0.24\linewidth,keepaspectratio]{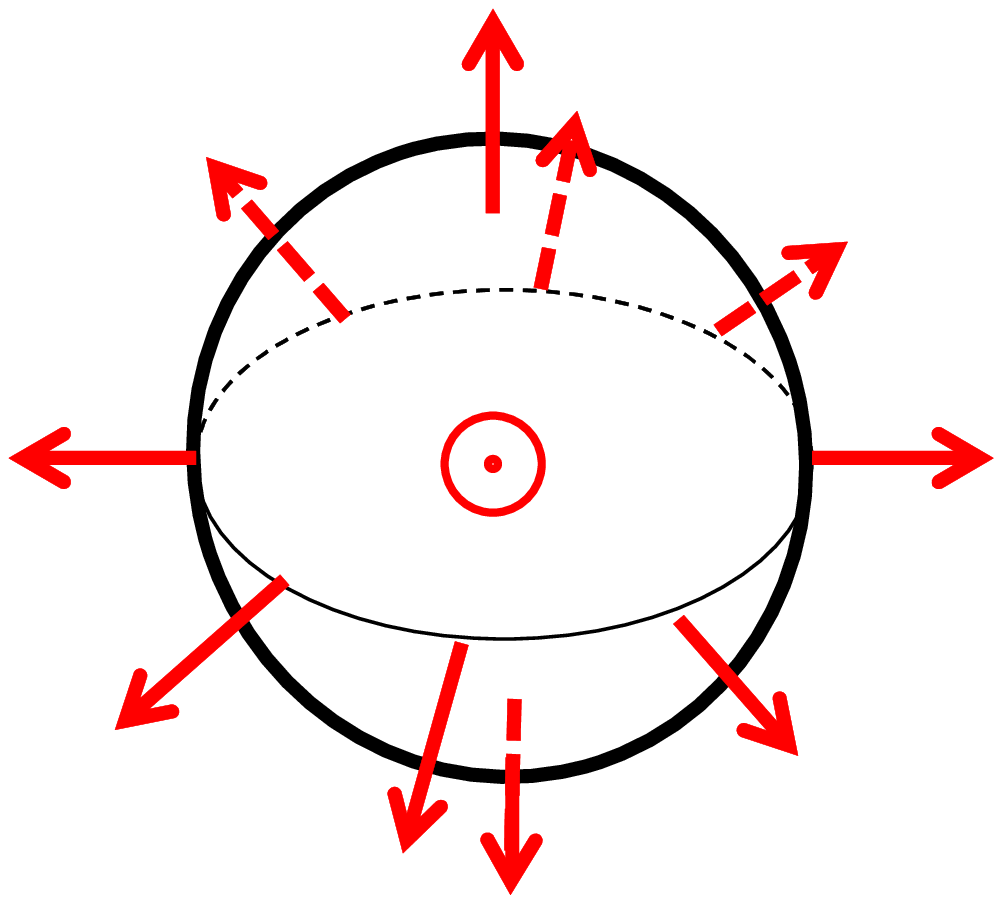} &
\includegraphics[width=0.14\linewidth,keepaspectratio]{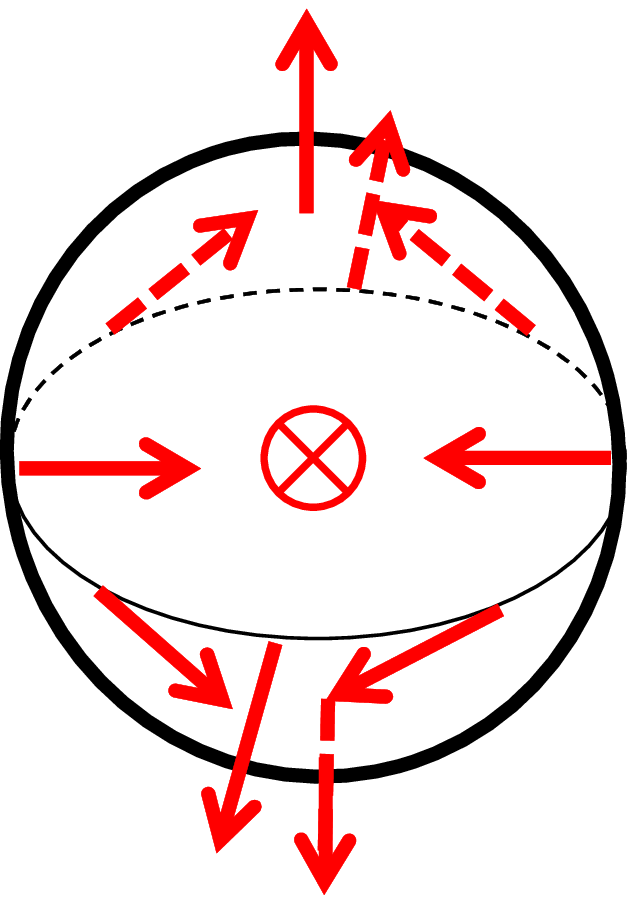} &
\includegraphics[width=0.14\linewidth,keepaspectratio]{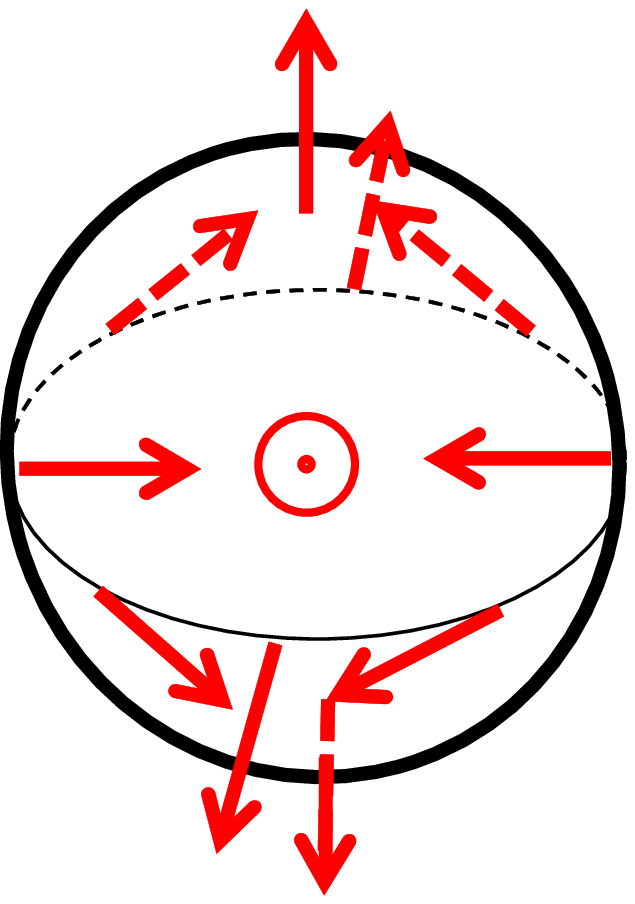} &
\includegraphics[width=0.24\linewidth,keepaspectratio]{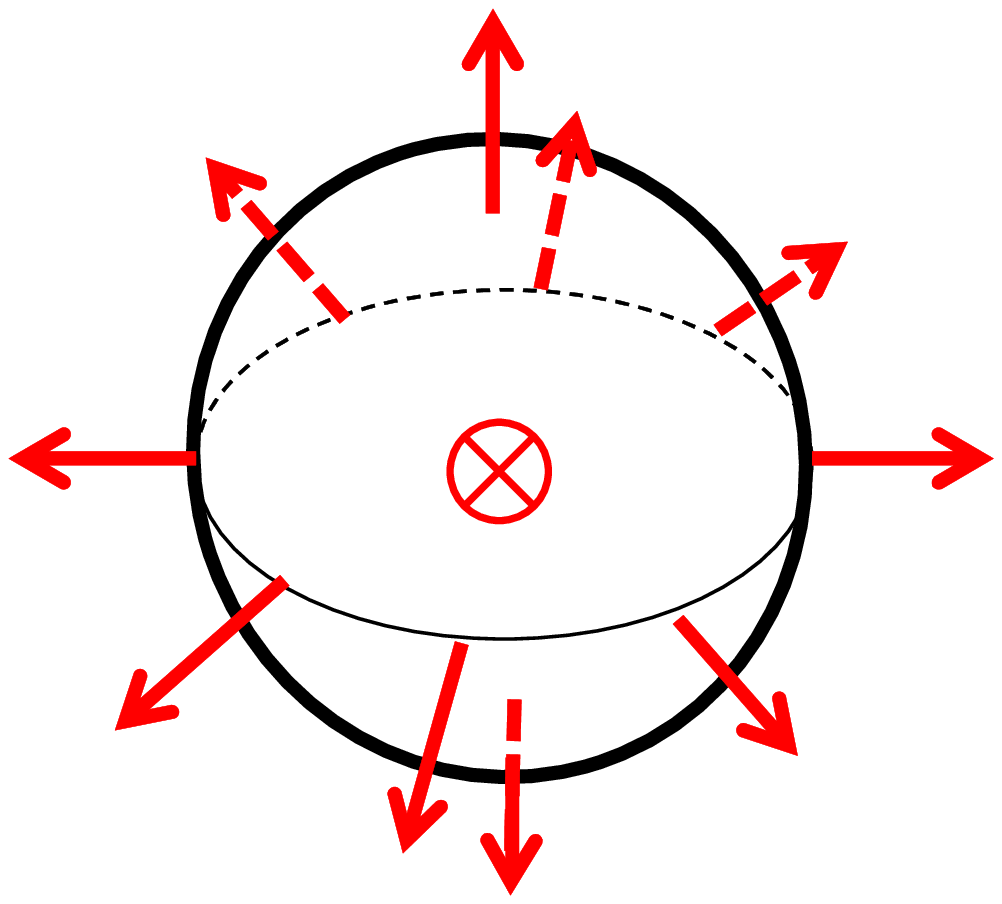} \\
&(4a) & (4b) & (4c) & (4d)
\end{tabular}
\end{center}
\caption{
Fractional instantons in the $O(4)$ model
with the twisted boundary conditions 
(1) $(-,+,+,+)$, (2) $(-,-,+,+)$, (3) $(-,-,-,+)$ and (4) $(-,-,-,-)$.
The notations of black and red arrows are the same with  
Fig.~\ref{fig:O(3)}.
The first lines indicate 
the topological charges (homotopy groups) 
characterizing (a host soliton, a daughter soliton, the total instanton charge) 
are 
$(\pi_{2},\pi_{0},\pi_3)$ for (1a)--(1d),
$(\pi_{1},\pi_{1},\pi_3)$ for (2a)--(2d), 
$(\pi_{0},\pi_{2},\pi_3)$ for (3a)--(3d), and
$(\pi_{-1},\pi_{3},\pi_3)$ for (4a)--(4d),
where $\pi_{-1}$ is merely formal.
For each boundary condition, fractional (anti-)instantons can make 
following composite structures:   
(a)+(b) instanton, (c)+(d) anti-instanton, 
(a)+(c), (b)+(d) bions.
\label{fig:O(4)}
}
\end{figure}
%%%%%%%%%%%%%%%%%%%%%%

In this paper, 
we classify fractional instantons and bions 
in the $O(N)$ nonlinear sigma model 
on ${\mathbb R}^{N-2} \times S^1$ with 
the twisted boundary conditions 
in which arbitrary number of fields change signs:
\beq
 (\underbrace{-,-,\cdots}_s,\underbrace{+,+,\cdots}_{N-s}): \quad
 (n_1,\cdots,n_N)(x+R) = 
(-n_1,\cdots,-n_s,+ n_{s+1},\cdots,+n_N)(x), \label{eq:tbc}
\eeq
where we have labeled the boundary condition by 
a set of $N$ signs as 
$(\underbrace{-,-,\cdots}_s,\underbrace{+,+,\cdots}_{N-s})$.
The $O(3)$ model is equivalent to the ${\mathbb C}P^1$ model 
for which 
the ${\mathbb Z}_2$ twisted boundary condition 
studied before \cite{Dunne:2012ae,Dunne:2012zk,Dabrowski:2013kba,
Bolognesi:2013tya,Misumi:2014jua} 
corresponds to $(-,-,+)$, while the cases of 
 $(-,+,+)$ and  $(-,-,-)$ have not been studied before. 
The $O(4)$ model is equivalent to a principal chiral model 
with a group $SU(2)$ or a Skyrme model if four derivative 
(Skyrme) term is added \cite{Skyrme:1962vh}, 
in which fractional instantons or bions were not studied before.
We find that general boundary conditions (\ref{eq:tbc}) induce fractional instantons as various types of composite solitons. 
Our results are summarized in Table \ref{table:summary} 
and Figs.~\ref{fig:O(3)} and \ref{fig:O(4)}.
Throughout the paper, 
red (black) arrows denote fields which are (not) twisted by 
the twisted boundary condition in Eq.~(\ref{eq:tbc}).
Depending on the boundary conditions, 
a fractional instanton in the $O(3)$ model 
is found to be 
a global vortex with an Ising spin (or a half-lump vortex) for 
the boundary condition $(-,+,+)$, 
a half sine-Gordon kink on a domain wall for $(-,-,+)$,
or a half lump on a ``space-filling brane" for $(-,-,-)$.
The second case was studied before.
In the third case we formally consider a space-filling brane 
for the situation that there is no localized host soliton.
A fractional instanton in the $O(4)$ model 
is found to be 
a global monopole with an Ising spin 
(or a half-Skyrmion monopole) for $(-,+,+,+)$, 
a half sine-Gordon kink on a global vortex for $(-,-,+,+)$,
a half lump on a domain wall for $(-,-,-,+)$,
or a half Skyrmion on a ``space-filling brane" for $(-,-,-,-)$. 

By using fractional instantons,
we can construct neutral bions in the $O(N)$ model.
On the other hand, charged bions are not possible 
in the $O(N)$ model. 
We note that constituent fractional instantons of 
bions in a principal chiral model  in Refs.~\cite{Cherman:2013yfa,Cherman:2014ofa} 
are not topological because they considered 
a space ${\mathbb R}^1 \times S^1$, 
while our case on ${\mathbb R}^2 \times S^1$ is topological.  
When fractional (anti-)instantons are (anti-)
Bogomol'nyi-Prasad-Sommerfield  (BPS) \cite{Bogomolny:1975de} 
or local solitons, 
the interaction between two of them does not exist or 
is suppressed exponentially $e^{-mr}$
with the distance $r$ between them, respectively. 
In either case, the interaction between 
fractional instanton and anti-instanton is 
 exponentially suppressed, and consequently 
neutral bions will play a role in resurgence because 
the energy (action value) of bions is the sum of 
individual fractional (anti-)instantons when they are well separated.
Most of fractional instantons are not BPS except for
those of the boundary condition $(-,-,+)$ 
in the $O(3)$ model, 
which is the case studied before.
We will summarize some modifications 
which may turn fractional instantons to be local or BPS 
so that they may play a role in resurgence.

This paper is organized as follows.
In Sec.~\ref{sec:model}, we first give the $O(N)$ model.
In Sec.~\ref{sec:general}, we provide
a general framework to construct 
fractional instantons as composite solitons 
in the $O(N)$ model 
with the twisted boundary conditions. 
In Secs.~\ref{sec:O(3)} and \ref{sec:O(4)}, 
we discuss fractional instantons and bions in 
the $O(3)$ model on ${\mathbb R}^1\times S^1$ and 
the $O(4)$ model on ${\mathbb R}^2\times S^1$, respectively, 
with the twisted boundary conditions.
Sec.~\ref{sec:summary} is devoted to a summary 
and discussion. 
We present a list of modification of the models 
which may make fractional instantons to be local or BPS. 

%%%%%%%%%%%%%%%%%%%%%%%%%%
\section{$O(N)$ model \label{sec:model}} 

We consider an $O(N)$ nonlinear sigma model, whose 
Lagrangian is given by
\beq
 \mathcal{L} = 
 \frac{1}{2}\p_\mu\mathbf{n}\cdot\p^\mu\mathbf{n}
+ \mathcal{L}_{\rm h.d.}  
- V(\mathbf{n})  , 
\label{eq:LO4} 
\eeq
with $N$-component scalar fields ${\bf n} = (n_1(x),n_2(x),\cdots,n_N(x))^T$ 
with a constraint ${\bf n}^2=1$. 
We have to consider higher derivative (or the Skyrme) term 
$\mathcal{L}_{\rm h.d.}$ to stabilize (fractional) instantons
in higher dimensions higher than two or three,
or two dimensions with a potential term.
In some cases, we also consider a potential term
$V(\mathbf{n})$ for the stability of fractional instantons. 
We compactify the $x^{N-1}$ coordinate to 
$S^1$ with a period $R$. 

The target space of the model is $M \simeq S^{N-1}$
\beq
\pi_{N-1}(S^{N-1}) \simeq \mathbb{Z}, \label{eq:total-homotopy}
\eeq
which admits topological textures, sine-Gordon kinks ($N=1$), 
lumps \cite{Polyakov:1975yp} or baby Skyrmions \cite{Piette:1994ug,Weidig:1998ii} ($N=2$), 
Skyrmions \cite{Skyrme:1962vh} ($N=3$).
The topological instanton charges $\pi_2(S^2), \pi_3(S^3)$ 
can be written as 
\beq
Q_2 
&=& -\frac{1}{8 \pi^2} \int d^2x \; \epsilon^{ABC} \epsilon^{ij} 
n_A \partial_i n_B \partial_j n_C = 
-\frac{1}{8 \pi^2} \int d^2x \epsilon^{ij} {\bf n} \cdot \del_i {\bf n} \times \del_j {\bf n},\\
Q_3 &=& -\frac{1}{12 \pi^2} \int d^3x \; \epsilon^{ABCD} \epsilon^{ijk} 
n_A \partial_i n_B \partial_j n_C \partial_k n_D ,
\eeq
respectively. 
The charge $\pi_3(S^3)$ is also called the baryon number 
in the context of the Skyrme model.
In general, the instanton charge in $\pi_{N-1} (S^{N-1})$ 
for the $O(N)$ model is given by 
(see, e.g., Ref.~\cite{Gudnason:2014uha})
\beq
Q_{N-1} = 
-\frac{\Gamma\left(\frac{N}{2}\right)}{2\pi^{\frac{N}{2}}} 
\int d^{N-1} x \;\frac{1}{(N-1)!}
\epsilon^{i_1\cdots i_{N-1}}\epsilon^{A_1\cdots A_N}
\p_{i_1}n_{A_1} \cdots \p_{i_{N-1}} n_{A_{N-1}} n_{A_N} .
\eeq

The $O(3)$ model is equivalent to the ${\mathbb C}P^1$ model.
Let  $\phi$  be a normalized complex two vector
($\phi^\dagger \phi =1$), and consider the Hopf map 
from $S^3$ to $S^2$ by
\beq 
   n_A \equiv \phi^\dagger {\sigma}_A  \phi 
\eeq 
with the Pauli matrices ${\sigma}_A$ $(A=1,2,3)$.
Let us define 
the stereographic coordinate $u$ of $S^2$ (projective coordinate 
of the ${\mathbb C}P^1$) by 
\beq 
   \phi^T = (1, u)^T / \sqrt{1 + |u|^2}.
\eeq 
In terms of $u$, the Lagrangian can be rewritten as
\beq
\mathcal{L} = 2 \frac{|\del_{\mu} u|^2}{(1 + |u|^2)^2}. 
\eeq
In this notation, the topological instanton charge 
can be rewritten as
\beq
 Q_2 = - \1{4 \pi^2}  \int d^2x 
{i \epsilon^{ij} \del_i u^* \del_j u 
\over (1+|u|^2)^2} .
\eeq
The boundary condition $(-,-,+)$ 
can be expressed in terms of $\phi$ and $u$ as 
\beq 
  (-,-,+):&& 
\quad 
  \phi (x+R) = W \phi (x), \quad 
  W \equiv \sigma_3= {\rm diag.} (1,-1) \\
&&\quad u(x+R)=-u(x).
\eeq

The $O(4)$ model is equivalent to a principal chiral model 
with a group $SU(2)$
or the Skyrme model if four derivative term is considered.
We define an $SU(2)$-valued field $U(x)\in SU(2)$ in terms of 
four reals scalar fields $n_A(x)$ ($A=1,2,3,4$):
\beq
U = i \sum_{a=1,2,3} n_a \sigma_a + n_4 \mathbf{1}_2 
\eeq
where $\sigma_a$ are the Pauli matrices and
$\mathbf{n}\cdot\mathbf{n}=1$ is equivalent to
$U^\dag U = \mathbf{1}_2$. 
In terms of  $U(x)$, the Lagrangian can be rewritten as
\beq
\mathcal{L} = 
\tr (\p_{\mu}U^{\dagger} \p^{\mu} U) .
\eeq
The symmetry of the Lagrangian  is 
$\tilde G = SU(2)_{\rm L} \times SU(2)_{\rm R}$ acting on $U$ as 
$U \to U'= g_{\rm L} U g_{\rm R}^\dag$.
This symmetry is spontaneously broken down to 
$\tilde H \simeq SU(2)_{\rm L+R}$, 
which in turn acts as $U \to U'= g U g^\dag$ so
that the target space is 
$\tilde G/\tilde H \simeq SU(2)_{\rm L-R} \simeq S^3$. 
The baryon number (the Skyrme charge) of $Q_3 \in \pi_3(S^3)$ 
can be rewritten as
\beq
Q_3 &=& -\1{24\pi^2} \int d^3x \; \epsilon^{ijk} 
\tr \left( U^\dag\p_i U U^\dag\p_j U U^\dag\p_k U\right) \non
&=& \1{24\pi^2} \int d^3x \; \epsilon^{ijk} 
\tr \left( U^\dag\p_i U\p_j U^\dag\p_k U\right) .
\eeq
The boundary condition $(-,-,+,+)$ can be expressed 
in terms of $U$ as 
\beq 
  (-,-,+,+): \quad U(x+R)=  W U(x) W^\dagger, \quad 
  W = \sigma_3= {\rm diag.} (1,-1)  \label{eq:tbcO4--++}
\eeq
so that the vacuum is center symmetric.

%%%%%%%%%%%%%%%%%%%%%%%%%%%%%%%%%%%%%%%%%%%%
\section{General framework for fractional instantons 
in the $O(N)$ model} \label{sec:general}
Here, we provide a general framework to 
construct fractional instantons  
in the $O(N)$ model with  the boundary condition (\ref{eq:tbc}). 
In general, the boundary condition (\ref{eq:tbc}) 
defines a fixed manifold 
\beq
&&
  {\cal N} = \left\{\sum_{A=s+1}^{N} (n_A)^2 = 1\right\} 
   \simeq S^{N-s-1} = S^n, \quad \non
&& 
  S^0 \simeq \{n_N = \pm 1\}, \quad n \equiv N-s-1
\eeq
as the fixed points 
of the action at the boundary. 
This is nothing but the moduli space of vacua,
since the boundary condition does not induce 
the gradient energy for the fields $n_A$ ($A=s+1,\cdots,N$) 
while it does for that of the rests  $n_A$ ($A=1,\cdots,s$) .
From the homotopy group of ${\cal N}$, 
\beq
 \pi_n ({\cal N}) \simeq {\mathbb Z},
\label{eq:host-homotopy}
\eeq
one finds the existence of a host soliton (defect) in the bulk.
Here, we have formally defined $\pi_{-1}$ for a space-filling brane 
in the case of $n=-1$ ($s=N$) 
for the situation that there is no localized defects.

At the core of the defect, 
the nonzero fields in the bulk must vanish,
and the relation $\sum_{A=s+1}^{N} (n_A)^2 = 0$ holds, 
which leads
\beq
 {\cal M} = \left\{\sum_{A=1}^{s} (n_A)^2 = 1\right\} \simeq S^{s-1}  = S^m, 
  \quad m \equiv s-1.
\eeq
This is nothing but the moduli 
localized on the host soliton's world volume 
(collective coordinates of the host soliton). 
This has a non-trivial homotopy group
\beq
 \pi_m ({\cal M}) \simeq {\mathbb Z}.  \label{eq:daughter-homotopy}
\eeq
The host soliton has world volume along 
the compact direction and the rests. 
Therefore, the moduli ${\cal M}$ must be twisted along the world volume 
in the compact direction with the twisted boundary condition.
It inevitably introduces a daughter soliton, 
which, we find,  belongs to a ``half" element of 
the homotopy group in Eq.~(\ref{eq:daughter-homotopy}).
In other words, 
a homotopy group in Eq.~(\ref{eq:daughter-homotopy})
is modified by the boundary condition 
to take a value in a half integer.  
While this should be explained by 
a relative homotopy group more rigorously, 
we do not do that in this paper. 
We denote it symbolically by 
\beq 
\pi_m^{\rm b.c.}  ({\cal M})\simeq {\mathbb Z} 
+ \1{2}. \label{eq:daughter-homotopy2}
\eeq 
We thus have a composite soliton. 
Each composite soliton 
consists of a daughter soliton, 
belonging to a half element of the homotopy group 
$\pi_m^{\rm b.c.}  ({\cal M})$ 
in Eq.~(\ref{eq:daughter-homotopy2}) modified by the boundary condition,
on a host soliton, belonging to 
the unit element of the homotopy group 
$\pi_n ({\cal N})$ 
in Eq.~(\ref{eq:host-homotopy}).
Consequently, 
the total homotopy group $\pi_{N-1}(M)$
in Eq.~(\ref{eq:total-homotopy}) 
is a product of the elements in 
$\pi_n ({\cal N})$ in Eq.~(\ref{eq:host-homotopy}) and 
$\pi_m^{\rm b.c.}  ({\cal M})$ 
in Eq.~(\ref{eq:daughter-homotopy2}),
and so it
belongs to a half element of the total homotopy group 
 $\pi_{N-1}(M)$ in Eq.~(\ref{eq:total-homotopy}), 
that is, a half instanton.

The sum of codimensions of a host soliton 
and of a daughter soliton 
is $N-1$, 
which is 2 or 3 for the  
$O(3)$ model on ${\mathbb R}^1 \times S^1$ or
the $O(4)$ model on ${\mathbb R}^2 \times S^1$, 
respectively.  
Equivalently,
there exists a certain relation between  
the dimensionality of the homotopy groups: 
\beq
 n+m+1= N-s-1 + (s-1) +1 = N-1
\eeq
which is $n+m +1 = 2$ for the $O(3)$ model and
$n+m +1 = 3$ for the $O(4)$ model.

In the following sections, we discuss 
fractional instantons and bions 
in more detail for each boundary condition 
in the $O(3)$ and $O(4)$ models.

%%%%%%%%%%%%%%%%%%%%%%%%%%%%%%%%%%%%
\section{Fractional instantons and bions in the $O(3)$ model}
\label{sec:O(3)}

%%%%%%%%%%%%%%%%%%%%%%%
\subsection{$(-,+,+)$: 
global vortex with an Ising spin or half lump-vortex}

%%%%%%%%%%%%%%%%%%%%%
\begin{figure}
\begin{center}
\includegraphics[width=0.1\linewidth,keepaspectratio]{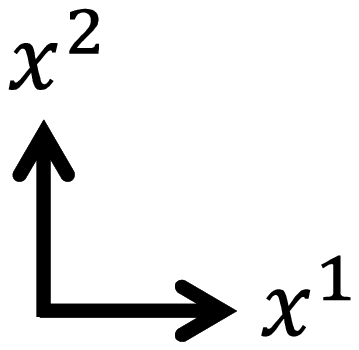}
\begin{tabular}{cc}
\includegraphics[width=0.40\linewidth,keepaspectratio]{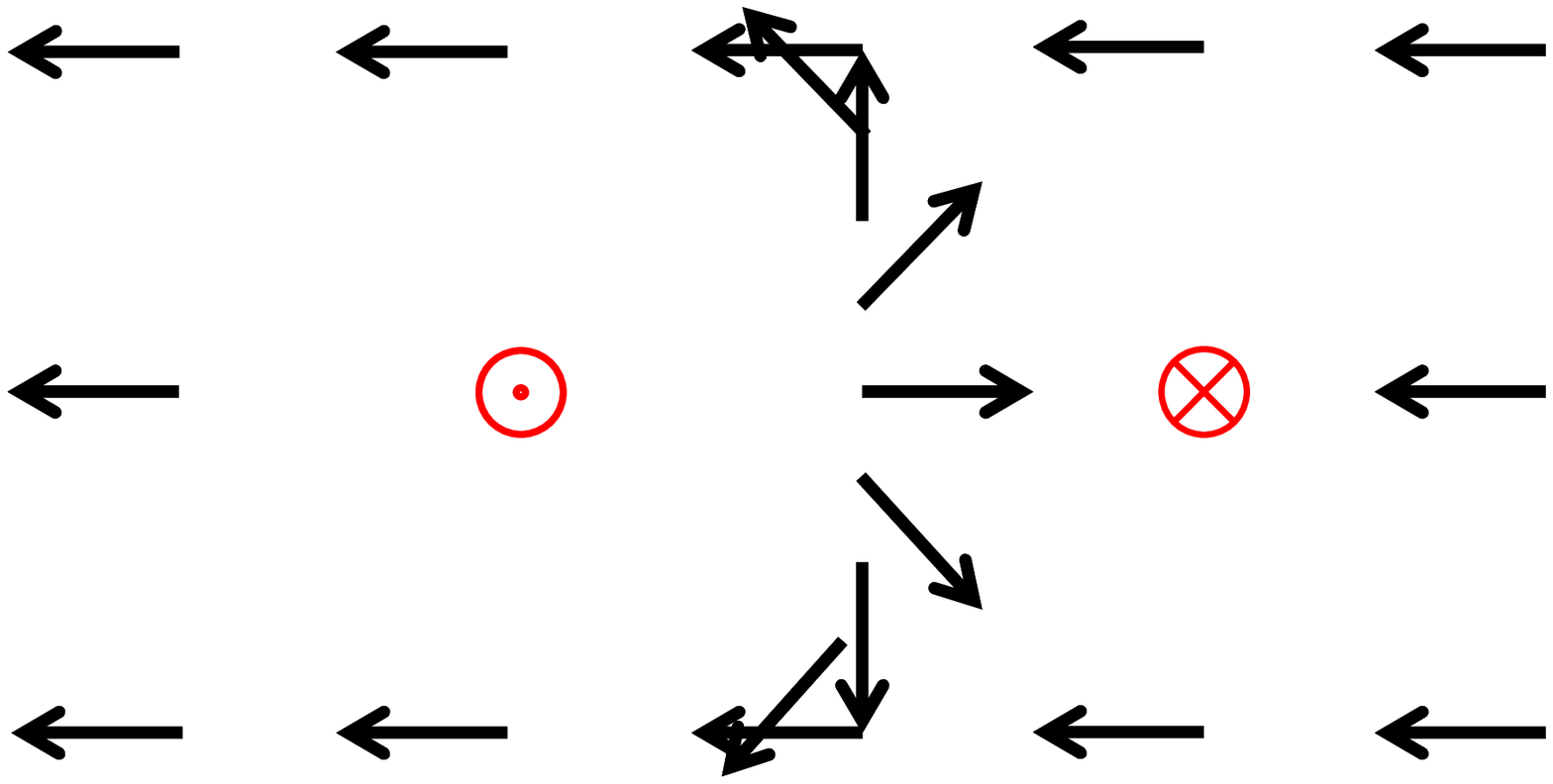} &
\includegraphics[width=0.40\linewidth,keepaspectratio]{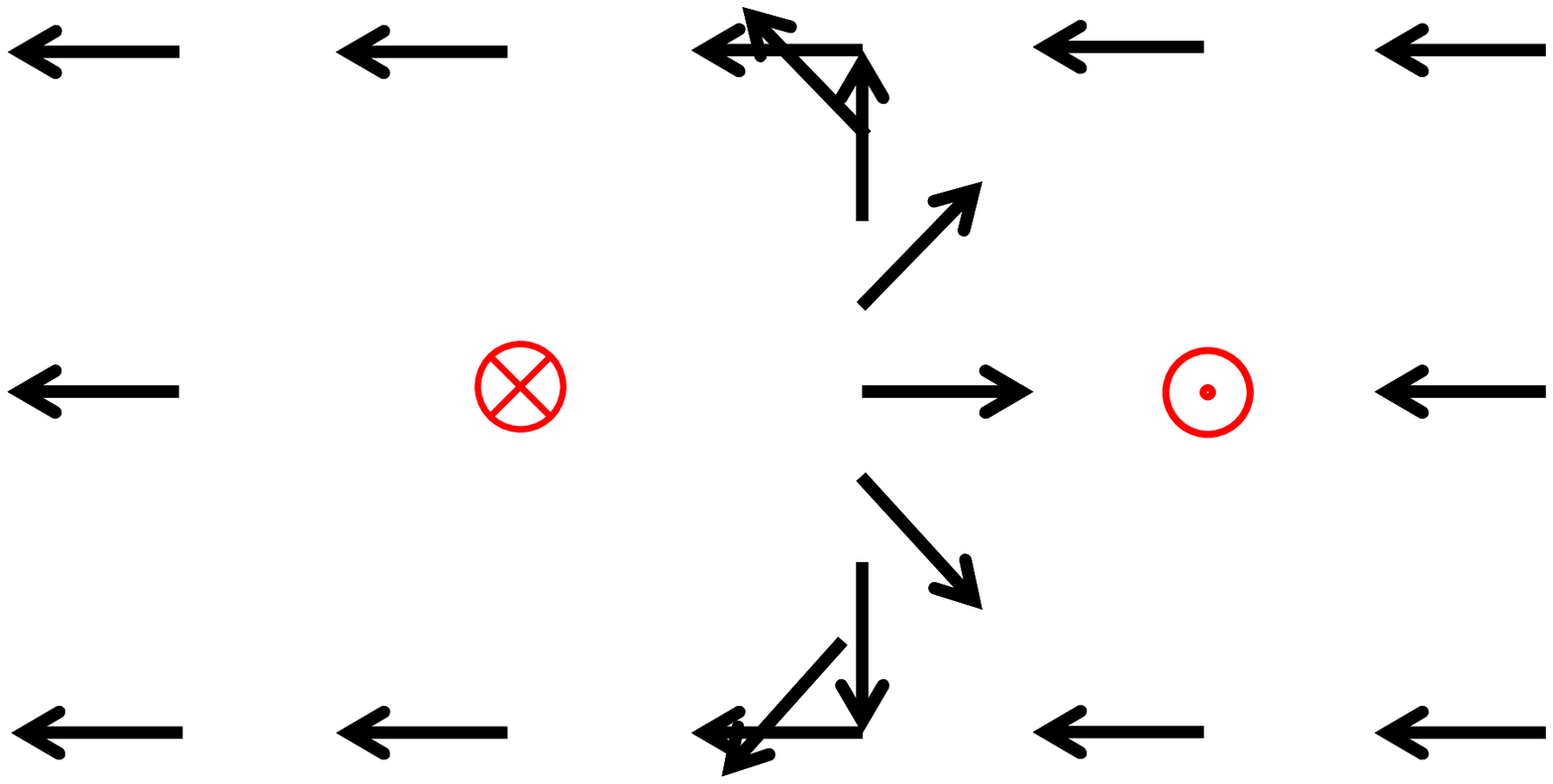} \\
(a) $(+1,+\1{2},+\1{2})+(-1,-\1{2},+\1{2})$ & 
(b) $(+1,-\1{2},-\1{2})+(-1,+\1{2},-\1{2})$ \\
\includegraphics[width=0.40\linewidth,keepaspectratio]{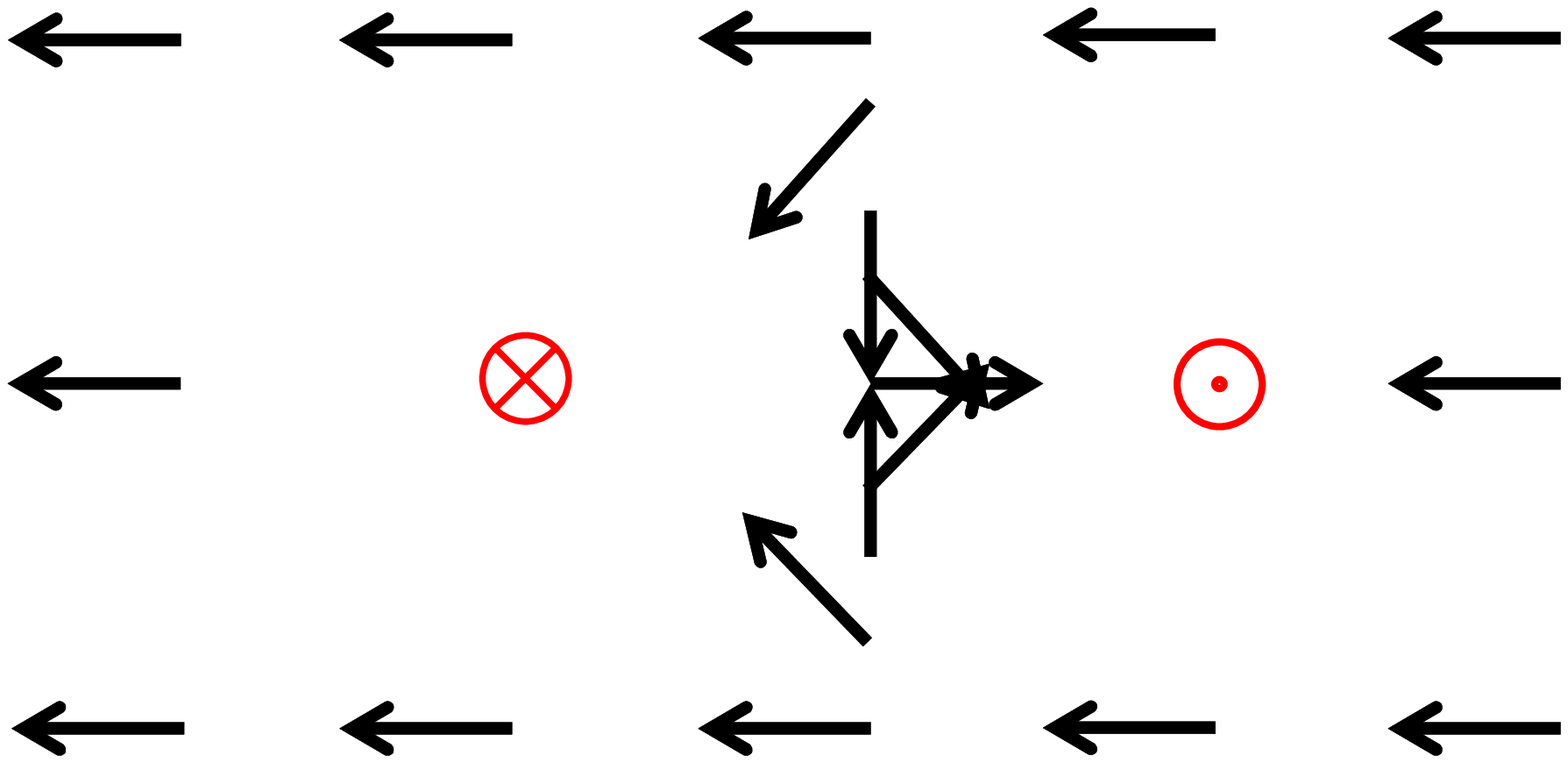} &
\includegraphics[width=0.40\linewidth,keepaspectratio]{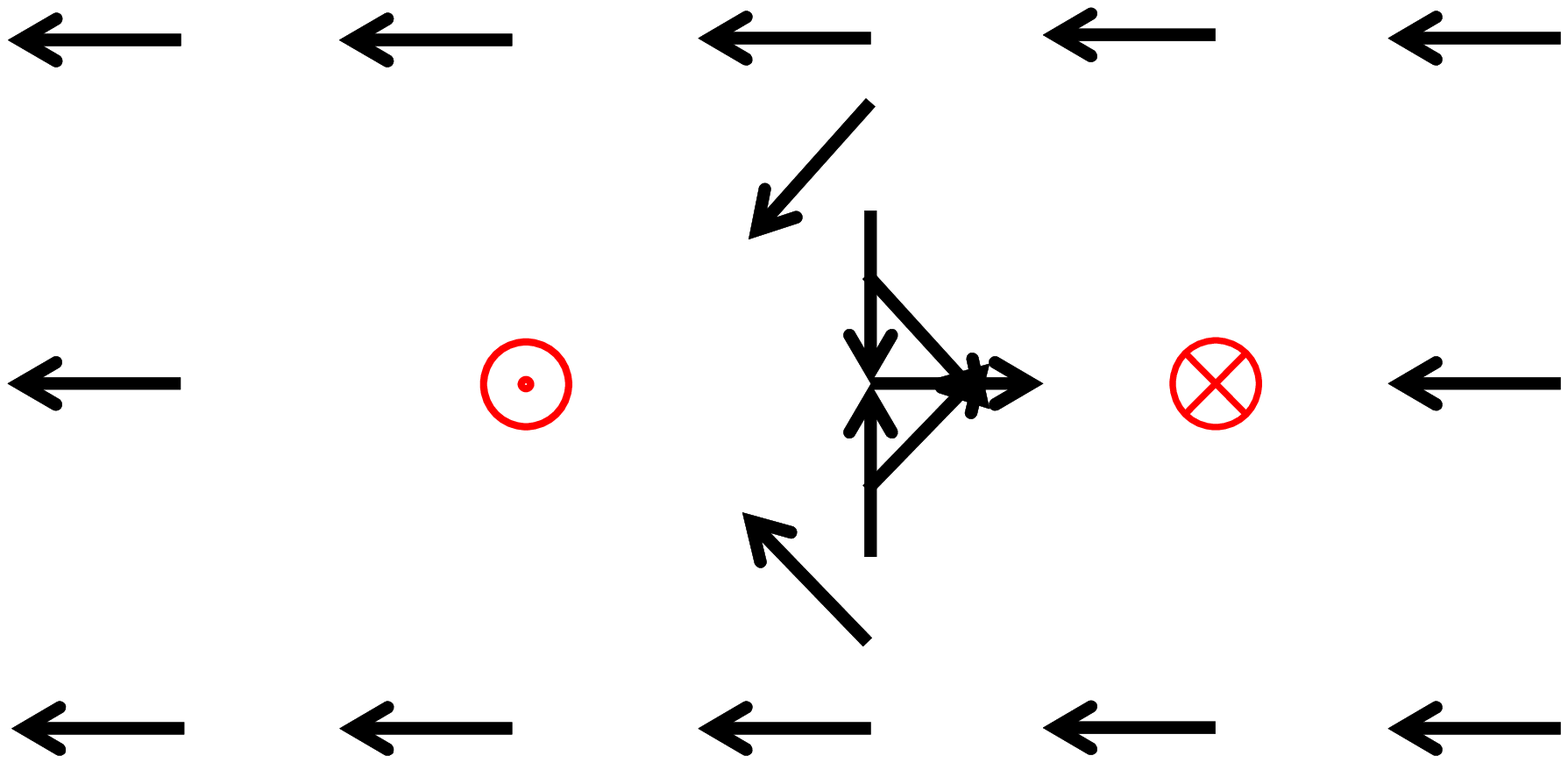} \\
(c) $(-1,-\1{2},+\1{2})+(+1,+\1{2},+\1{2})$ & 
(d) $(-1,+\1{2},-\1{2})+(+1,-\1{2},-\1{2})$ 
\end{tabular}
\end{center}
\caption{Fractional instantons in the $O(3)$ model 
with the boundary condition $(-,+,+)$. 
$\odot$ and $\otimes$ correspond to 
$n_1=+1$ and $n_1 = -1$, respectively.
Black arrows represent $(n_2,n_3)$  with $n_2^2+n_3^2=1$ ($n_1=0$) 
parameterizing the moduli space 
of vacua ${\cal N} \simeq S^1$:
$\leftarrow$, $\rightarrow$, $\uparrow$, $\downarrow$ correspond to 
$n_3 = +1$, $n_3 = -1$, 
$n_2 = +1$, $n_2 = -1$, respectively. 
We chose the vacuum $n_3=+1$ at the boundary.
Topological charges $(*,*,*)$ denote 
a host vortex charge $\pi_1$, 
an Ising spin $\pi_0$ in its core,
and the total instanton charge $\pi_2$, 
respectively.  
(a) An instanton is split into two fractional instantons 
$(+1,+\1{2},+\1{2})$ and $(-1,-\1{2},+\1{2})$ 
separated into the $x^1$ direction by a sine-Gordon domain wall. 
(b) An anti-instanton is split into 
two fractional anti-instantons 
$(+1,-\1{2},-\1{2})$ and $(-1,+\1{2},-\1{2})$ 
separated into the $x^1$ direction by a sine-Gordon anti-domain wall.
(c) and (d) are isomorphic to (a) and (b), respectively, 
by a $2\pi$ rotation along an axis 
at the center of the sine-Gordon (anti-)domain wall, 
which exchanges two fractional instantons.
\label{fig:fractional-O3-1}}
\end{figure}
%%%%%%%%%%%%%%%%%%%%%%%
%%%%%%%%%%%%%%%%%%%%%
\begin{figure}
\begin{center}
\begin{tabular}{ccc}
\includegraphics[width=0.33\linewidth,keepaspectratio]{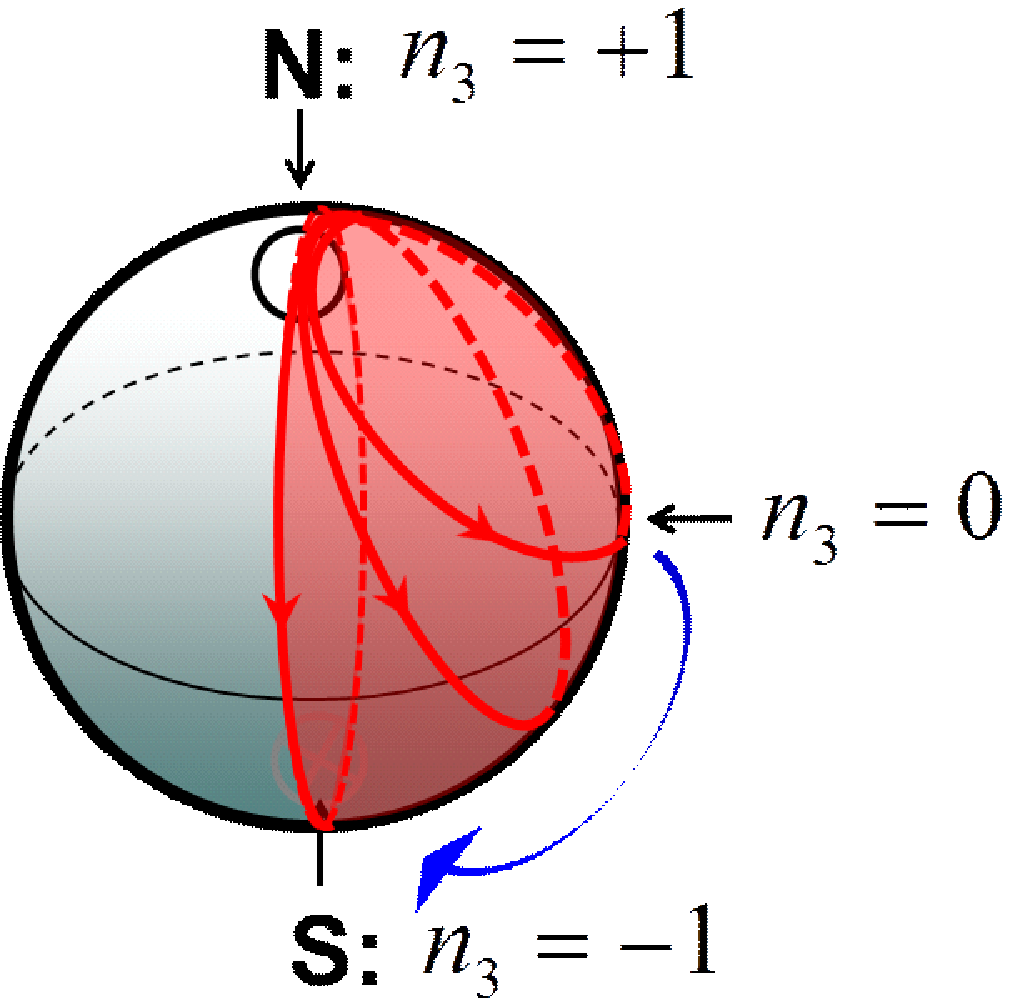} &
\includegraphics[width=0.33\linewidth,keepaspectratio]{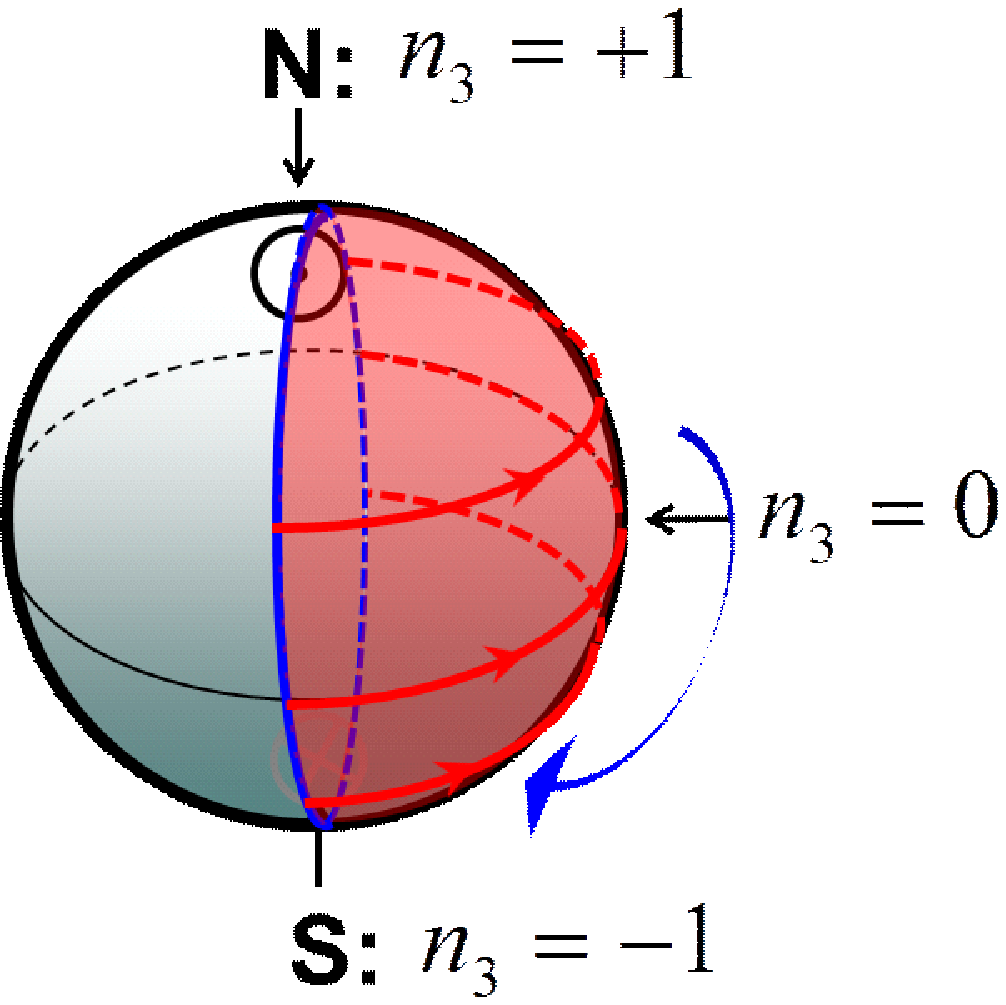} &
\includegraphics[width=0.33\linewidth,keepaspectratio]{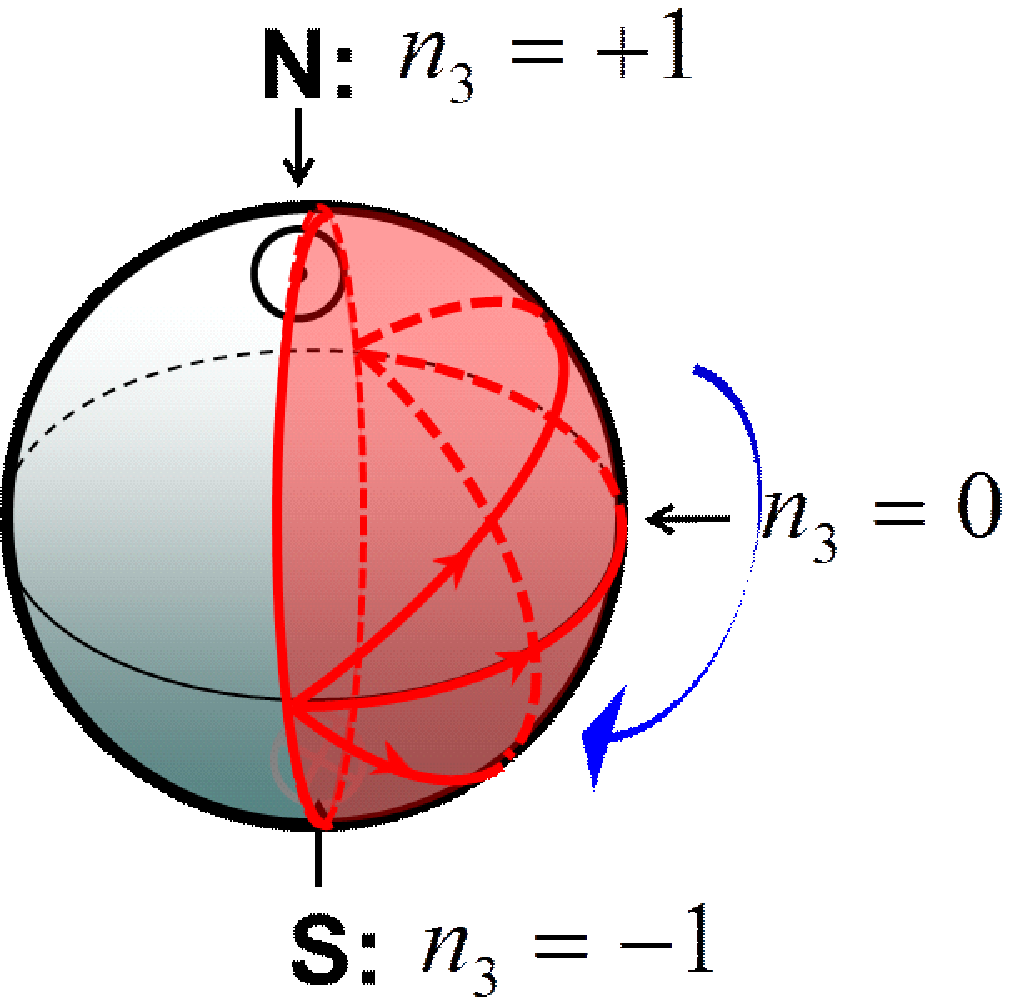} \\
(a) $(-,+,+)$ & (b) $(-,-,+)$ & (c) $(-,-,-)$\\
\end{tabular}
\end{center}
\caption{Images of fractional instantons in the target space $S^2$ of the $O(3)$ model 
with the boundary conditions (a) $(-,+,+)$, 
(b) $(-,-,+)$ and (c) $(-,-,-)$. 
Each path represents an image of $x_1=$ constant 
with $x_2=0$ to $x_2 = R$,
where an arrow indicates a direction.  
With changing $x_1$ from $x_1=-\infty$ to $x_1=+\infty$, 
the path moves following the blue arrow to 
cover a half sphere. 
\label{fig:fractional-path-O3}}
\end{figure}
%%%%%%%%%%%%%%%%%%%%%%%%%%%%%%%%%%%%%%%%%

%%%%%%%%%%%%%%%%%%%%%
\begin{figure}
\begin{center}
\includegraphics[width=0.1\linewidth,keepaspectratio]{2d-frame}
\begin{tabular}{cc}
\includegraphics[width=0.4\linewidth,keepaspectratio]{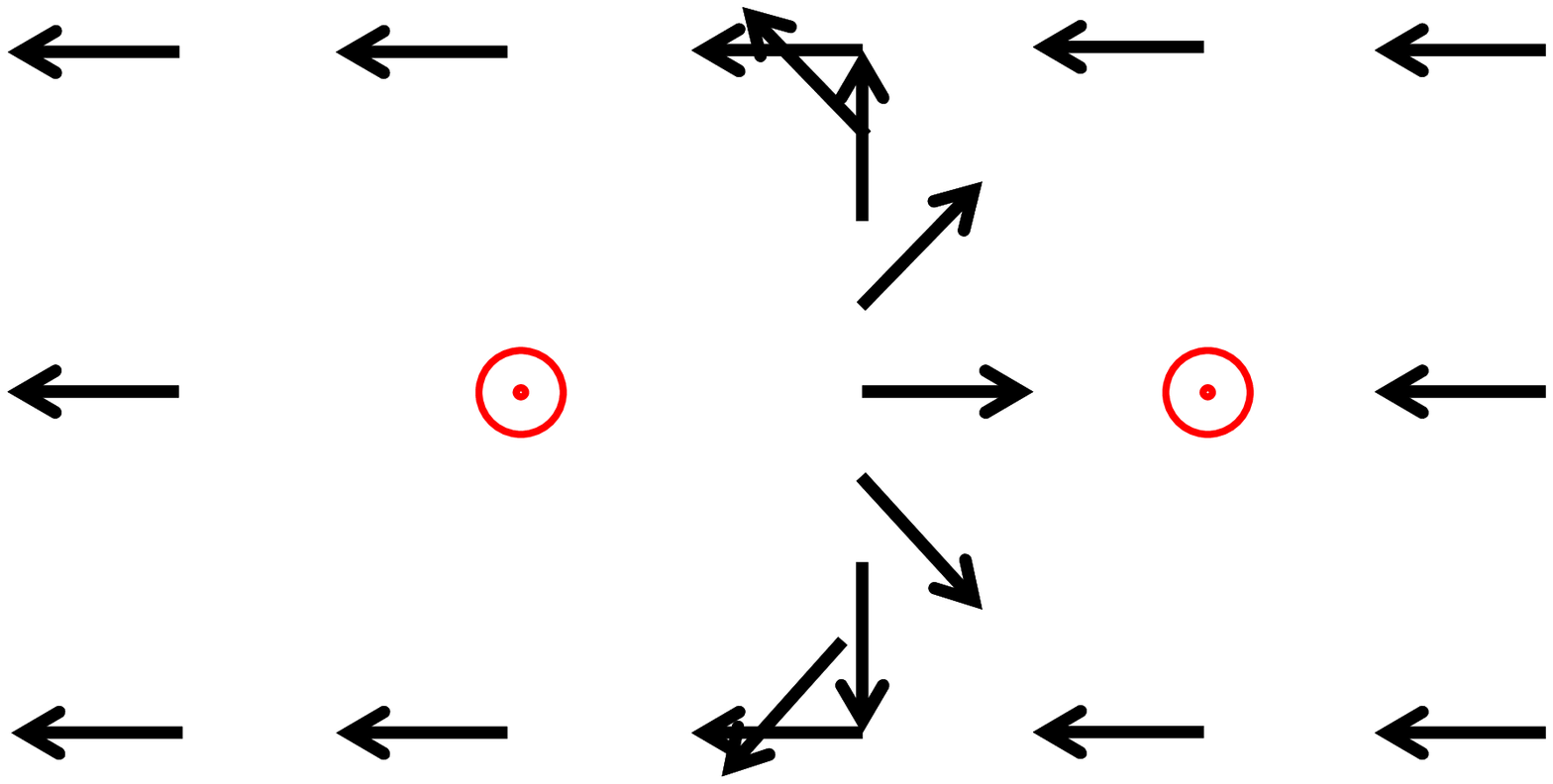} &
\includegraphics[width=0.4\linewidth,keepaspectratio]{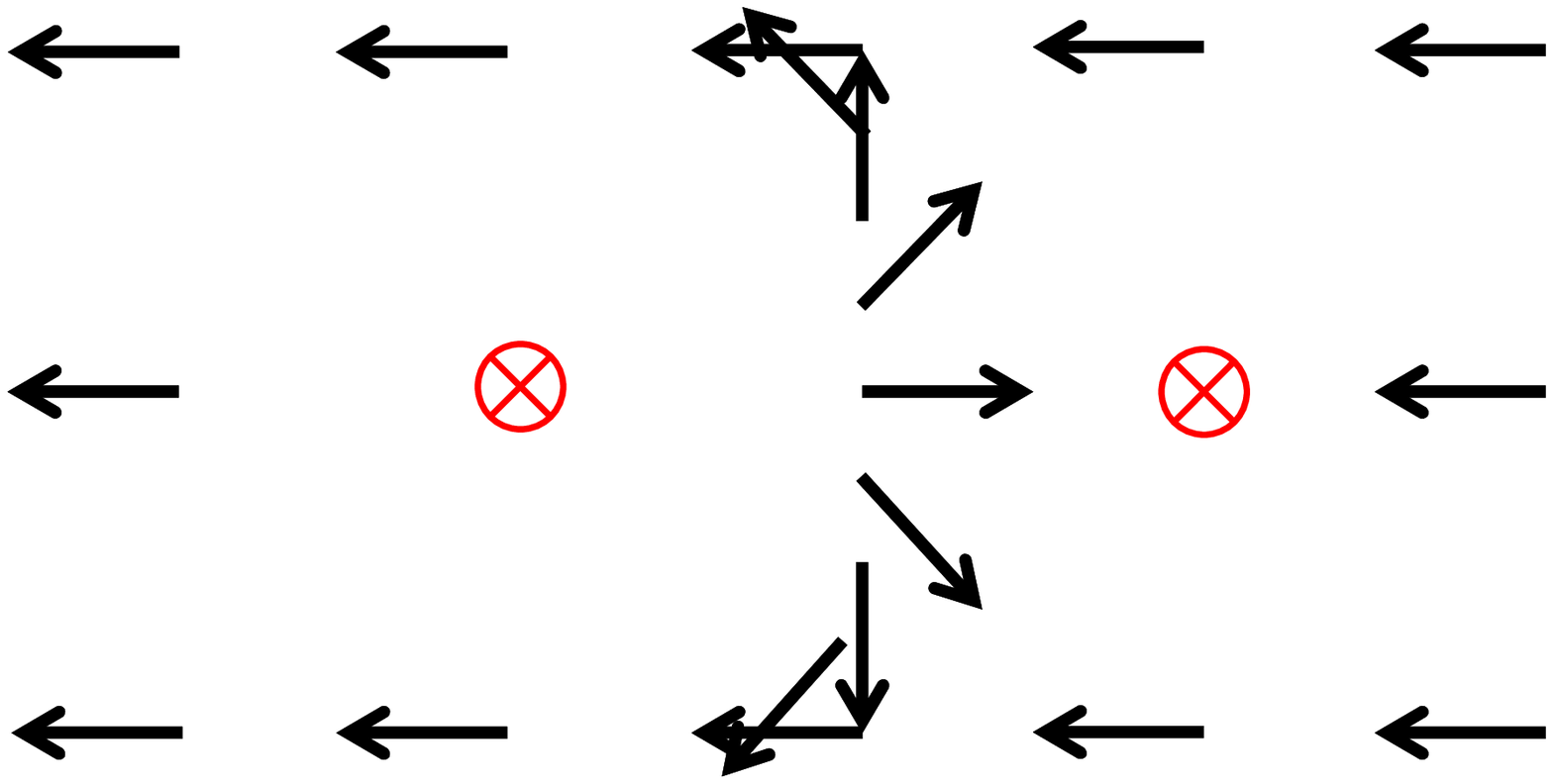} \\
(a) $(+1,+\1{2},+\1{2})+(-1,+\1{2},-\1{2})$ & 
(b) $(+1,-\1{2},-\1{2})+(-1,-\1{2},+\1{2})$ \\
\includegraphics[width=0.4\linewidth,keepaspectratio]{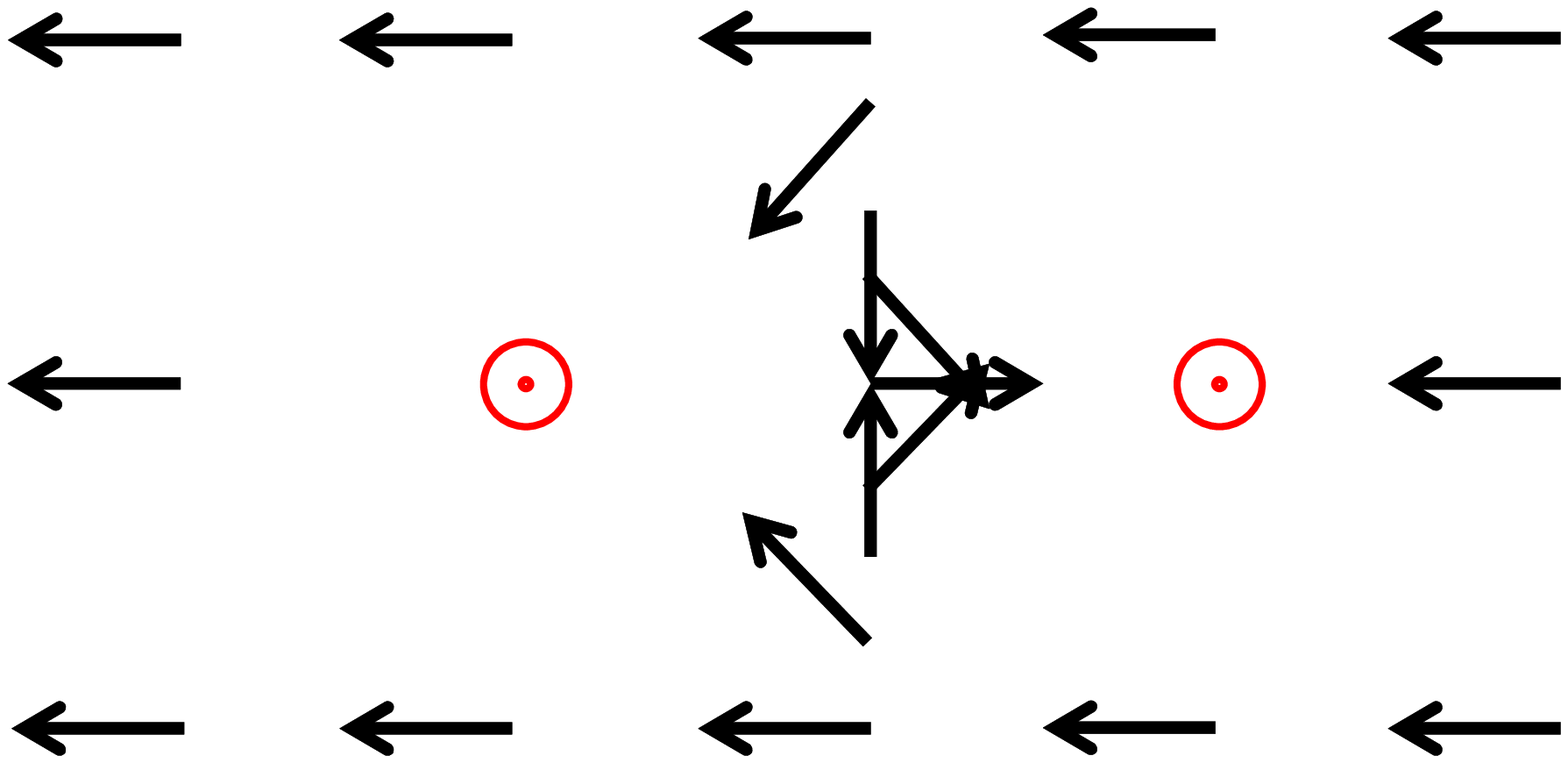} &
\includegraphics[width=0.4\linewidth,keepaspectratio]{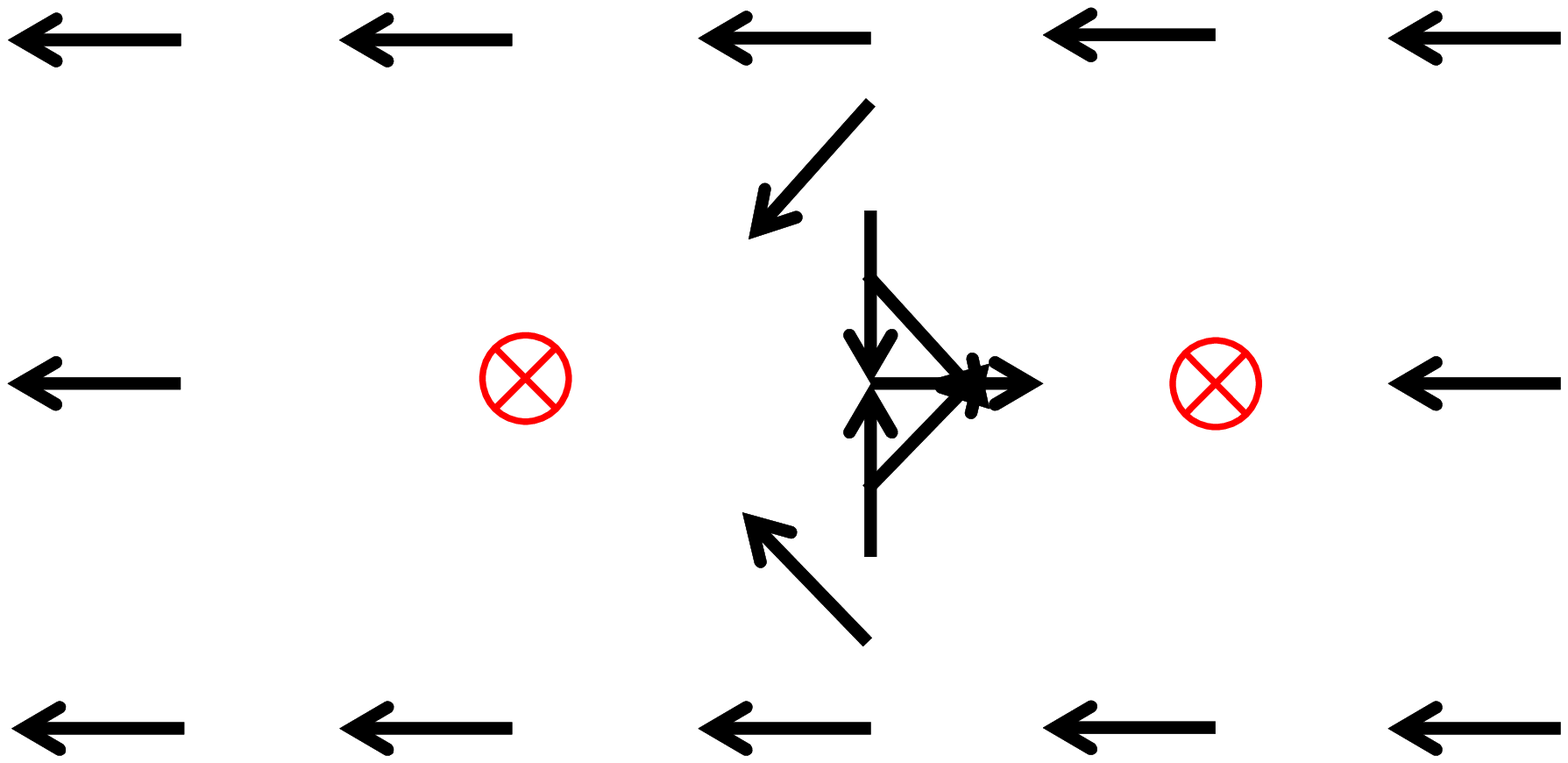} \\
(c) $(-1,+\1{2},-\1{2})+(+1,+\1{2},+\1{2})$ & 
(d) $(-1,-\1{2},+\1{2})+(+1,-\1{2},-\1{2})$ 
\end{tabular}
\end{center}
\caption{Bions in the $O(3)$ model 
with the boundary condition $(-,+,+)$. 
The notations are the same with Fig.~\ref{fig:fractional-O3-1}.
(c) and (d) are isomorphic to (a) and (b), respectively, 
by a $2\pi$ rotation along an axis 
at the center of the sine-Gordon (anti-)domain wall.
\label{fig:bion-O3-1}}
\end{figure}
%%%%%%%%%%%%%%%%%%%%%%

The fixed manifold is characterized by 
$n_1=0$, equivalently $(n_2)^2+(n_3)^2=1$, 
which is ${\cal N} \simeq S^1$. 
This is the moduli space of vacua as explained in the last section. 
It has a nontrivial homotopy 
$\pi_1(S^1) \simeq {\mathbb Z}$, allowing a global vortex 
having the winding in $n_2 + in_3$. 
In the vortex core, the two fields 
constituting the vortex must vanish $n_2= n_3=0$ 
and the rest field $n_1$ appears taking a value $n_1 = \pm 1$,  
giving an Ising spin degree of freedom to the vortex.
Therefore, the moduli space of the vortex is ${\cal M} \simeq \{\pm 1\}$. 
This is a fractional (anti-)instanton.
Depending on the vortex winding and the vortex 
moduli, there are four possibilities 
for fractional (anti-)instantons 
in the boundary condition 
$(-,+,+)$, as shown in Fig.~\ref{fig:O(3)} (1a)--(1d).  
A unit (anti-)instanton (lump) can be decomposed 
into two fractional (anti-)instantons 
as illustrated in Fig.~\ref{fig:fractional-O3-1}. 
Each fractional instanton wraps a half of the target space 
$S^2$. 
For instance, the left half of Fig.~\ref{fig:fractional-O3-1}(a) 
wraps a half sphere as in Fig.~\ref{fig:fractional-path-O3}(a).
If a fractional (anti-)instanton is well separated from the rest 
and is isolated, 
it becomes one of Fig.~\ref{fig:O(3)} (1a)--(1d). 

In order to write down explicit configurations, 
it is useful to define a complex coordinate by 
$z \equiv x_1 + i x_2$. 
Then, asymptotic forms near fractional instantons located at $z=0$ 
can be given by 
\beq
&&\mbox{(1a)}:\quad u \sim z, 
\quad 
\mbox{(1b)}:\quad u \sim 1/ z, \\
&&\mbox{(1c)}:\quad u \sim \bar z, 
\quad
\mbox{(1d)}:\quad u \sim 1/\bar z.
\eeq
This expression is also good for large compactification radius $R$. 
The topological charges of fractional (anti-)instantons 
with the boundary condition $(-,+,+)$ 
are summarized in Table \ref{table:homotopy-O3-1}.
Here, we have defined the value of $\pi_0$ for the Ising spin to be $\pm 1/2$ 
to be consistent with the other boundary conditions 
discussed below. 
%%%%%%%%%%%%%%%%%%%%%%%%%
\begin{table}[h]
\begin{tabular}{c|c|c|c} 
     & $\pi_1$ & $\pi_0$ & $\pi_2$ \\ \hline
Fig.~\ref{fig:O(3)} (1a) &$ +1$     & $+1/2$ & $+1/2$ \\
Fig.~\ref{fig:O(3)} (1b) &$ -1$     & $-1/2$ & $+1/2$ \\
Fig.~\ref{fig:O(3)} (1c) &$ -1$     & $+1/2$ & $-1/2$ \\
Fig.~\ref{fig:O(3)} (1d) &$ +1$     & $-1/2$ & $-1/2$  
\end{tabular}
\caption{Homotopy groups of fractional instantons in the 
$O(3)$ model with the boundary condition $(-,+,+)$.
The columns represent the homotopy groups  
of a host soliton $\pi_1$, a daughter soliton $\pi_0$, 
and the total instanton $\pi_2$ from left to right. 
\label{table:homotopy-O3-1}}
\end{table}
%%%%%%%%%%%%%%%%%%%%%%

Let us discuss the interaction between fractional instantons.
When constituent fractional instantons are well separated 
at distance $r$ in a large compactification radius $R$, 
the interaction between them is  $E_{\rm int} \sim \pm \log r$ 
(the force is $F \sim \pm 1/r$)
because they are global vortices. 
Here, the interaction is repulsive for a pair of (anti-)vortices, 
and attractive for a pair of a vortex and an anti-vortex. 
Therefore, it is attractive for a pair of fractional (anti-)instantons 
constituting an (anti-)instanton. 
On the other hand,
when the compactification radius $R$ is 
as the same as the size of fractional instantons 
as in Fig.~\ref{fig:bion-O3-1}, 
a sine-Gordon (anti-)kink connects 
a fractional instanton and anti-instanton so that they are confined by 
a linear interaction energy $E_{\rm int} \sim r$ with distance $r$ 
and the force between them is constant.

Next let us discuss bion configurations. 
Configurations near a bion can be written as 
\beq
&&\mbox{(1b)+(1d)}:\quad
 u \sim \frac{\alpha_1}{z-z_1} + \frac{\bar \alpha_1}{\bar z-\bar z_1} +\beta_1\\
&&\mbox{(1a)+(1c)}:\quad
 u \sim \frac{(z-z_1)(\bar z - \bar z_2) }{\alpha z + \beta \bar z + \gamma}.
\eeq
This is good for large compactification radius $R$.
Bions for small compactification radius $R$ 
are schematically drawn in Fig.~\ref{fig:bion-O3-1}.  
Each of 
Fig.~\ref{fig:bion-O3-1} shows a sine-Gordon (anti-)kink connecting 
two fractional (anti-)instantons for small compactification radius $R$,
and consequently they are confined by 
a linear potential $E_{\rm int}  \sim r$ with distance $r$ 
for large separation.

Before going to the other boundary conditions, 
let us make a comment on fractional instantons 
in related models.
There exist topologically the same fractional instantons on ${\mathbb R}^2$
{\it without} twisted boundary condition.
One is baby Skyrmions \cite{Piette:1994ug,Weidig:1998ii} 
in an $O(3)$ sigma model with 
a potential term $V=m^2 n_1^2$ and 
a four derivative (baby Skyrme) term
\cite{Jaykka:2010bq,Kobayashi:2013aja,Kobayashi:2013wra}. 
The other is a vortex in a $U(1)$ gauged $O(3)$ sigma model 
with a potential term $V=m^2 n_1^2$, 
in which the $U(1)$ acting on $n_2 +i n_3$ is gauged 
\cite{Schroers:1995he,Schroers:1996zy,Baptista:2004rk,
Nitta:2011um,Alonso-Izquierdo:2014cza}. 
Vortices in this case are local, that is, 
of Abrikosov-Nielsen-Olesen (ANO) 
type \cite{Abrikosov:1956sx}.
In both cases, the potential term plays an alternative role 
of the twisted boundary condition.
Interactions between fractional instantons 
are rather different from our case of the twisted boundary condition.
In the former, the interaction between fractional instantons 
constituting an instanton is attractive 
at large distance and repulsive at short distance, 
resulting in a stable molecule
\cite{Jaykka:2010bq,Kobayashi:2013aja,Kobayashi:2013wra}. 
In the latter, the interaction between them is exponentially suppressed 
which is either repulsive or attractive for 
type-II or type-I superconductor, 
and non-interactive for the critical limit, which is BPS  
\cite{Nitta:2011um}.

%%%%%%%%%%%%%%%%%%%%%%%
\subsection{$(-,-,+)$: a half sine-Gordon kink inside a domain wall}
This is only the case studied in the literature.
This case is equivalent to 
the ${\mathbb C}P^1$ model with ${\mathbb Z}_2$ symmetric boundary 
condition, allowing
fractional instantons \cite{Eto:2004rz,Eto:2006mz,Eto:2006pg}
and bions \cite{Dunne:2012ae,Dunne:2012zk,Dabrowski:2013kba,
Bolognesi:2013tya,Misumi:2014jua}.

%%%%%%%%%%%%%%%%%%%%%
\begin{figure}
\begin{center}
\includegraphics[width=0.1\linewidth,keepaspectratio]{2d-frame}
\includegraphics[width=0.70\linewidth,keepaspectratio]{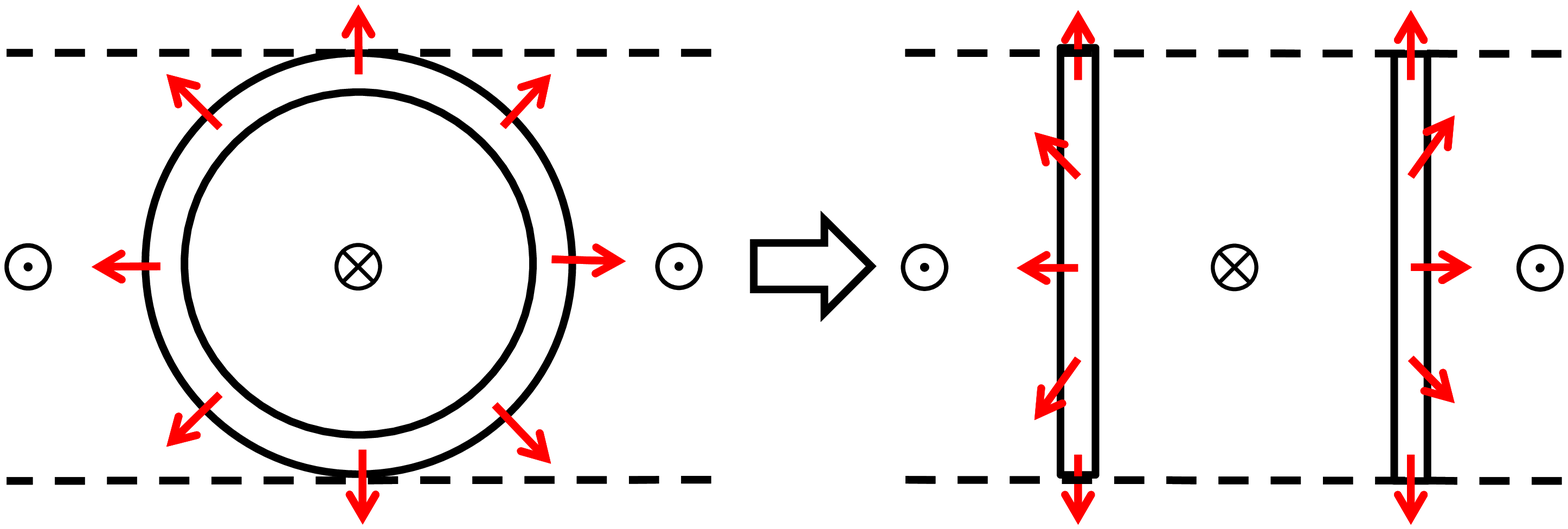} 
\end{center}
\caption{Twisted domain wall ring decaying into two fractional 
instantons 
in the $O(3)$ model 
with the boundary condition $(-,-,+)$. 
$\odot$ and $\otimes$ correspond to 
$n_3=+1$ and $n_3 = -1$, respectively,
representing vacua.  
Red arrows represent 
$(n_1,n_2)$ with $n_1^2+n_2^2=1$ ($n_3=0$) parameterizing 
the moduli space of a domain wall ${\cal M}\simeq S^1$:  
$\leftarrow$, $\rightarrow$, $\uparrow$, $\downarrow$ correspond to 
$n_1 = -1$, $n_1 = +1$, 
$n_2 = +1$, $n_2 = -1$, respectively. 
The dotted lines denote the boundary at $x^2=0$ and $x^2 =R$.
When the size of a domain wall ring is of that of the compact direction, 
it decays thorough a reconnection into 
two fractional instantons, 
which are domain walls with half twisted $U(1)$ moduli.
\label{fig:decay-ring-O3}}
\end{figure}
%%%%%%%%%%%%%%%%%%%%%%%

%%%%%%%%%%%%%%%%%%%%%
\begin{figure}
\begin{center}
\includegraphics[width=0.1\linewidth,keepaspectratio]{2d-frame}
\begin{tabular}{cc}
\includegraphics[width=0.40\linewidth,keepaspectratio]{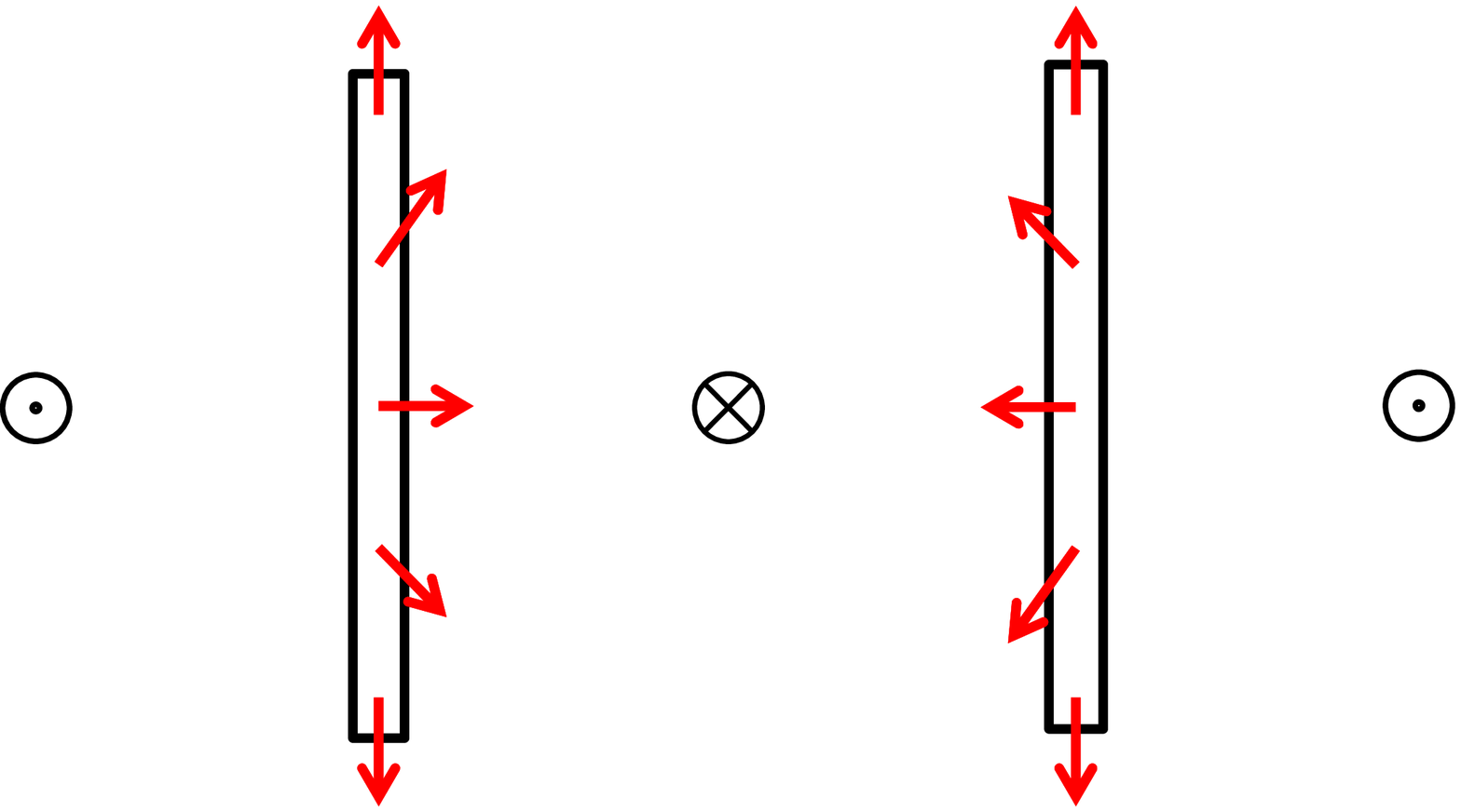} &
\includegraphics[width=0.40\linewidth,keepaspectratio]{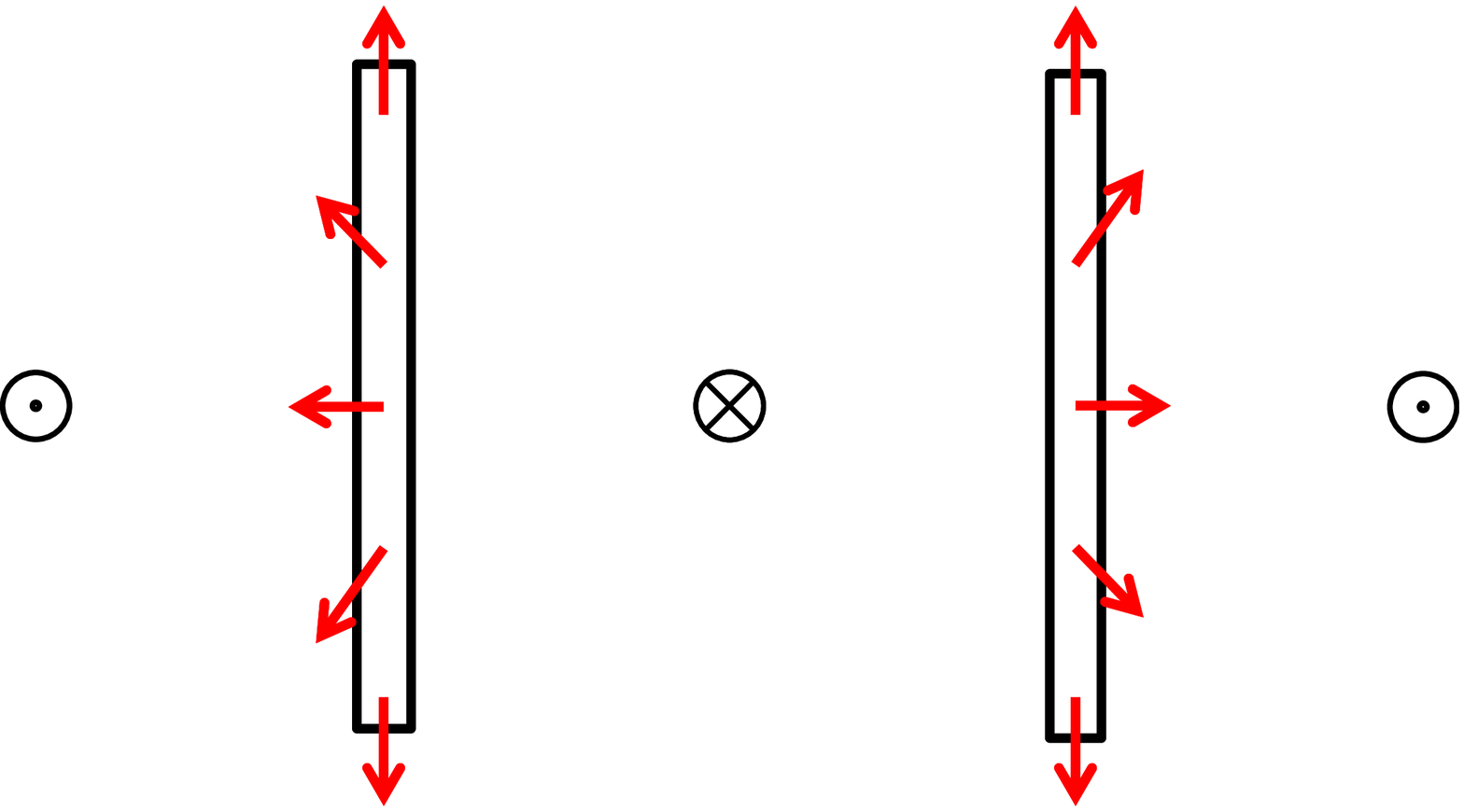} \\
(a) $(+1,+\1{2},+\1{2})+(-1,-\1{2},+\1{2})$ & 
(b) $(+1,-\1{2},-\1{2})+(-1,+\1{2},-\1{2})$ \\
\includegraphics[width=0.40\linewidth,keepaspectratio]{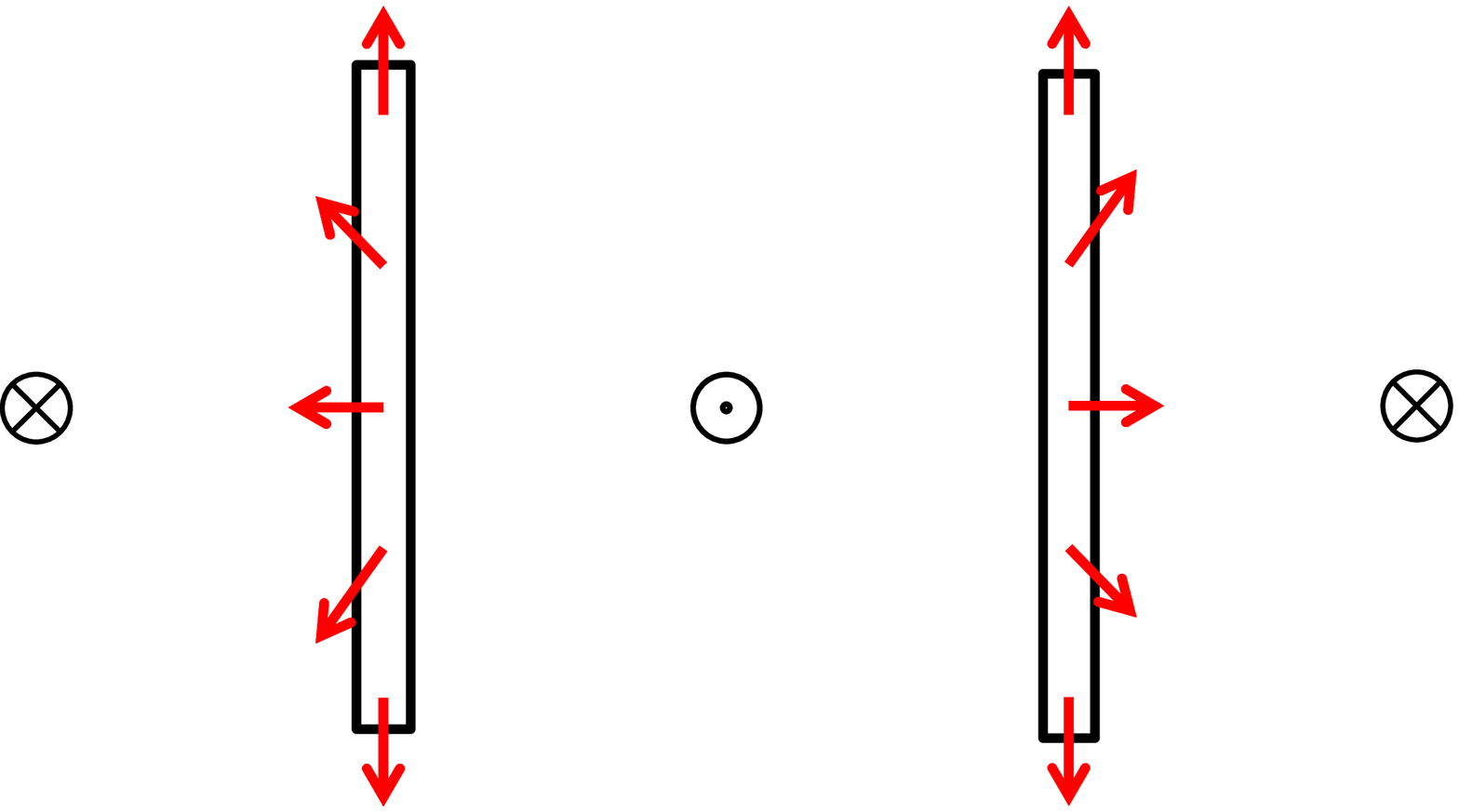} &
\includegraphics[width=0.40\linewidth,keepaspectratio]{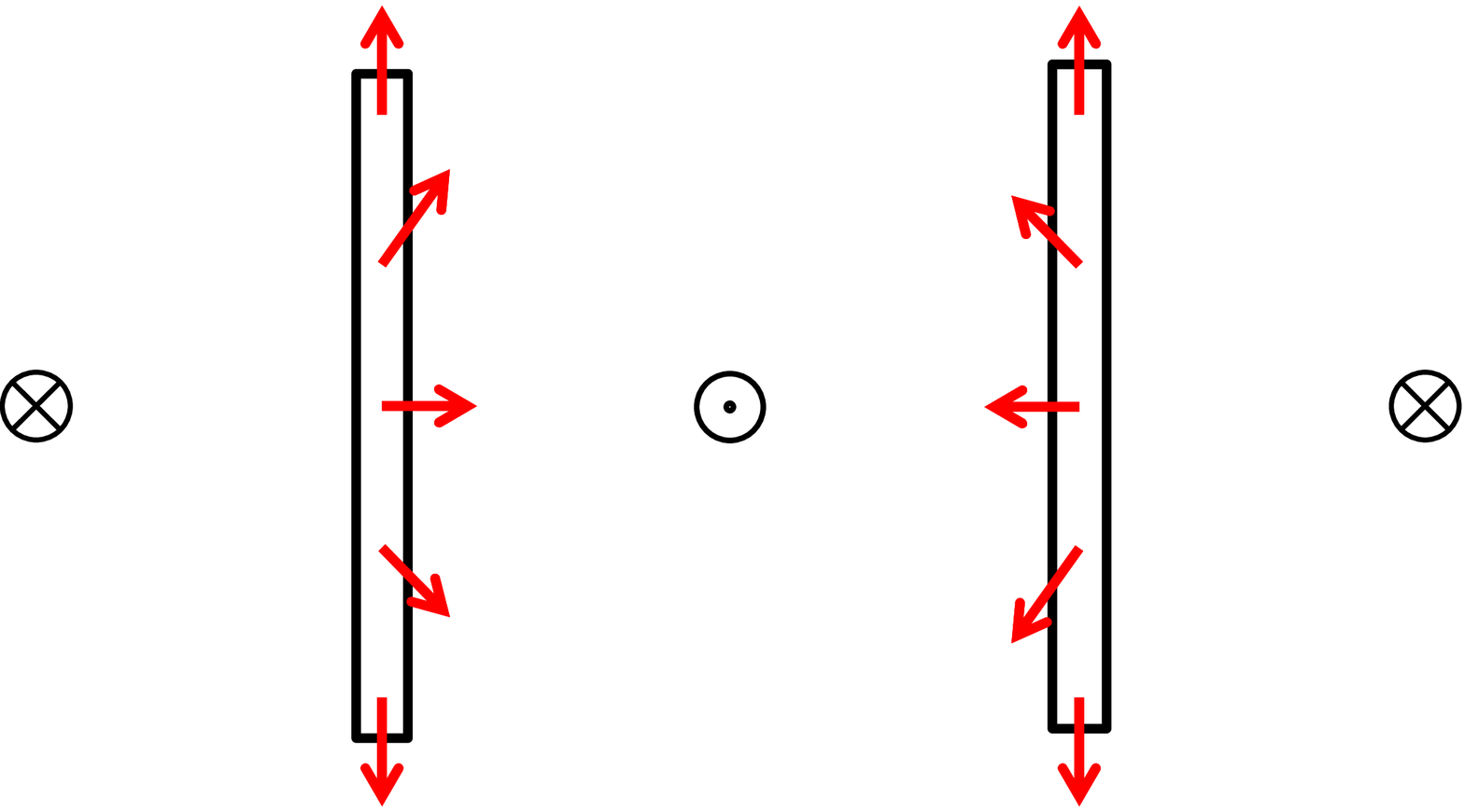} \\
(c) $(-1,-\1{2},+\1{2})+(+1,+\1{2},+\1{2})$ & 
(d) $(-1,+\1{2},-\1{2})+(+1,-\1{2},-\1{2})$ 
\end{tabular}
\end{center}
\caption{Fractional instantons in the $O(3)$ model 
with the boundary condition $(-,-,+)$. 
The notations are the same with Fig.~\ref{fig:decay-ring-O3}. 
Topological charges $(*,*,*)$ denote 
a host domain wall charge $\pi_0$, 
a sine-Gordon kink charge $\pi_1$ on it,
and the total instanton charge $\pi_2$, 
respectively.
(a) An instanton is split into two fractional instantons 
$(+1,+\1{2},+\1{2})$ and $(-1,-\1{2},+\1{2})$ 
separated by the vacuum $\otimes$. 
(b) An anti-instanton is split into 
two fractional anti-instantons 
$(+1,-\1{2},-\1{2})$ and $(-1,+\1{2},-\1{2})$ 
separated by the vacuum $\otimes$. 
(c) and (d) are obtained from (a) and (b), respectively, 
by exchanging the positions of the fractional instanton and anti-instanton.
\label{fig:fractional-O3-2}}
\end{figure}
%%%%%%%%%%%%%%%%%%%%%%%
%%%%%%%%%%%%%%%%%%%%%
\begin{figure}
\begin{center}
\includegraphics[width=0.1\linewidth,keepaspectratio]{2d-frame}
\begin{tabular}{cc}
\includegraphics[width=0.4\linewidth,keepaspectratio]{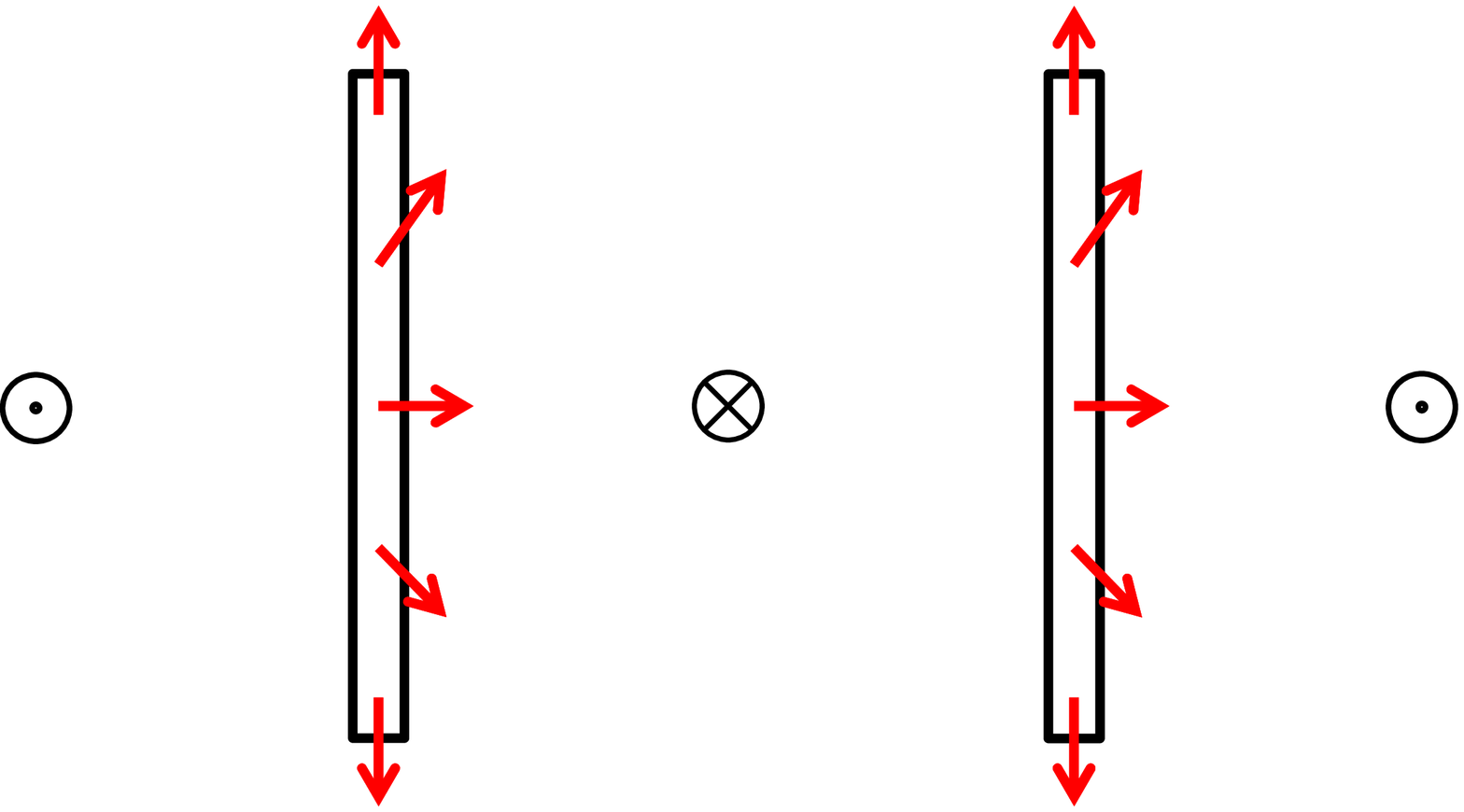} &
\includegraphics[width=0.4\linewidth,keepaspectratio]{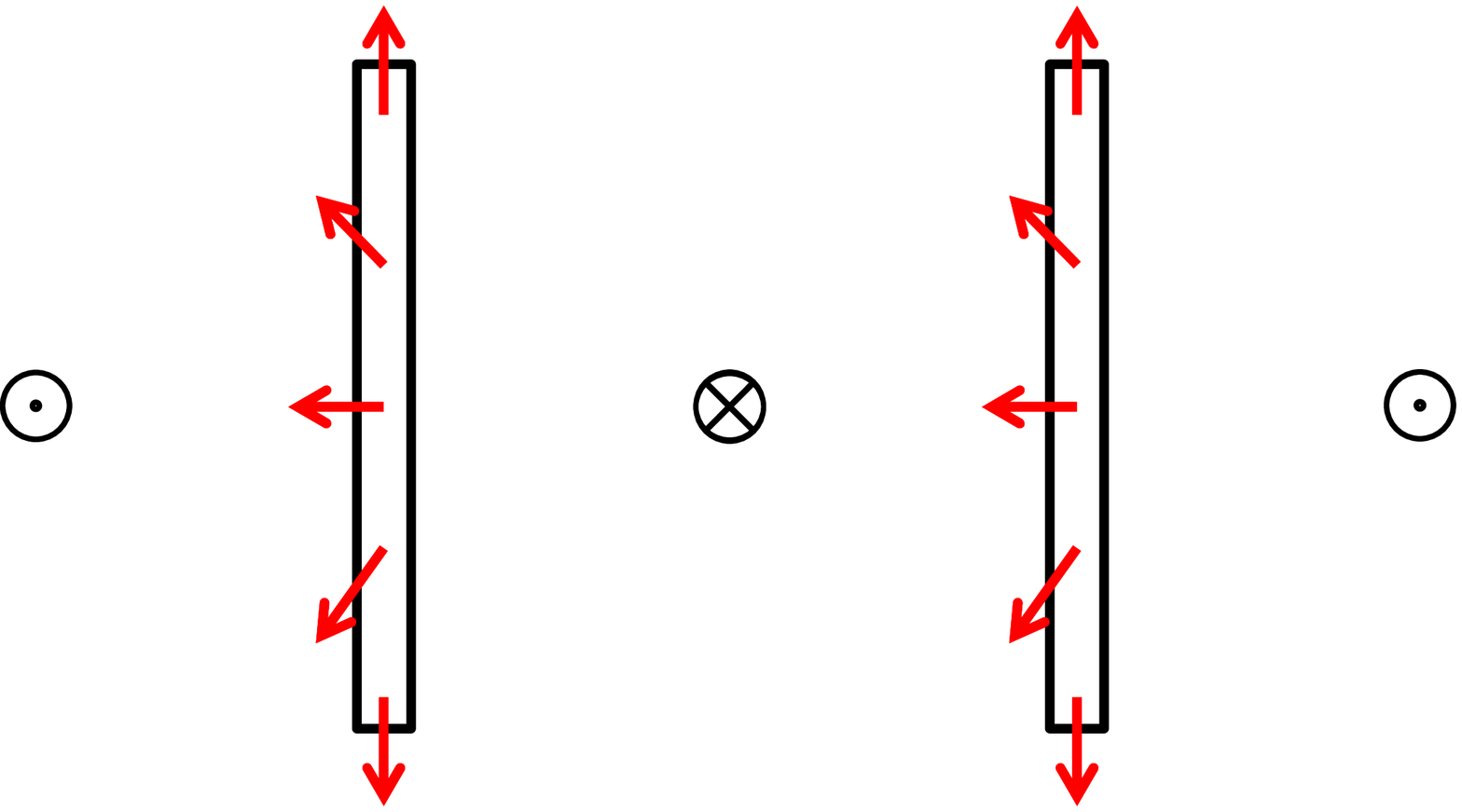} \\
(a) $(+1,+\1{2},+\1{2})+(-1,+\1{2},-\1{2})$ & 
(b) $(+1,-\1{2},-\1{2})+(-1,-\1{2},+\1{2})$ \\
\includegraphics[width=0.4\linewidth,keepaspectratio]{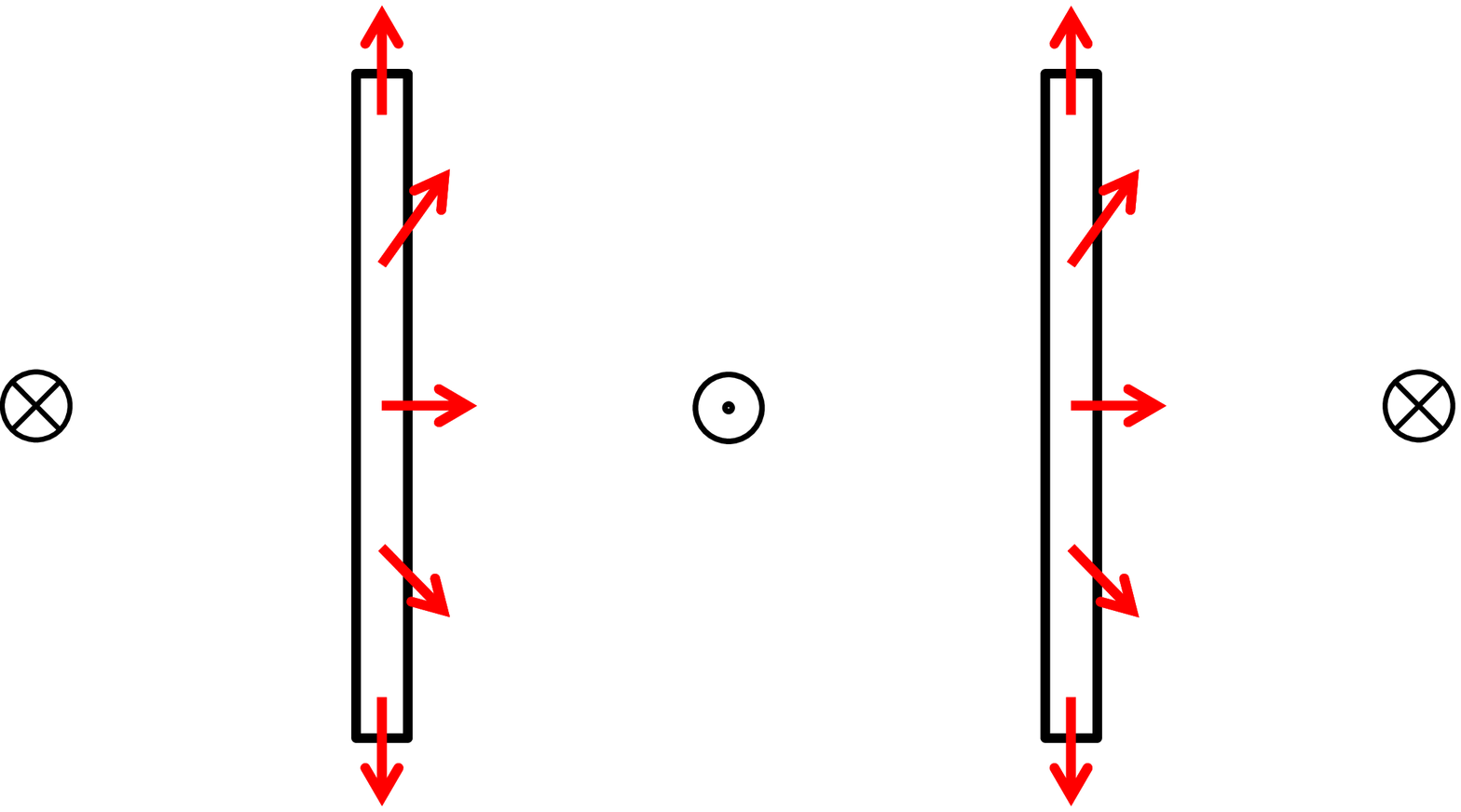} &
\includegraphics[width=0.4\linewidth,keepaspectratio]{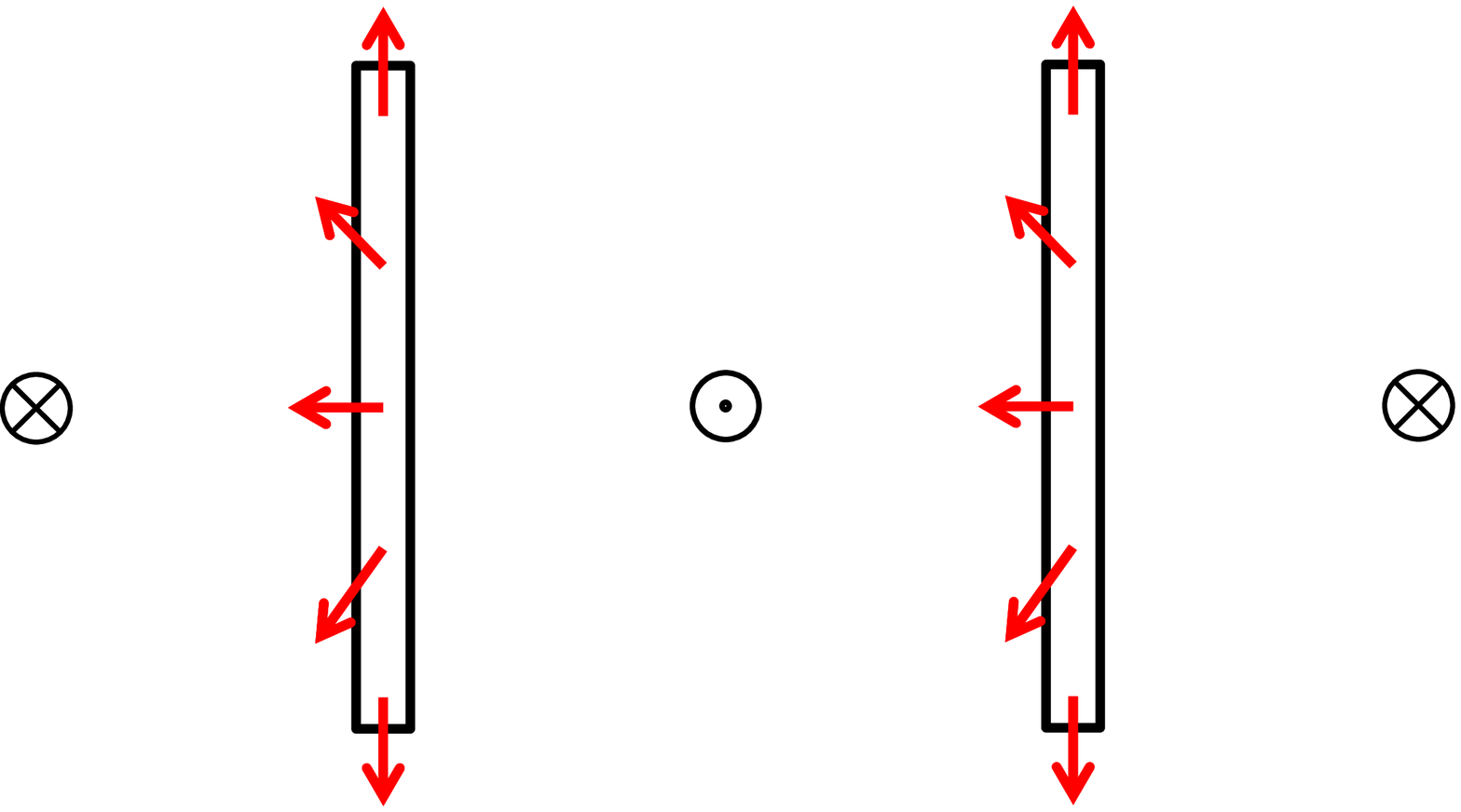} \\
(c) $(-1,+\1{2},-\1{2})+(+1,+\1{2},+\1{2})$ & 
(d) $(-1,-\1{2},+\1{2})+(+1,-\1{2},-\1{2})$ 
\end{tabular}
\end{center}
\caption{Bions in the $O(3)$ model 
with the boundary condition $(-,-,+)$.
The notations are the same with Fig.~\ref{fig:decay-ring-O3}. 
(c) and (d) are obtained from (a) and (b), respectively, 
by exchanging the positions of the fractional instanton and anti-instanton.
\label{fig:bion-O3-2}}
\end{figure}
%%%%%%%%%%%%%%%%%%%%%%

Instantons in the $O(3)$ model can be represented as 
a domain wall ring along which a $U(1)$ modulus is twisted 
\cite{Nitta:2012kj,Nitta:2012xq,Kobayashi:2013ju}. 
When the size of the domain wall ring is that of the compactification 
radius $R$, 
the top and bottom of the domain wall ring touch each other 
through the compact direction $x^2$ with 
the twisted boundary condition. 
Then, a reconnection of two parts of the ring occurs, 
and it can be split into two domain wall lines 
separated into the $x^1$ direction,
 as shown in Fig.~\ref{fig:decay-ring-O3}.
The $U(1)$ modulus is twisted half along 
the domain lines extending to the $x^2$ direction, 
resulting in fractional (anti-)instantons.
We have two pairs for instanton and anti-instanton respectively 
as seen in Fig.~\ref{fig:fractional-O3-2}.
We thus find 
four kinds of fractional (anti-)instantons 
shown in  Fig.~\ref{fig:O(3)}  (2a)--(2d). 
Each fractional instanton wraps a half sphere of the target space 
$S^2$. 
For instance, the left half of Fig.~\ref{fig:fractional-O3-2}(a) 
wraps a half sphere as in Fig.~\ref{fig:fractional-path-O3}(b).

Explicit configurations of isolated fractional (anti-)instantons 
can be given as
\beq
&&\mbox{(2a)}:\quad u = e^{\pi z}, \quad
   \mbox{(2b)}:\quad u = e^{-\pi z}, \\
&&\mbox{(2c)}:\quad u = e^{\pi \bar z}, \quad
   \mbox{(2d)}:\quad u = e^{-\pi \bar z}.  
\eeq
The topological charges of fractional (anti-)instantons 
with the boundary condition $(-,-,+)$ 
are summarized in Table \ref{table:homotopy-O3-2}.
%%%%%%%%%%%%%%%%%%%%%%%%%
\begin{table}[h]
\begin{tabular}{c|c|c|c} 
     & $\pi_0$ & $\pi_1$ & $\pi_2$ \\ \hline
Fig.~\ref{fig:O(3)} (2a) &$ +1$     & $+1/2$ & $+1/2$ \\
Fig.~\ref{fig:O(3)} (2b) &$ -1$     & $-1/2$ & $+1/2$ \\
Fig.~\ref{fig:O(3)} (2c) &$ -1$     & $+1/2$ & $-1/2$ \\
Fig.~\ref{fig:O(3)} (2d) &$ +1$     & $-1/2$ & $-1/2$    
\end{tabular}
\caption{Homotopy groups of fractional instantons in the 
$O(3)$ model with the boundary condition $(-,-,+)$.
The columns represent the homotopy groups  
of a host soliton $\pi_0$, a daughter soliton $\pi_1$, 
and the total instanton $\pi_2$ from left to right. 
\label{table:homotopy-O3-2}}
\end{table}
%%%%%%%%%%%%%%%%%%%%%%

Bions with the boundary condition 
$(-,-,+)$ are shown in 
Fig.~\ref{fig:bion-O3-2}. 
Explicit bion ansatz can be constructed as
\beq
 &&\mbox{(2d) + (2a)}: \quad u = e^{-\pi (\bar z-\bar z_1)} + e^{\pi (z-z_2)}, \\
 &&\mbox{(2b) + (2c)}: \quad  u = e^{-\pi (z-z_1)} +  e^{\pi (\bar z-\bar z_2)}.
\eeq
These ansatz are different from Ref.~\cite{Misumi:2014jua}, 
but their asymptotic behaviors are the same. 
The interactions between fractional (anti-)instantons 
are exponentially suppressed so that 
the total action becomes a sum of each action 
when they are well separated.

Before going to the next case,
let us give a brief comment on a relation to 
dimensional reduction in this case.
In the zero radius limit of the compact direction, 
the theory is dimensionally reduced. 
By assuming the dependence  of the fields 
on the compact direction $x^2$ as 
\beq 
(n_1,n_2) 
= \left(\hat n_1(x^1)\cos {\pi \over R} x^2, 
\hat n_2(x^1) \sin {\pi \over R} x^2 \right)
\eeq
in the presence of the twisted boundary condition, 
we see that a potential term is effectively induced from the gradient 
term of the fields:
\beq 
 V  = \int_0^R d x^2 \left[(\del_2 n_1)^2 + (\del_2 n_2)^2 \right] 
      = m^2 (\hat n_1^2 + \hat n_2^2) = m^2 (1-\hat n_3^2),
   \quad
  m^2 \equiv {\pi^2 \over 4 R}  . \label{eq:SS-dim-red}
\eeq 
This is known as the Scherk-Schwarz dimensional reduction 
in the context of supersymmetric theories 
in which 
the induced mass is called a twisted mass.
The dimensionally reduced ${\mathbb C}P^1$ model is often called 
as the massive ${\mathbb C}P^1$ model in the context of supersymmetry.
Lumps (instantons) are reduced to 
(a pair of) domain walls in the massive ${\mathbb C}P^1$ model 
\cite{Abraham:1992vb,Kobayashi:2014xua}.  
This case was generalized to the ${\mathbb C}P^{N-1}$ and Grassmann sigma models, for which domain walls \cite{Isozumi:2004jc,Eto:2004vy}, 
instantons (lumps) \cite{Shifman:2006kd,Eto:2007yv},  
fractional instantons  \cite{Eto:2006mz}, 
bions \cite{Misumi:2014bsa} were studied.

%%%%%%%%%%%%%%%%%%%%%%%
\subsection{$(-,-,-)$}

%%%%%%%%%%%%%%%%%%%%%
\begin{figure}
\begin{center}
\includegraphics[width=0.1\linewidth,keepaspectratio]{2d-frame}
\begin{tabular}{cc}
\includegraphics[width=0.40\linewidth,keepaspectratio]{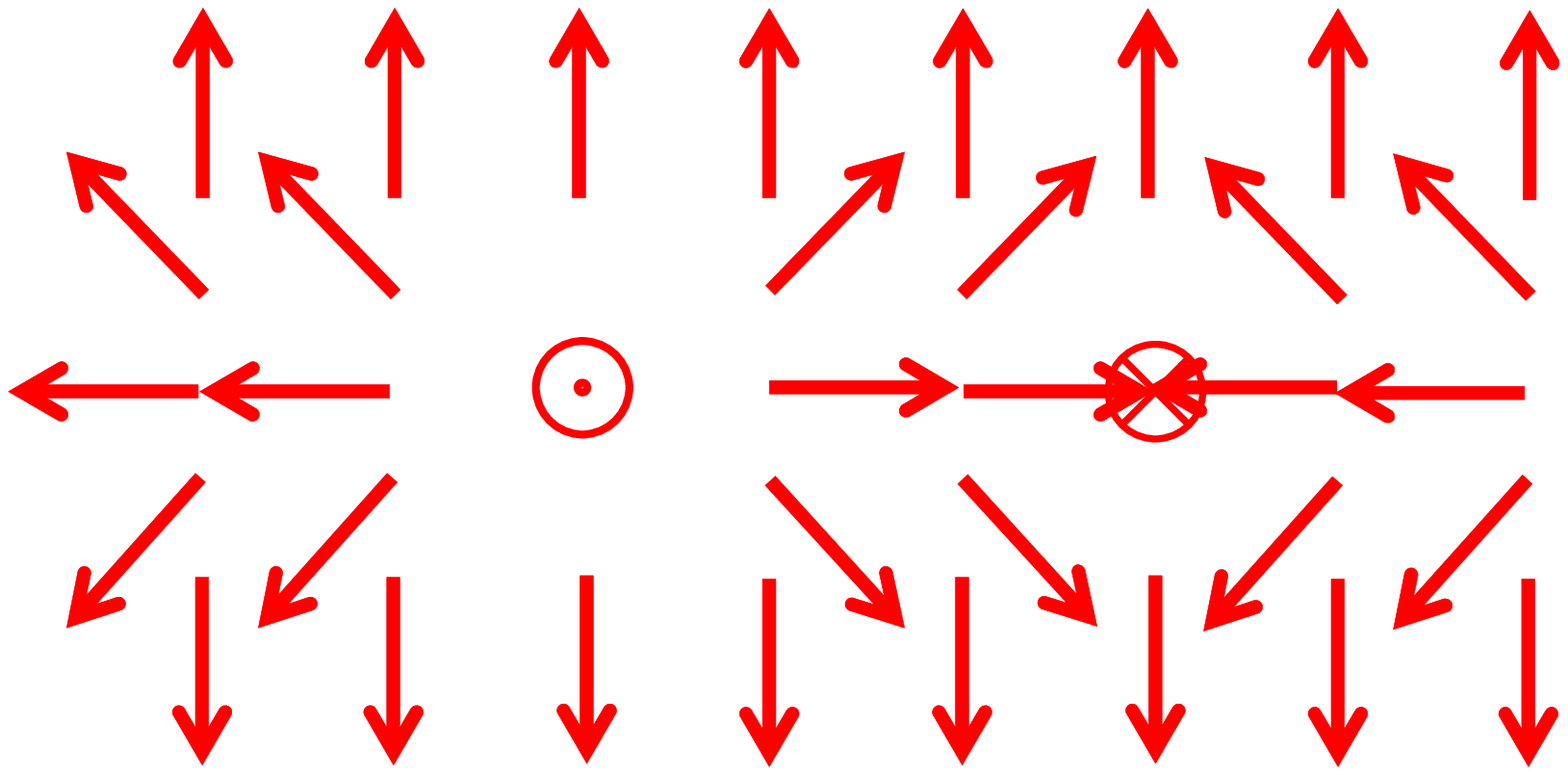} &
\includegraphics[width=0.40\linewidth,keepaspectratio]{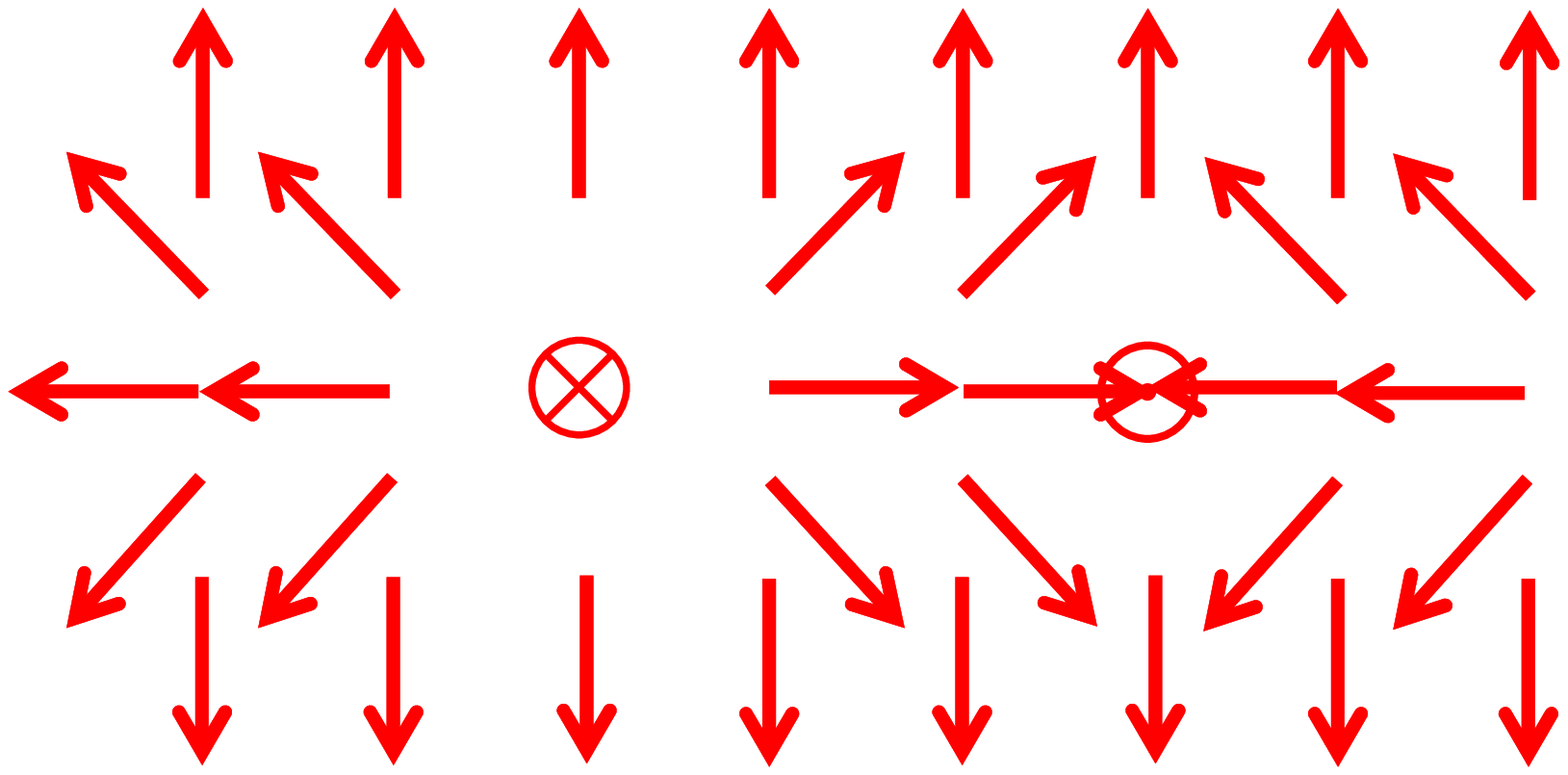} \\
(a) $(+1,+\1{2},+\1{2})+(-1,-\1{2},+\1{2})$ & 
(b) $(+1,-\1{2},-\1{2})+(-1,+\1{2},-\1{2})$ \\
\includegraphics[width=0.40\linewidth,keepaspectratio]{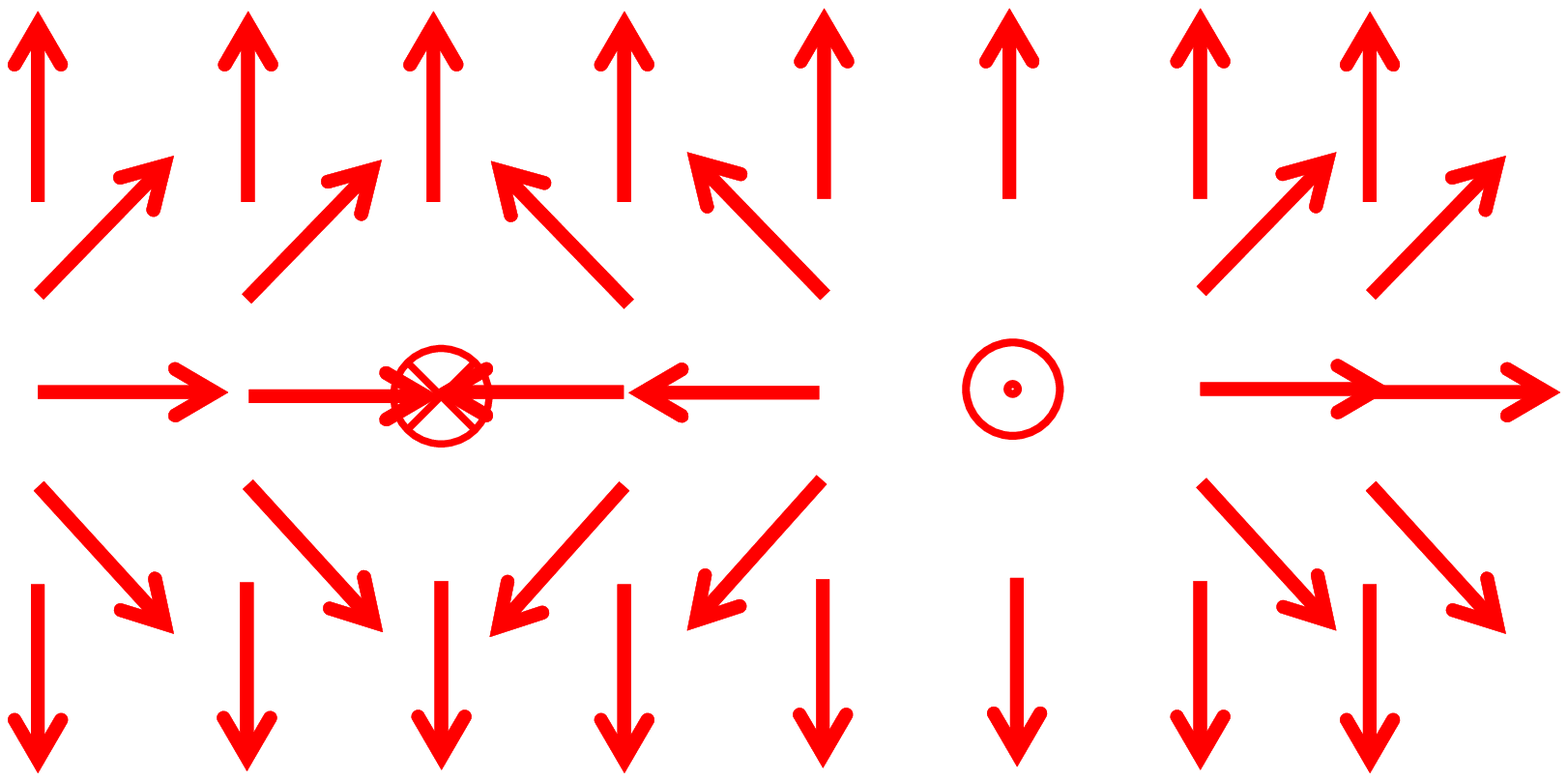} &
\includegraphics[width=0.40\linewidth,keepaspectratio]{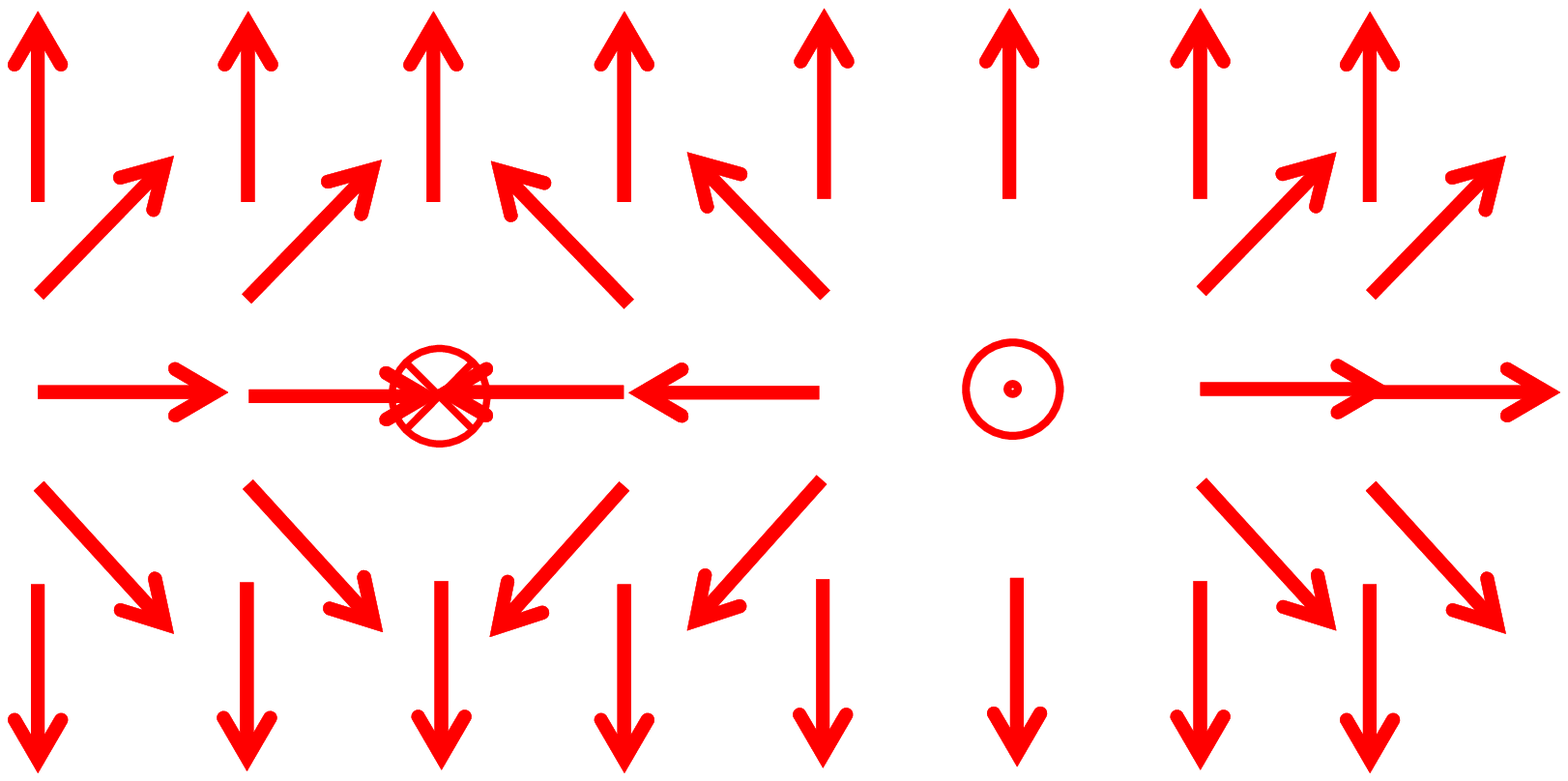} \\
(c) $(-1,-\1{2},+\1{2})+(+1,+\1{2},+\1{2})$ & 
(d) $(-1,+\1{2},-\1{2})+(+1,-\1{2},-\1{2})$ 
\end{tabular}
\end{center}
\caption{Fractional instantons in the $O(3)$ model 
with the boundary condition $(-,-,-)$. 
$\odot$ and $\otimes$ correspond to 
$n_3=+1$ and $n_3 = -1$, respectively, 
and
$\leftarrow$, $\rightarrow$, $\uparrow$, $\downarrow$ correspond to 
$n_1 = -1$, $n_1 = +1$, 
$n_2 = +1$, $n_2 = -1$, respectively. 
Topological charges $(*,*,*)$ denote 
a host space-filling soliton charge $\pi_{-1}$ which is merely formal, 
a lump charge $\pi_{2}$ on it,
and the total instanton charge $\pi_2$, 
respectively.  
(a) An instanton is split into two fractional instantons 
$(+1,+\1{2},+\1{2})+(-1,-\1{2},+\1{2})$ and 
$(+1,-\1{2},-\1{2})+(-1,+\1{2},-\1{2})$ 
separated by a half sine-Gordon domain wall with opposite 
orientation with the boundary at $x^1 = \pm \infty$. 
(b) An anti-instanton is split into 
two fractional anti-instantons 
$(-1,-\1{2},+\1{2})+(+1,+\1{2},+\1{2})$  
and $(-1,+\1{2},-\1{2})+(+1,-\1{2},-\1{2})$ 
separated
by a half sine-Gordon domain wall with opposite 
orientation with the boundary $x^1 = \pm \infty$.
(c) and (d) are isomorphic to (a) and (b), respectively, 
by a $2\pi$ rotation along an axis at the center.
\label{fig:fractional-O3-3}}
\end{figure}
%%%%%%%%%%%%%%%%%%%%%%%%%

%%%%%%%%%%%%%%%%%%%%%
\begin{figure}
\begin{center}
\includegraphics[width=0.1\linewidth,keepaspectratio]{2d-frame}
\begin{tabular}{cc}
\includegraphics[width=0.4\linewidth,keepaspectratio]{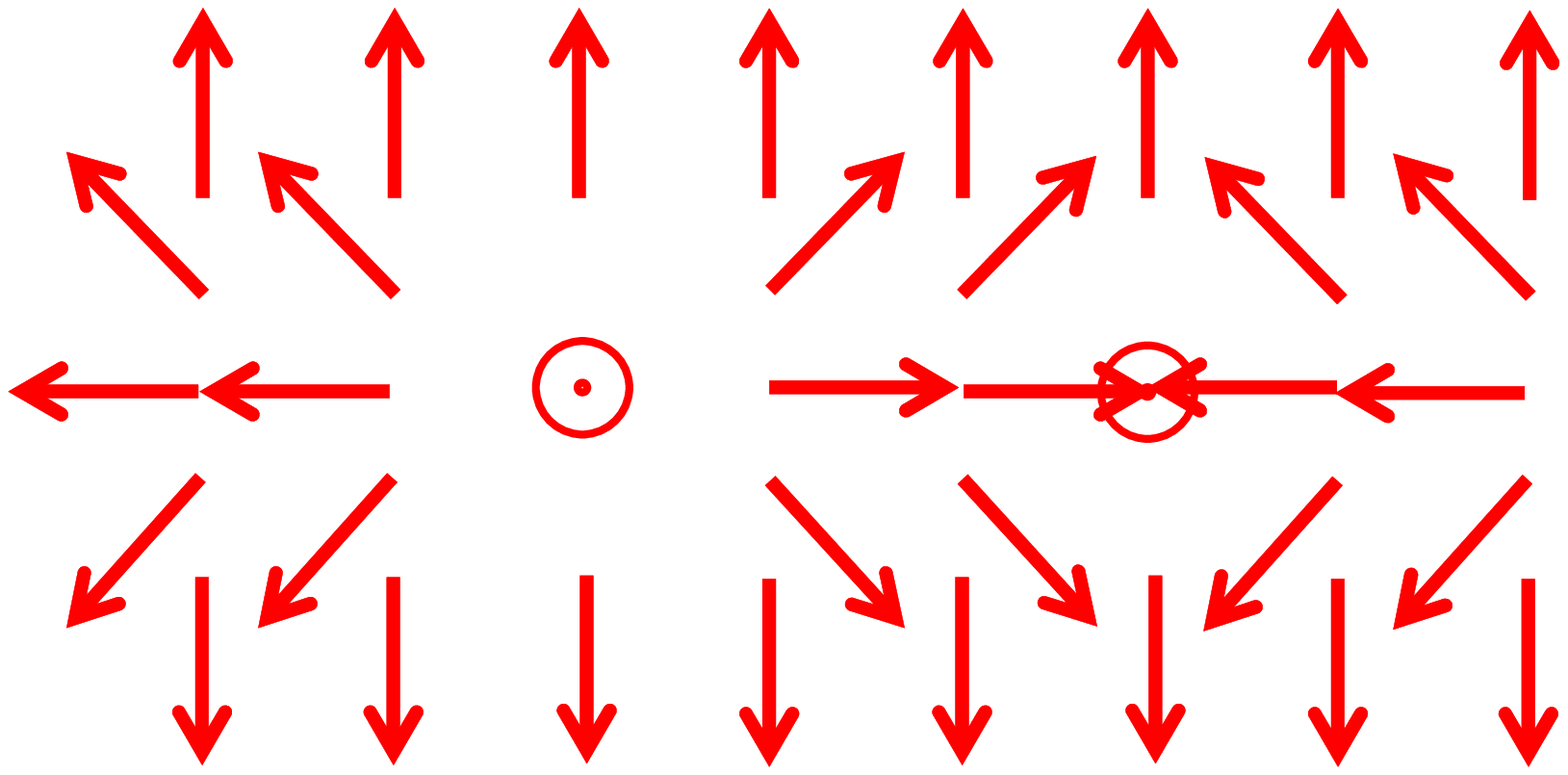} &
\includegraphics[width=0.4\linewidth,keepaspectratio]{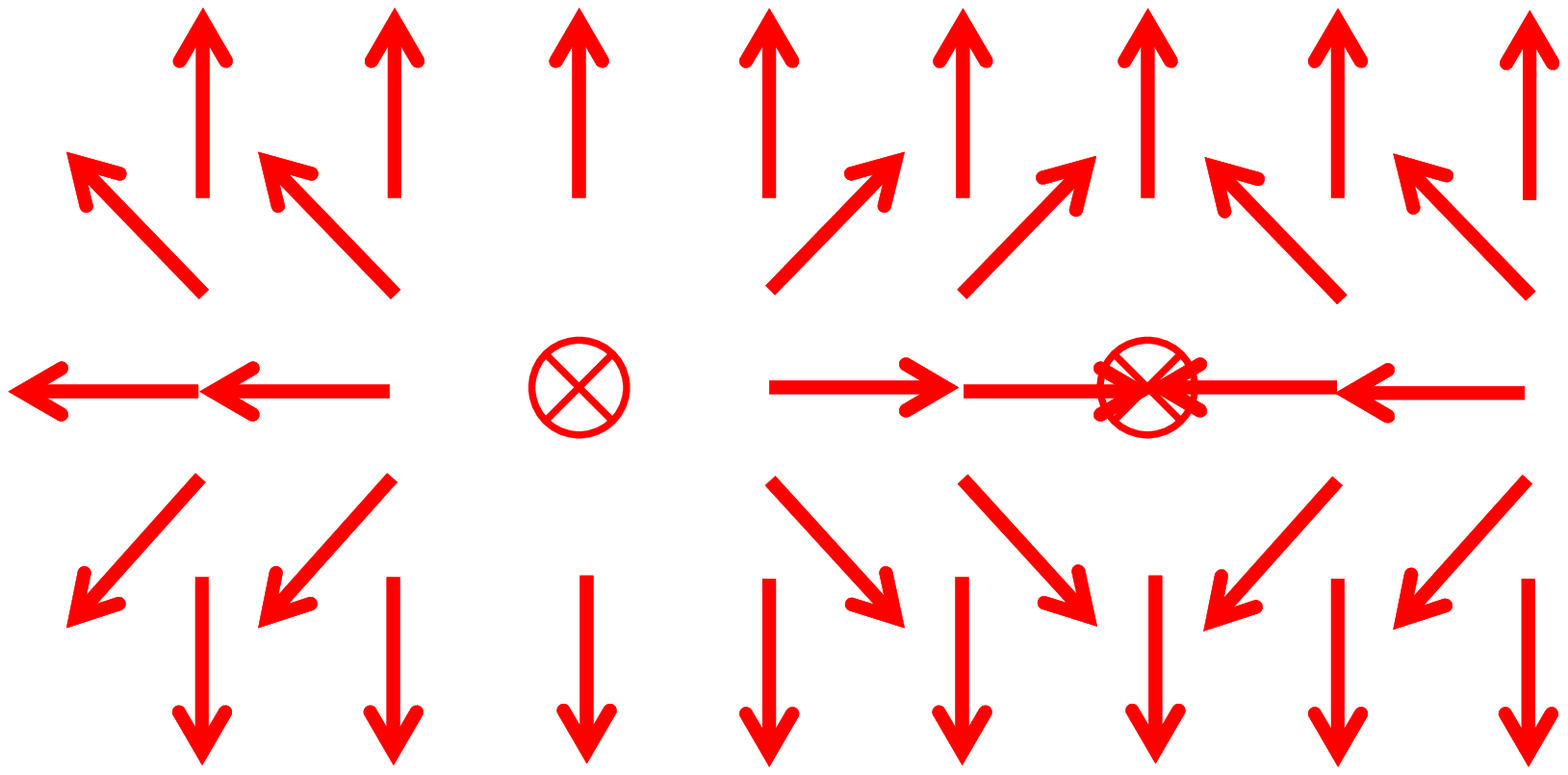} \\
(a) $(+1,+\1{2},+\1{2})+(-1,+\1{2},-\1{2})$ & 
(b) $(+1,-\1{2},-\1{2})+(-1,-\1{2},+\1{2})$ \\
\includegraphics[width=0.4\linewidth,keepaspectratio]{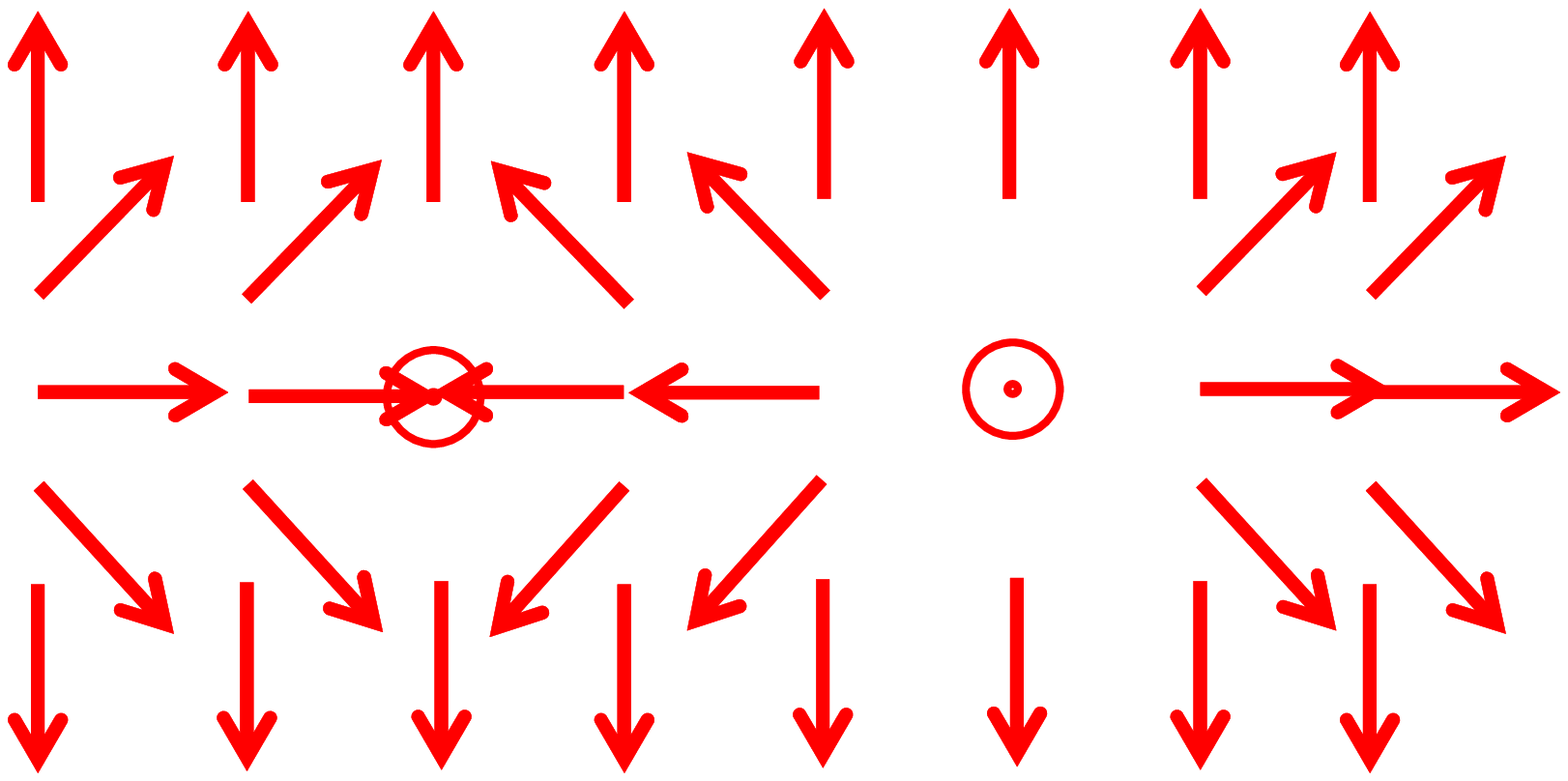} &
\includegraphics[width=0.4\linewidth,keepaspectratio]{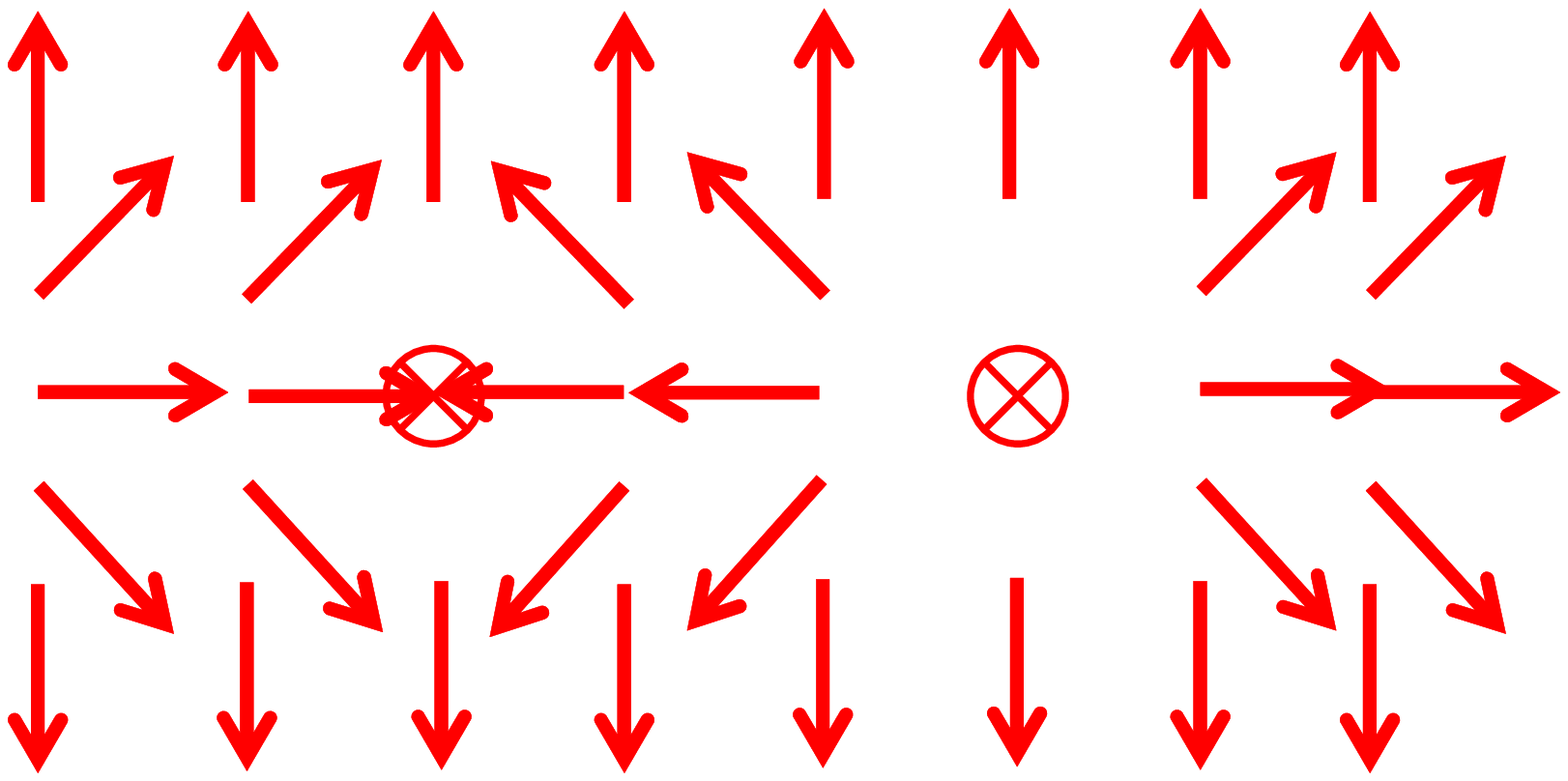} \\
(c) $(-1,+\1{2},-\1{2})+(+1,+\1{2},+\1{2})$ & 
(d) $(-1,-\1{2},+\1{2})+(+1,-\1{2},-\1{2})$ 
\end{tabular}
\end{center}
\caption{Bions in the $O(3)$ model 
with the boundary condition $(-,+,+)$.
The notations are the same with Fig.~\ref{fig:fractional-O3-3}.
(c) and (d) are isomorphic to (a) and (b), respectively, 
by a $2\pi$ rotation along an axis 
at the center of the domain wall.
\label{fig:bion-O3-3}}
\end{figure}
%%%%%%%%%%%%%%%%%%%%%%

There are no fixed points for 
the boundary condition $(-,-,-)$ unlike the above cases. 
This implies that there is no vacuum. In fact, 
even in the least energy configuration, 
the fields must be twisted because of the boundary condition, 
and there exist gradient energy. 
We do not have localized solitons wrapping 
around a fixed manifold. 
We interpret this situation that 
there is a space-filling soliton of codimension zero
to be consistent with the other cases.
Then, we interpret the original target space $S^2$ 
as moduli of the space-filling soliton.

An (anti-)instanton is separated into two 
fractional (anti-)instantons with the boundary condition 
$(-,-,-)$ as shown in 
Fig.~\ref{fig:fractional-O3-3}. 
The ansatz for an isolated fractional (anti-)instanton can be given as
\beq
&& \left(\begin{array}{c}
  n_1 \\ n_2 \\ n_3
  \end{array}\right)
= \left(\begin{array}{ccc}
   1 & 0 & 0\\
   0 &       \cos f(x^1) & \mp \sin f(x^1) \\
   0 & \pm \sin f(x^1) &    \cos f(x^1) \\ 
\end{array}\right)
\left(\begin{array}{c}
-\cos {\pi \over R} x^2 \\ 
  \sin {\pi \over R} x^2 \\
 0
\end{array}\right)
= \left(\begin{array}{c}
- \cos f(x^1) \\ 
    \cos f(x^1) \sin {\pi \over R} x^2  \\ 
   \pm \sin f(x^1)  \sin {\pi \over R} x^2 
\end{array}\right)
, \quad\quad\label{eq:O(3)---}\\
&& f(x^1=-\infty) = 0, \quad f(x^1=+\infty) = \pi.   \label{eq:O(3)---bc}
\eeq
Each fractional (anti-)instanton wraps a half sphere of the target space 
$S^2$. 
For instance, the left half of Fig.~\ref{fig:fractional-O3-3}(a) 
wraps a half sphere as in Fig.~\ref{fig:fractional-path-O3}(c).
The topological charges of fractional (anti-)instantons 
are summarized in Table~\ref{table:homotopy-O3-3}.
Here, we formally use $\pi_{-1}$ for 
space-filling solitons of codimension zero, 
to be consistent with the other cases.  
%%%%%%%%%%%%%%%%%%%%%%%%%
\begin{table}[h]
\begin{tabular}{c|c|c|c} 
     & $\pi_{-1}$ & $\pi_2$ & $\pi_2$ \\ \hline
Fig.~\ref{fig:O(3)} (3a) &$ +1$     & $+1/2$ & $+1/2$ \\
Fig.~\ref{fig:O(3)} (3b) &$ -1$     & $-1/2$ & $+1/2$ \\
Fig.~\ref{fig:O(3)} (3c) &$ -1$     & $+1/2$ & $-1/2$ \\
Fig.~\ref{fig:O(3)} (3d) &$ +1$     & $-1/2$ & $-1/2$    
\end{tabular}
\caption{Homotopy groups of fractional instantons in the 
$O(3)$ model with the boundary condition $(-,-,-)$.
The columns represent the homotopy groups  
of a host soliton $\pi_{-1}$, a daughter soliton $\pi_2$, 
and the total instanton $\pi_2$ from left to right.
$\pi_{-1}$ is merely formal.
\label{table:homotopy-O3-3}}
\end{table}
%%%%%%%%%%%%%%%%%%%%%%

It may be worth to mention that 
the existence of a daughter soliton is not required 
from the boundary condition, 
because 
the $x^1$-dependent rotation in Eq.~(\ref{eq:O(3)---})
is not necessary and 
a configuration 
$(n_1,n_2,n_3) =
(-\cos {\pi \over R} x_2,
  \sin {\pi \over R} x_2,
 0)$ 
is in fact a minimum energy state. 
This is in contrast to the other boundary conditions 
in which the existence of a daughter soliton 
is required in the presence of a host soliton.

In this case, the boundary condition is not enough to stabilize 
fractional instantons, 
unlike the other two boundary conditions. 
One needs to 
add a potential term 
\beq 
 V = m^2 n_3^2 \label{eq:O(3)add-pot}
\eeq 
to the original Lagrangian 
for the stability of half instantons.
For this particular potential term,
the function $f$ in Eq.~(\ref{eq:O(3)---}) 
is a sine-Gordon kink, 
$f = \arctan \exp (m x^1)$.

For one instanton, one chooses the boundary condition of 
$f$ as $f = 0$ at $x^1 \to -\infty $ and $f = 2\pi$ at $x^1 \to +\infty$, 
instead of Eq.~(\ref{eq:O(3)---bc}).
The interaction energy between two fractional instantons 
would be suppressed exponentially because 
the energy density between the two fractional instantons 
is the same with that outside them, 
and there is no confining force between them. 
The detail form depends on the choice of a potential term 
to stabilize half instantons.
For the potential in Eq.~(\ref{eq:O(3)add-pot}), 
the interaction is that of two sine-Gordon kinks, 
which is repulsive.

Bions with the boundary condition 
$(-,-,-)$ are shown in 
Fig.~\ref{fig:bion-O3-3}.
The function $f$ in Eq.~(\ref{eq:O(3)---}) behaves as 
as $f = 0$ at $x^1 -\infty $, $f \sim \pi$ in some intermediate region 
and back to $f=0$ at $x^1 \to +\infty$. 
The interaction energy between two fractional instantons 
constituting a bion
would be suppressed exponentially because of the same reason 
with the above while the detailed form depends on the 
choice of the potential term.
For the potential term in Eq.~(\ref{eq:O(3)add-pot}), 
the interaction is that between a sine-Gordon kink and an anti-kink.

%%%%%%%%%%%%%%%%%%%%%%%%%%%%%%%%
\section{Fractional instantons and bions in the $O(4)$ model}
\label{sec:O(4)}

\subsection{$(-,+,+,+)$: global monopole with an Ising spin 
or half Skyrmion-monopole}

The fixed manifold is characterized by 
$n_1=0$, equivalently $(n_2)^2+(n_3)^2+(n_4)^2=1$, 
which is  the moduli space of vacua ${\cal N} \simeq S^2$. 
Therefore, it has a nontrivial homotopy 
$\pi_2(S^2) \simeq {\mathbb Z}$, allowing a global monopole.
In the monopole core $n_2= n_3=n_4=0$, 
the field $n_1$ appears taking a value $n_1 = \pm 1$ 
in the center, giving an Ising spin degree of freedom to the monopole, 
that is, 
the moduli space of the monopole is ${\cal M} \simeq \{\pm 1\}$.
This is a fractional (anti-)instanton 
with the boundary condition 
$(-,+,+,+)$ 
as drawn in Fig.~\ref{fig:O(4)} (1a)--(1d). 
%%%%%%%%%%%%%%%%%%%%%
\begin{figure}
\begin{center}
\begin{tabular}{cc}
\includegraphics[width=0.49\linewidth,keepaspectratio]{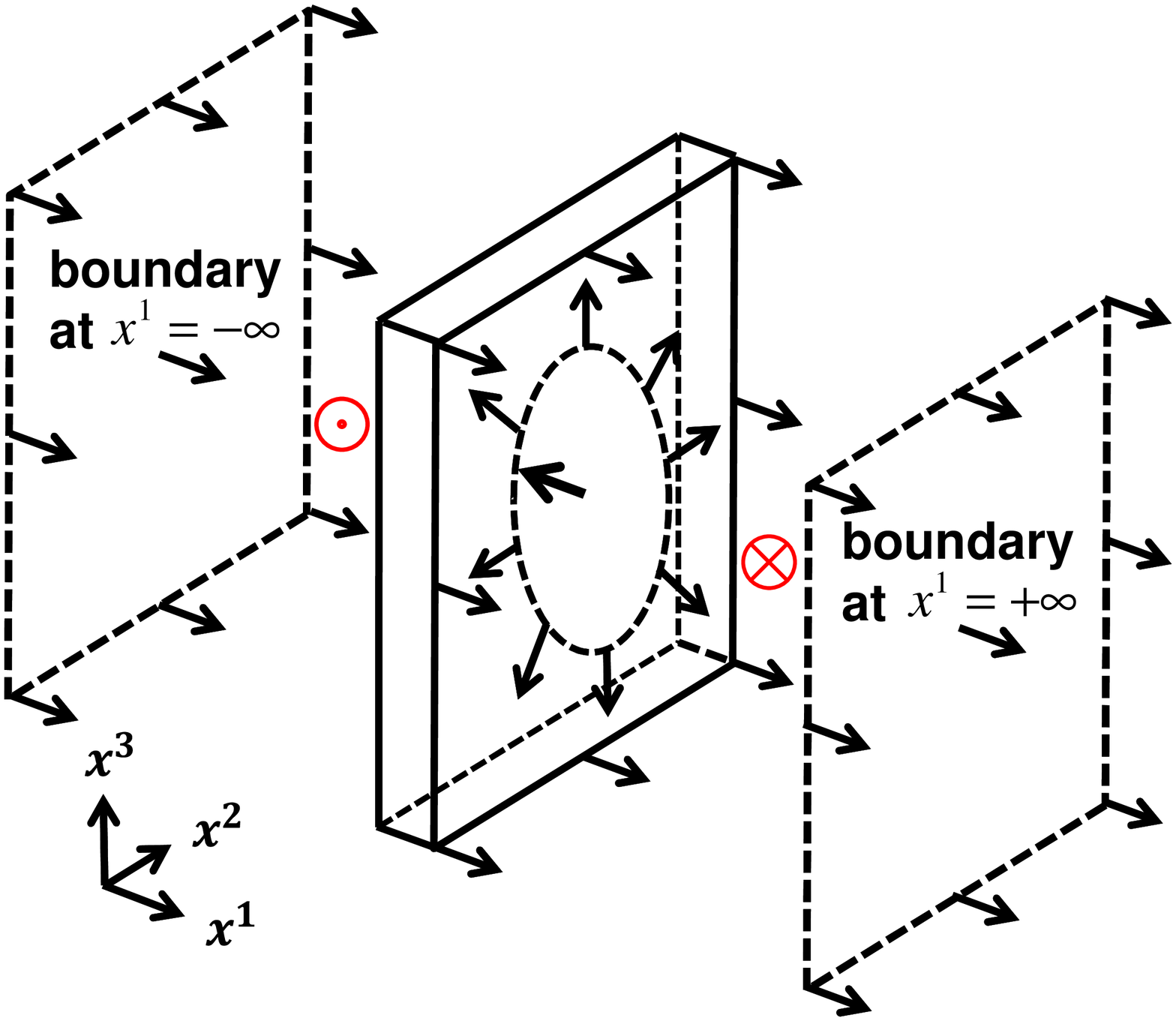} &
\includegraphics[width=0.49\linewidth,keepaspectratio]{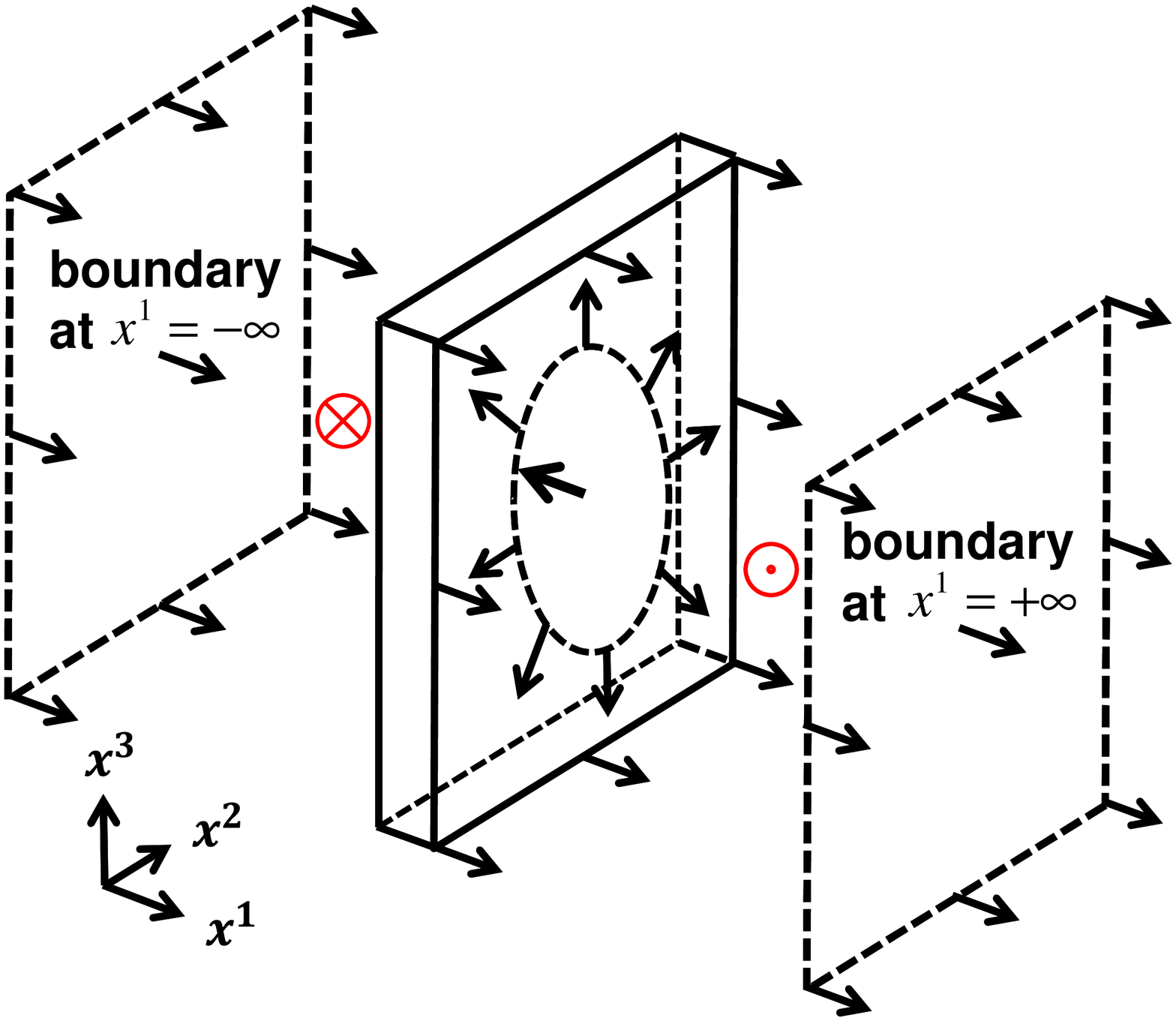} \\
(a) $(+1,+\1{2},+\1{2})+(-1,-\1{2},+\1{2})$ & 
(b) $(+1,-\1{2},-\1{2})+(-1,+\1{2},-\1{2})$ \\
\includegraphics[width=0.49\linewidth,keepaspectratio]{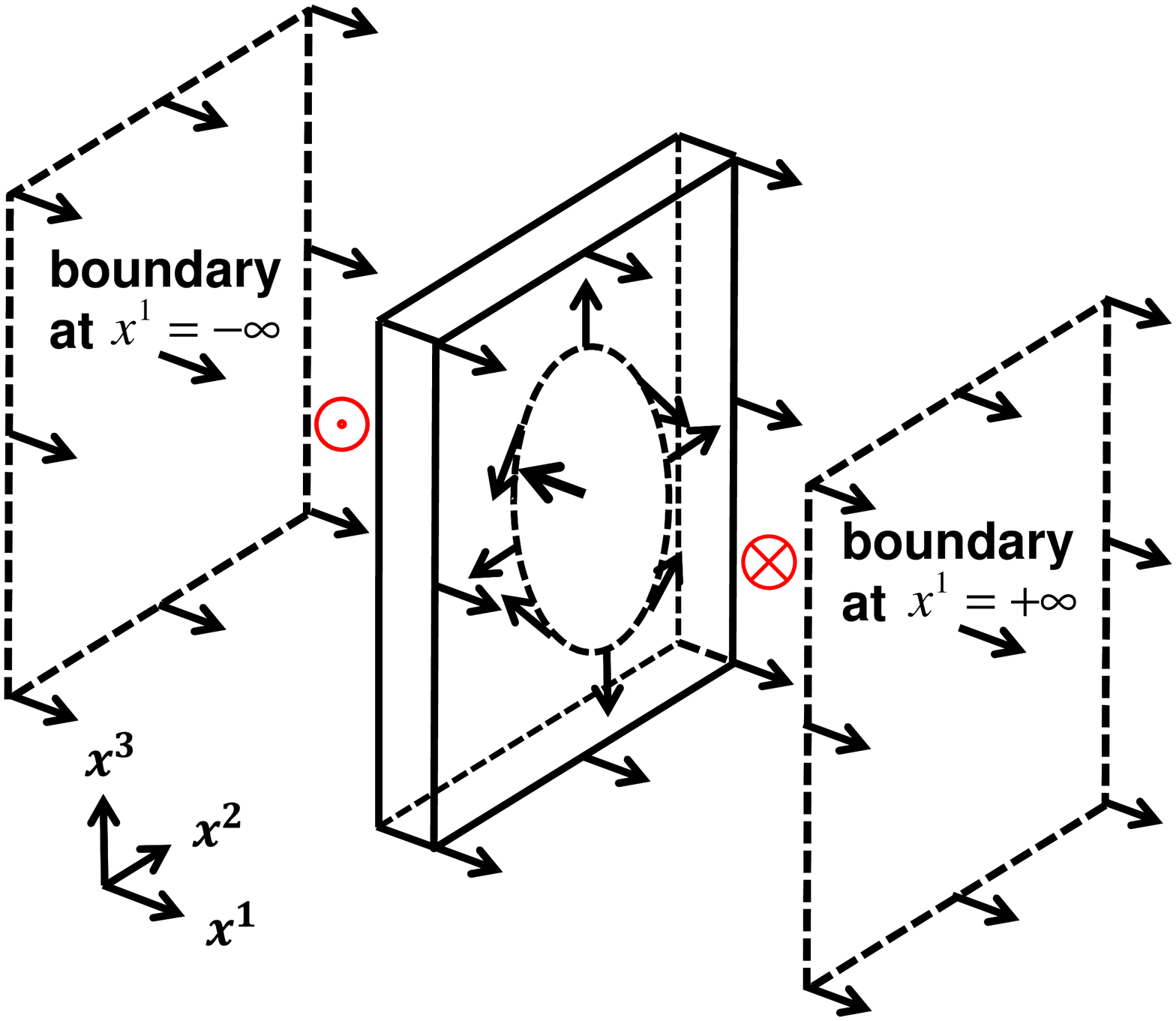} &
\includegraphics[width=0.49\linewidth,keepaspectratio]{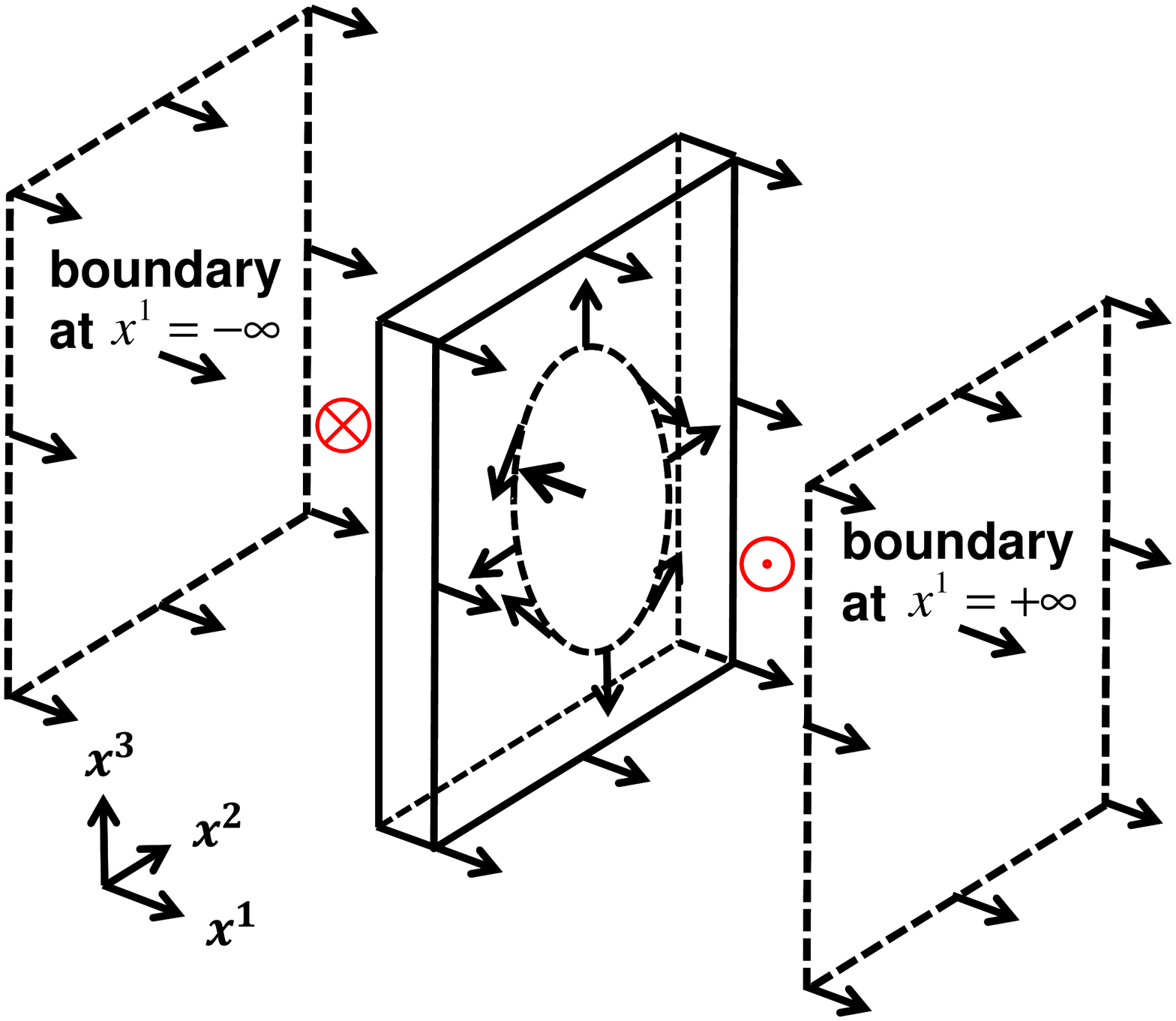} \\
(c) $(-1,-\1{2},+\1{2})+(+1,+\1{2},+\1{2})$ & 
(d) $(-1,+\1{2},-\1{2})+(+1,-\1{2},-\1{2})$ \end{tabular}
\end{center}
\caption{Fractional instantons in the $O(4)$ model 
with the boundary condition $(-,+,+,+)$. 
$\odot$ and $\otimes$ correspond to 
$n_1=+1$ and $n_1 = -1$, respectively, 
representing the moduli space (an Ising spin) ${\cal M}$ of a monopole 
Black arrows represent $(n_2,n_3,n_4)$ with $n_2^2+n_3^2+n_3^2=1$ 
($n_1=0$) parameterizing the moduli space of vacua ${\cal N}\simeq S^2$. 
Brackets  $(*,*,*)$ denote
topological charges for a host monopole characterized by $\pi_2$, 
that for an Ising spin characterized by $\pi_0$, 
and that for an instanton characterized by $\pi_3$. 
(a) An instanton is split into two fractional instantons 
$(+1,+\1{2},+\1{2})$ and $(-1,-\1{2},+\1{2})$ 
separated by a lump.
(b) An anti-instanton is split into 
two fractional anti-instantons 
$(+1,-\1{2},-\1{2})$ and $(-1,+\1{2},-\1{2})$ 
separated by an anti-lump. 
(c) and (d) are isomorphic to (a) and (b), respectively, 
by a $2\pi$ rotation along the $x^1$ axis.
\label{fig:fractional-O4-1}}
\end{figure}
%%%%%%%%%%%%%%%%%%%%%%%% 
Apparently, 
each fractional instanton wraps a half of the target space 
$S^3$.
A unit (anti-)instanton can be separated 
into two fractional (anti-)instantons  
as shown in 
Fig.~\ref{fig:fractional-O4-1}. 
Again, each fractional instanton wraps a half of the target space 
$S^3$. 
If one is well separated from the rests,
it becomes one of Fig.~\ref{fig:O(4)} (1a)--(1d). 
The topological charges of fractional (anti-)instantons 
with the boundary condition $(-,+,+,+)$ 
are summarized in Table \ref{table:homotopy-O4-1}.
Here, we have defined the value of $\pi_0$ for the Ising spin to be $\pm 1/2$ 
to be consistent with the other boundary conditions 
discussed below. 
%%%%%%%%%%%%%%%%%%%%%%%%%
\begin{table}[h]
\begin{tabular}{c|c|c|c} 
     & $\pi_2$ & $\pi_0$ & $\pi_3$ \\ \hline
Fig.~\ref{fig:O(4)} (1a) &$ +1$     & $+1/2$ & $+1/2$ \\
Fig.~\ref{fig:O(4)} (1b) &$ -1$     & $-1/2$ & $+1/2$ \\
Fig.~\ref{fig:O(4)} (1c) &$ -1$     & $+1/2$ & $-1/2$ \\
Fig.~\ref{fig:O(4)} (1d) &$ +1$     & $-1/2$ & $-1/2$  
\end{tabular}
\caption{Homotopy groups of fractional instantons in the 
$O(4)$ model with the boundary condition $(-,+,+,+)$.
The columns represent the homotopy groups  
of a host soliton $\pi_2$, a daughter soliton $\pi_0$, 
and the total instanton $\pi_3$ from left to right. 
\label{table:homotopy-O4-1}}
\end{table}
%%%%%%%%%%%%%%%%%%%%%%

We need higher derivative (Skyrme) term 
for the stability of fractional instantons (Skyrmions) 
\cite{Gudnason:2014xxx}, 
as is so for usual Skyrmions. 

When the compactification radius $R$ is large,
the interaction between two well-separated fractional instantons 
is the same with that of global monopoles at large distance. 
For a small compactification radius $R$ of the order of 
fractional instanton size, the interaction 
between two well-separated fractional instantons 
at distance $r$  
is $E_{\rm int} \sim r$ because of 
a lump string connecting them.

\begin{figure}
\begin{center}
\begin{tabular}{cc}
\includegraphics[width=0.49\linewidth,keepaspectratio]{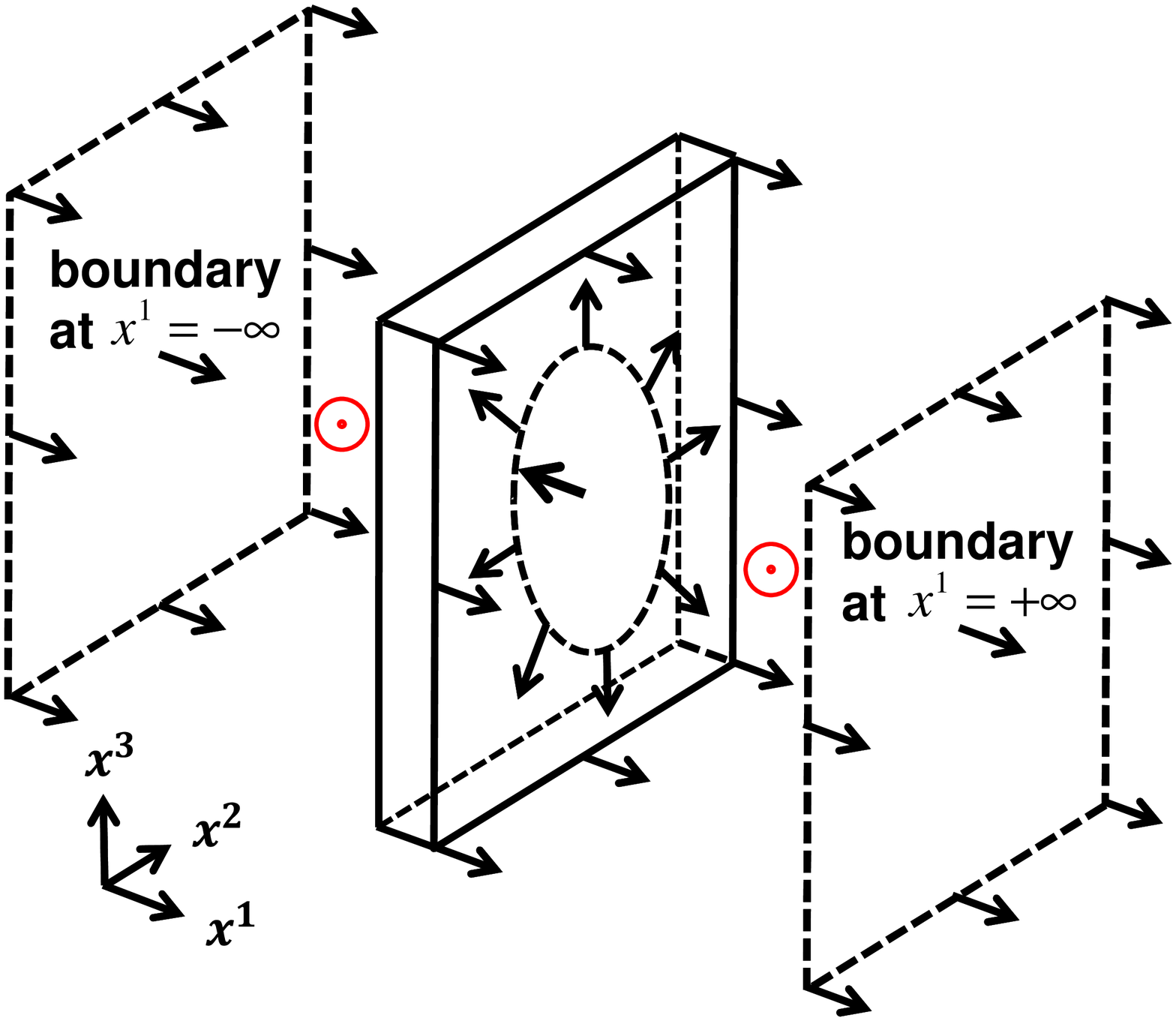} &
\includegraphics[width=0.49\linewidth,keepaspectratio]{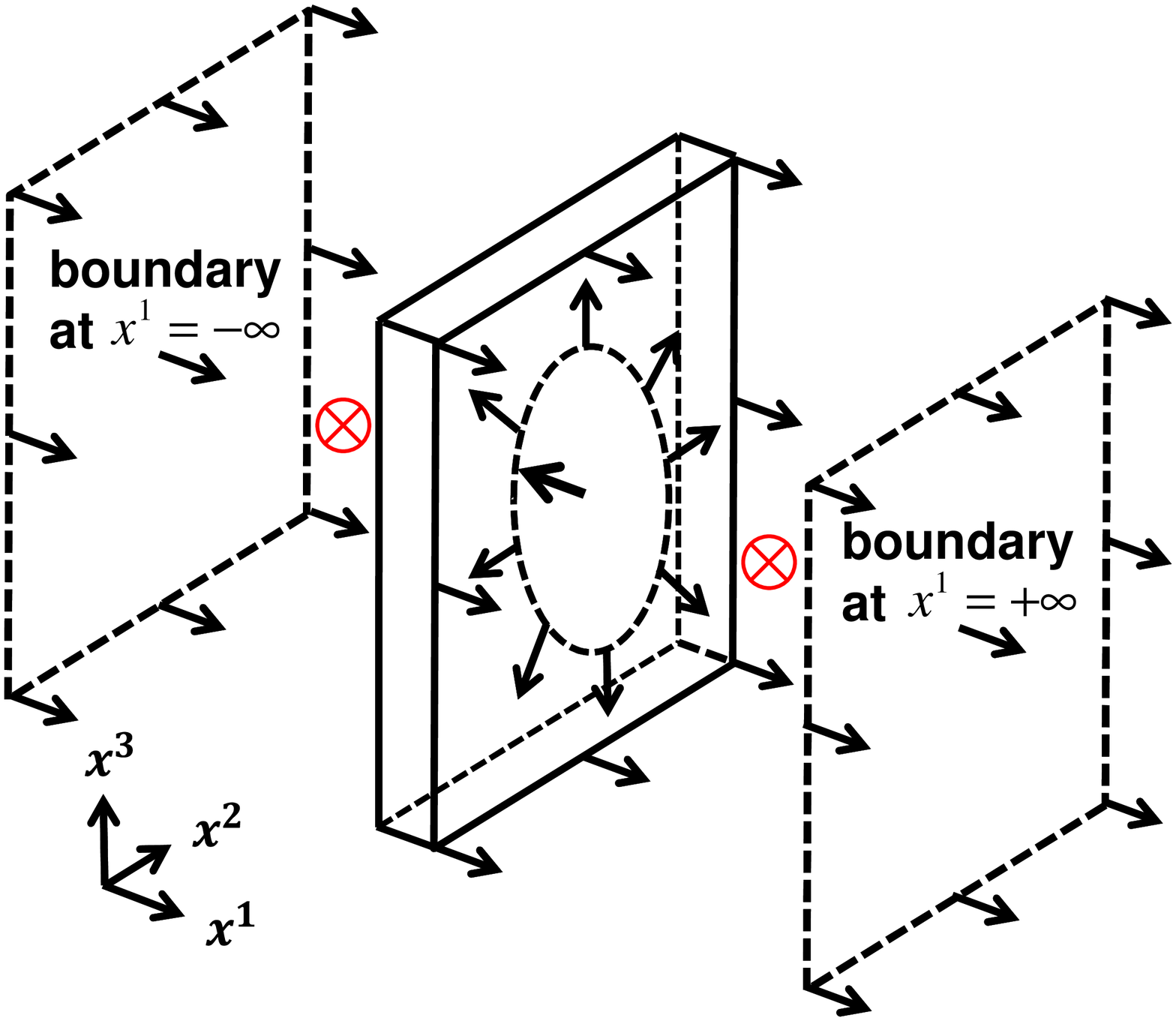} \\
(a) $(+1,+\1{2},+\1{2})+(-1,+\1{2},-\1{2})$ & 
(b) $(+1,-\1{2},-\1{2})+(-1,-\1{2},+\1{2})$ \\
\includegraphics[width=0.49\linewidth,keepaspectratio]{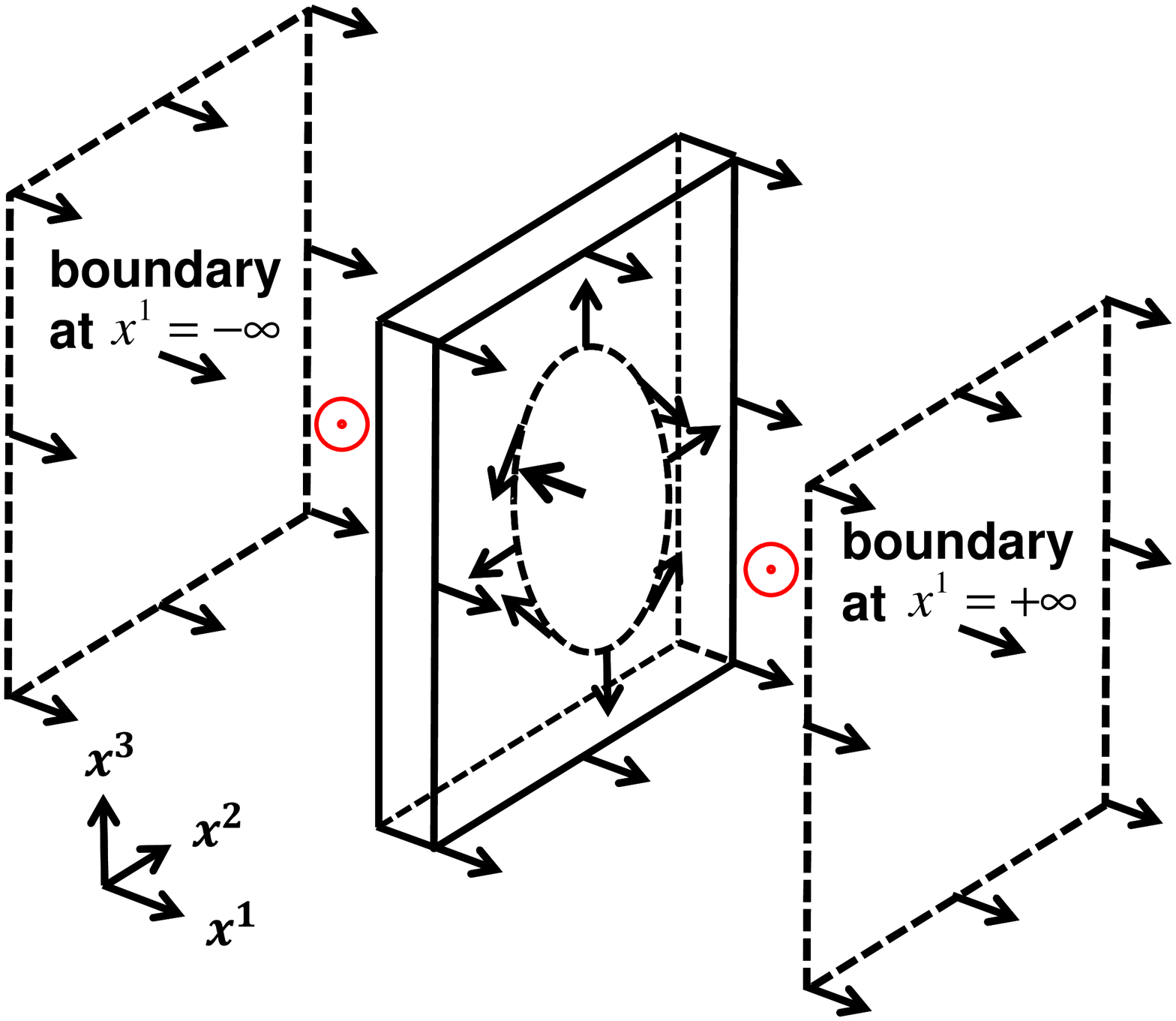} &
\includegraphics[width=0.49\linewidth,keepaspectratio]{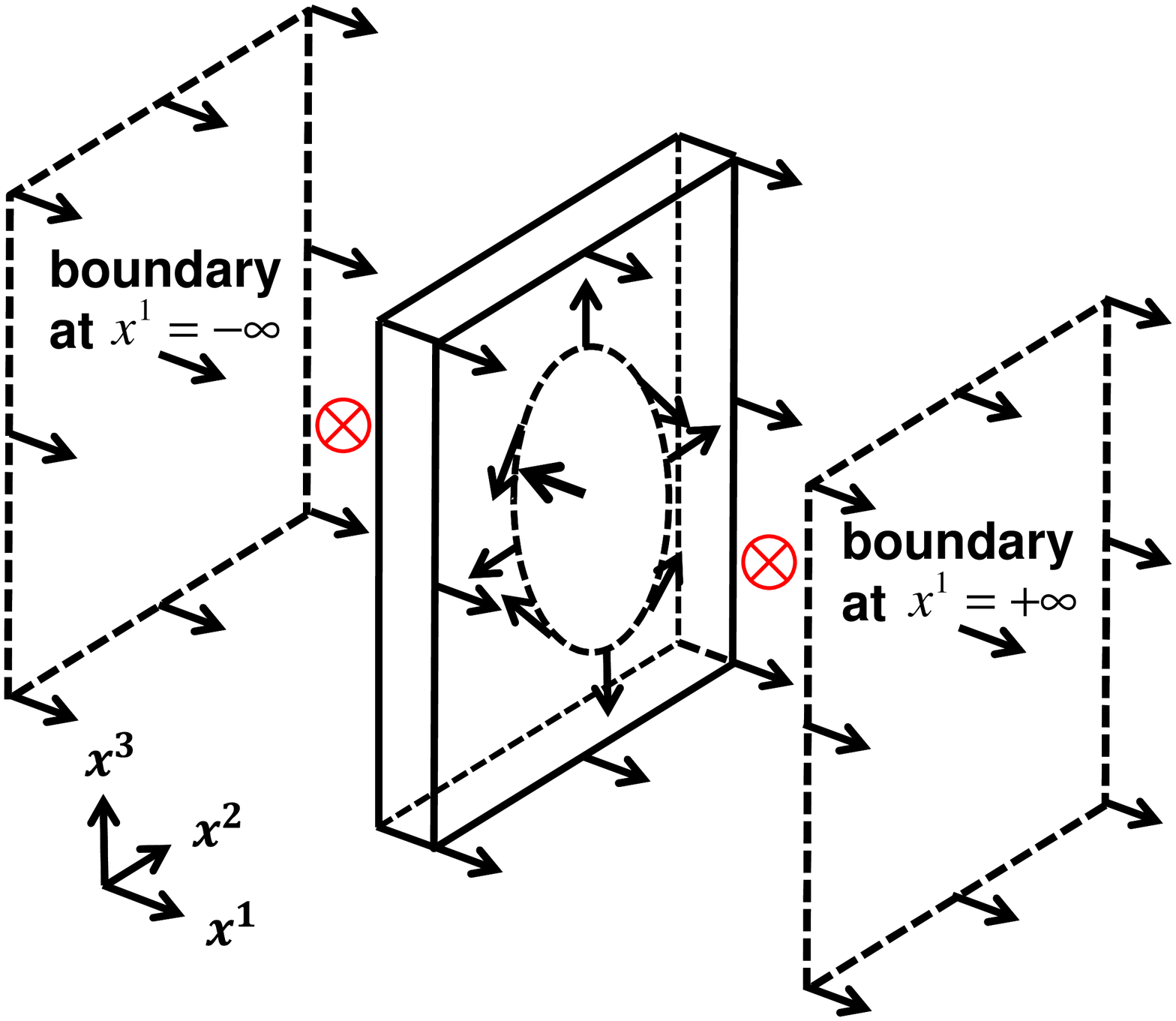} \\
(c) $(-1,+\1{2},-\1{2})+(+1,+\1{2},+\1{2})$ & 
(d) $(-1,-\1{2},+\1{2})+(+1,-\1{2},-\1{2})$ 
\end{tabular}
\end{center}
\caption{Bions in the $O(4)$ model 
with the boundary condition $(-,+,+,+)$. 
The notations are the same with Fig.~\ref{fig:fractional-O4-1}.
(c) and (d) are isomorphic to (a) and (b), respectively, 
by a $2\pi$ rotation along an axis parallel to the $x^1$ axis 
at the center of the domain wall.
\label{fig:bion-O4-1}}
\end{figure}

Bions with the boundary condition 
$(-,+,+,+)$ are schematically drawn in 
Fig.~\ref{fig:bion-O4-1}.  
While each domain separated by a lump 
an instanton charge,
the total instanton charges are canceled out.
Again, the two well separated fractional instantons at distance $r$ 
are confined by a linear potential 
$E_{\rm int} \sim r$  
for a compactification radius $R$ of the order of 
fractional instanton size.
%%%%%%%%%%%%%%%%%%%%%
%%%%%%%%%%%%%%%%%%%%%%%%

If one gauges the $SO(3)$ symmetry acting on $(n_2,n_3,n_4)$, 
a half-Skyrmion monopole
becomes local, that is, of `t Hooft-Polyakov type 
\cite{'tHooft:1974qc} 
having finite energy. 
This is in fact the case of 
the $SO(3)$ gauged model 
with a potential term $V = m^2 n_1^2$
defined on ${\mathbb R}^3$ {\it without} twisted boundary condition
\cite{Brihaye:1998}.

%%%%%%%%%%%%%%%%%%%
\subsection{$(-,-,+,+)$: a half sine-Gordon kink inside a vortex}

%%%%%%%%%%%%%%%%%%%%%
\begin{figure}
\begin{center}
\begin{tabular}{cc}
\includegraphics[width=0.35\linewidth,keepaspectratio]{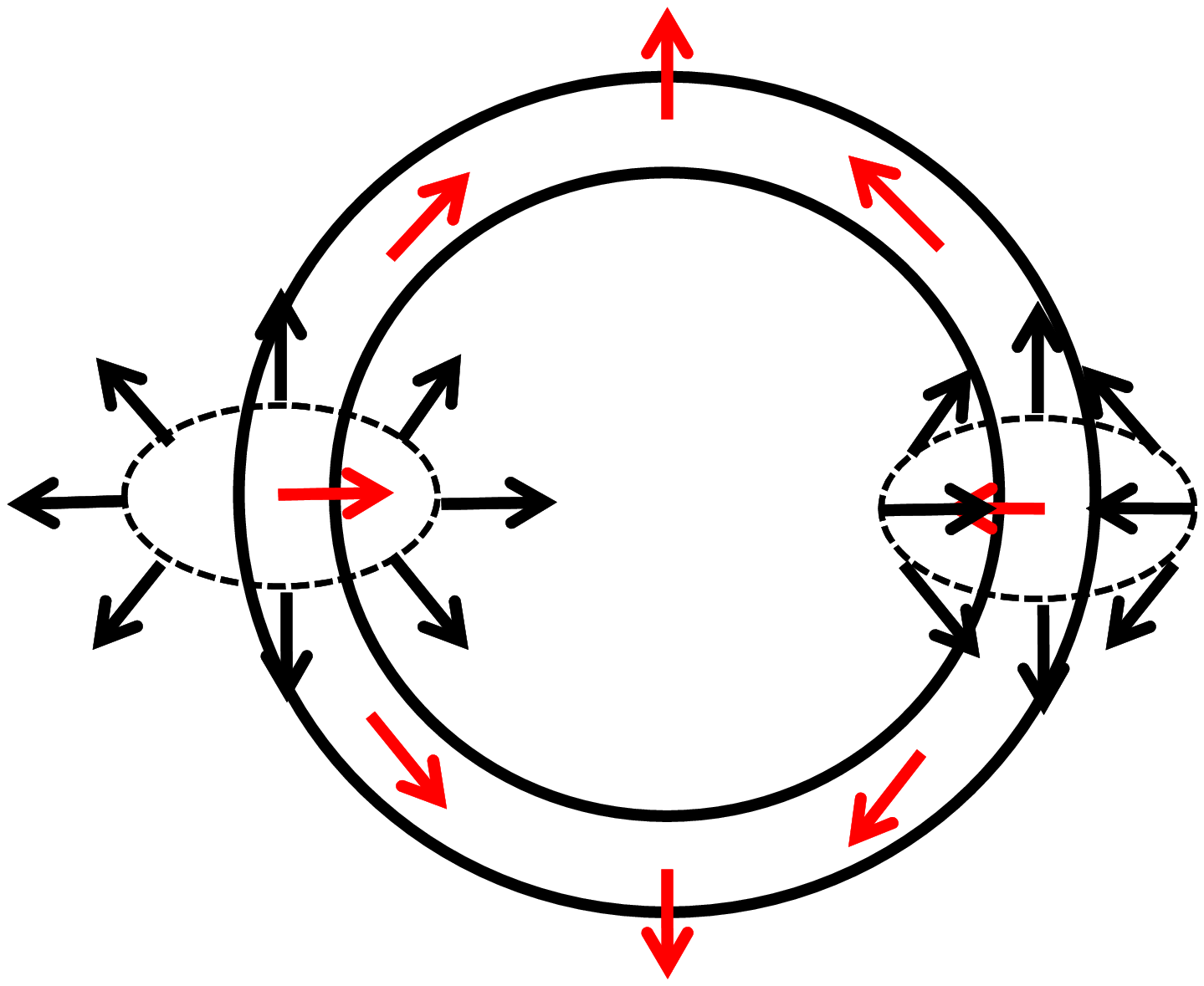} &
\includegraphics[width=0.35\linewidth,keepaspectratio]{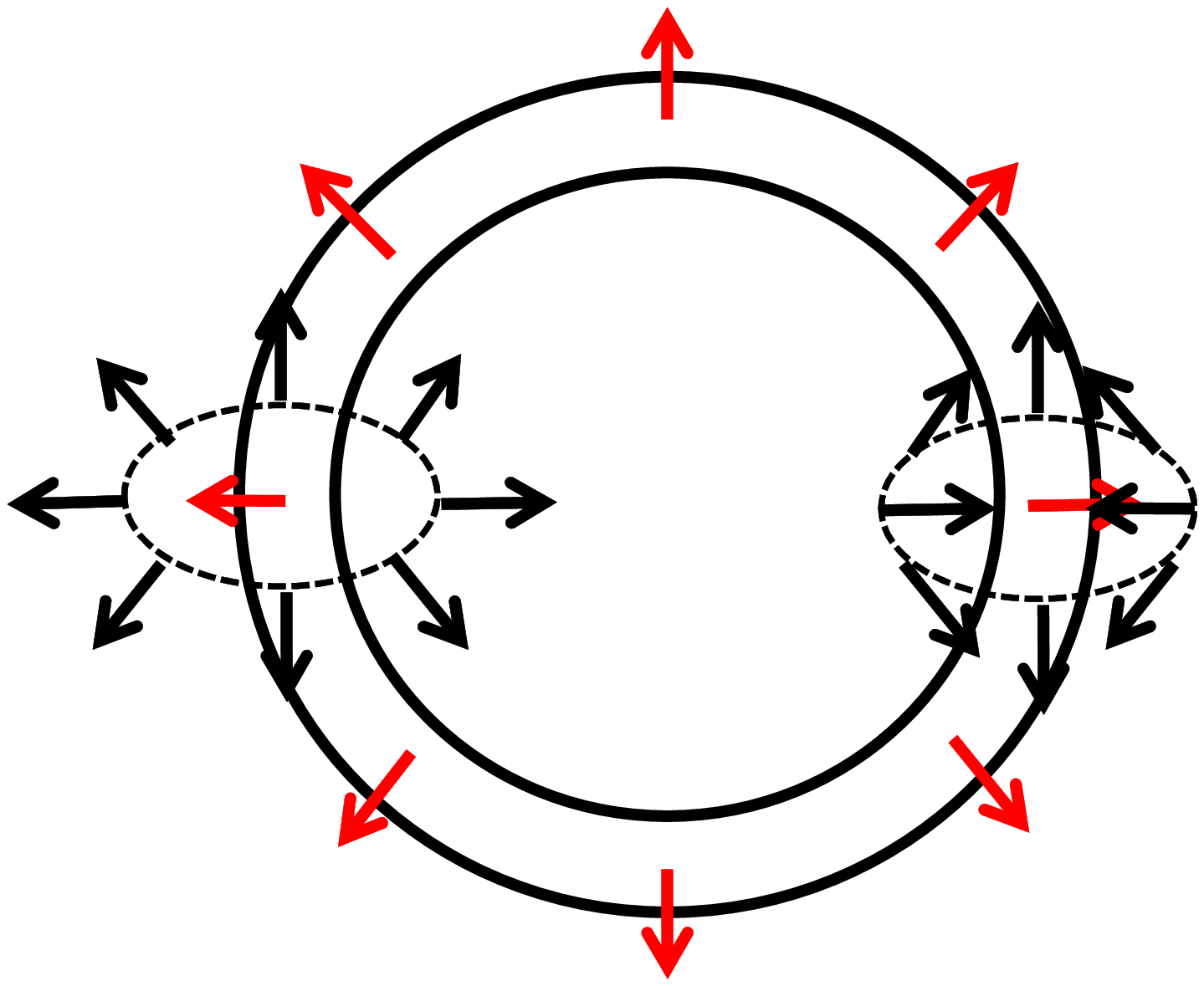} \\
(a) $(+1,+\1{2},+\1{2})+(-1,-\1{2},+\1{2})$ & 
(b) $(+1,-\1{2},-\1{2})+(-1,+\1{2},-\1{2})$ \\
\hspace{1.7cm}
\includegraphics[width=0.35\linewidth,keepaspectratio]{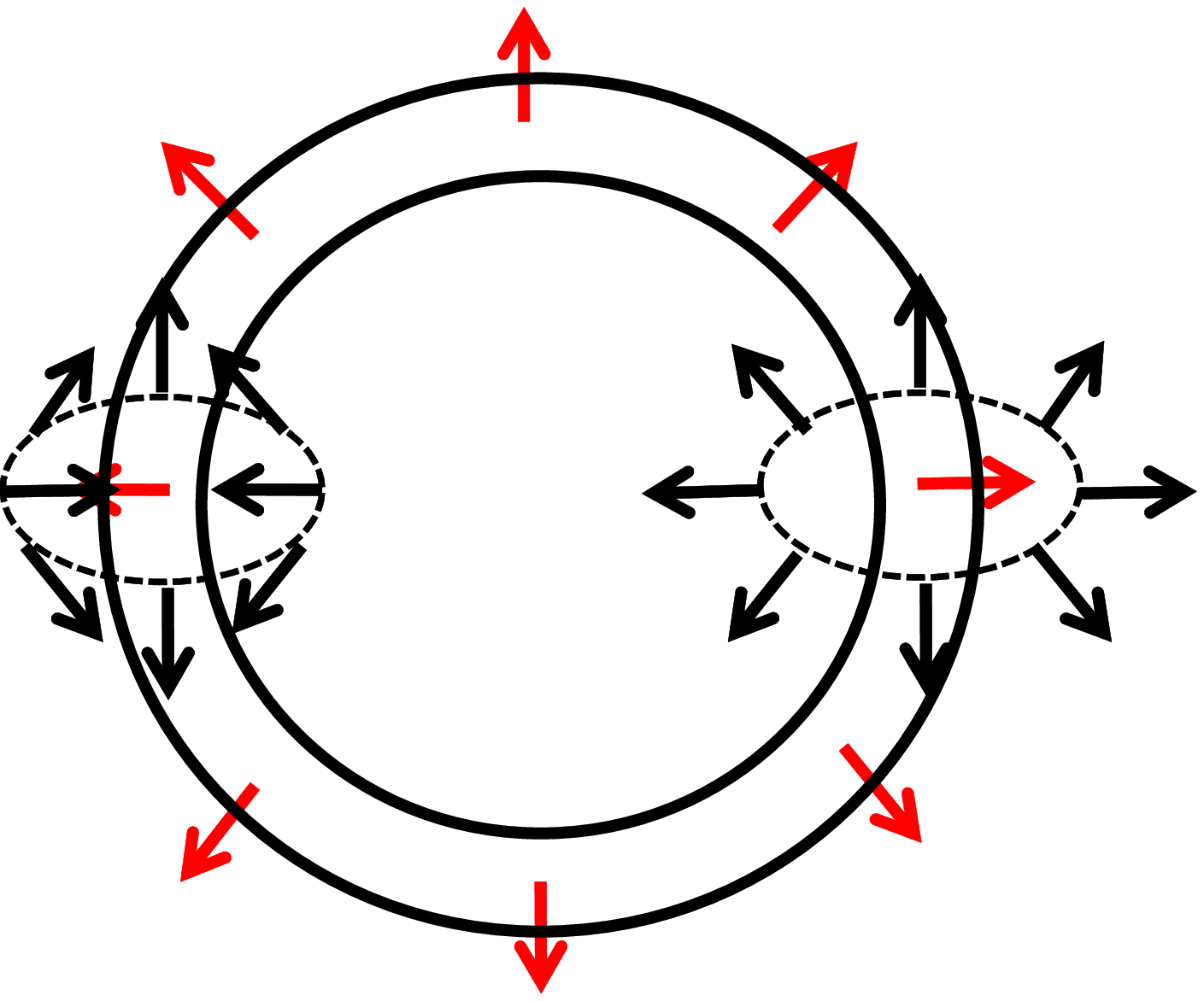} &
\hspace{1.7cm}
\includegraphics[width=0.35\linewidth,keepaspectratio]{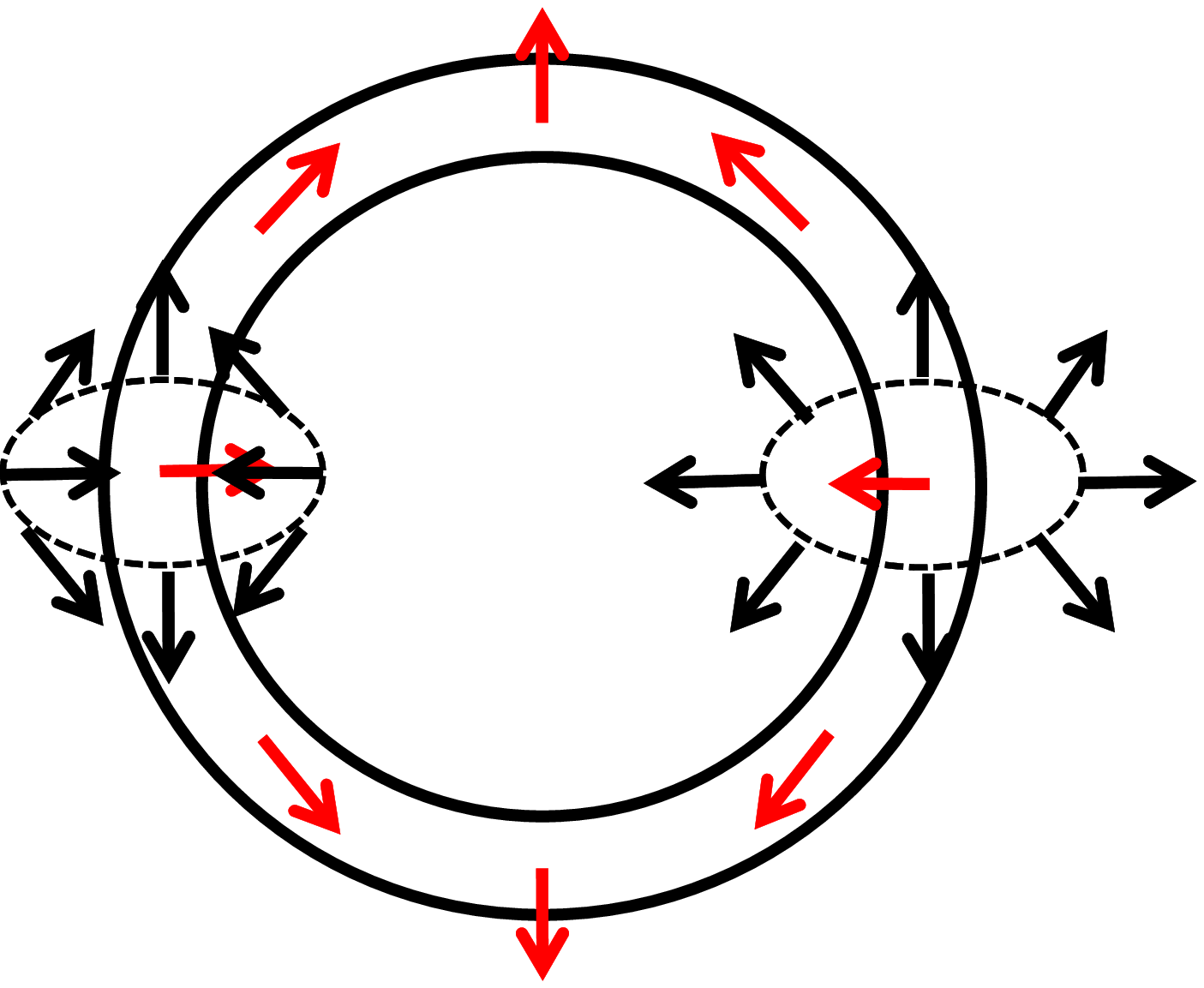} \\
(c) $(-1,-\1{2},+\1{2})+(+1,+\1{2},+\1{2})$ & 
(d) $(-1,+\1{2},-\1{2})+(+1,-\1{2},-\1{2})$ 
\end{tabular}
\includegraphics[width=0.1\linewidth,keepaspectratio]{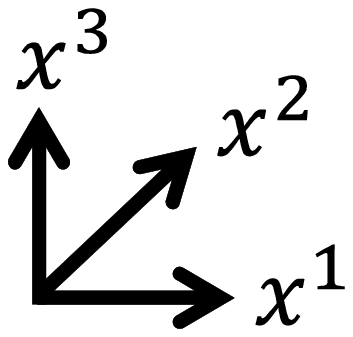} 
\end{center}
\caption{Fractional instantons from instantons in the $O(4)$ model 
with the boundary condition $(-,-,+,+)$. 
Black arrows represent 
$(n_1,n_2)$ with $n_1^2+n_2^2=1$ ($n_3=n_4=0$) parameterizing 
the moduli space of vacua ${\cal N}\simeq S^1$, 
while red arrows represent $(n_3,n_4)$ with $n_3^2+n_4^2=1$ 
($n_1=n_2=0$)
parameterizing 
the moduli space of a vortex ${\cal M}\simeq S^1$. 
An instanton can be represented as a vorton, that is, 
a vortex ring along which the $U(1)$ modulus is twisted once. 
Brackets  $(*,*,*)$ denote
topological charges for a host global vortex characterized by $\pi_1$, 
that for a sine-Gordon kink characterized by $\pi_1$, 
and that for an instanton characterized by $\pi_3$. 
(a) An instanton (vorton) can be split into two fractional instantons 
$(+1,+\1{2},+\1{2})$ and $(-1,-\1{2},+\1{2})$.
(b) An anti-instanton (anti-vorton) can be split into 
two fractional anti-instantons 
$(+1,-\1{2},-\1{2})$ and $(-1,+\1{2},-\1{2})$.  
(c) and (d) are isomorphic to (a) and (b), respectively, 
by a $2\pi$ rotation along an axis parallel to the $x^1$ axis.
\label{fig:fractional-O4-2}}
\end{figure}
%%%%%%%%%%%%%%%%%%%%%%%%
%%%%%%%%%%%%%%%%%%%%%
\begin{figure}
\begin{center}
\includegraphics[width=0.09\linewidth,keepaspectratio]{3d-frame} 
\includegraphics[width=0.90\linewidth,keepaspectratio]{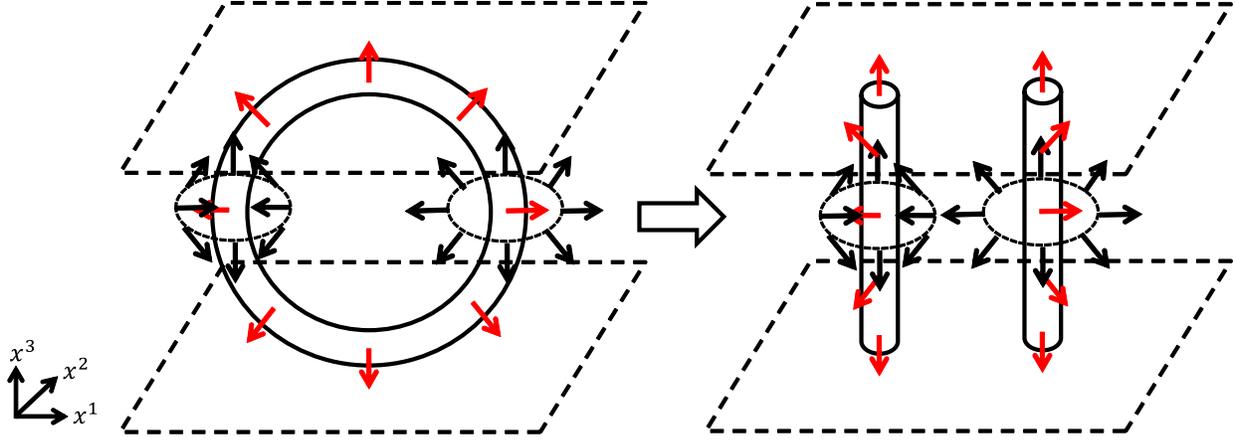} 
\end{center}
\caption{A twisted vortex ring of the size of the compact direction 
decays into two fractional 
instantons through a reconnection in the $O(4)$ model 
with the boundary condition $(-,-,+,+)$. 
The notations are the same with Fig.~\ref{fig:fractional-O4-2}.
The dotted planes denote the boundary at $x^3=0$ and $x^3=R$.
When the top and bottom of the ring touch each other through 
the compact direction $x^3$ with the twisted boundary condition, 
a reconnection of the two parts of the string occurs 
and
the ring is split into two fractional (anti-)instantons, 
vortices with the half twisted $U(1)$ moduli.
\label{fig:decay-ring-O4}}
\end{figure}
%%%%%%%%%%%%%%%%%%%%%%%

The twisted boundary condition $(-,-,+,+)$ is equivalent to 
Eq.~(\ref{eq:tbcO4--++}) in terms of 
the $SU(2)$-valued field $U(x)$. 
Here we use the original notation of the four real scalar fields 
$n_A(x)$.
The fixed manifold is characterized by 
$n_1=n_2=0$, equivalently $(n_3)^2+(n_4)^2=1$, 
which is the moduli space of vacua ${\cal N} \simeq S^1$. 
It has a nontrivial homotopy 
$\pi_1(S^1) \simeq {\mathbb Z}$, 
allowing a global vortex having a winding in $n_3 + i n_4$.
In the vortex core, the winding field must vanish $n_3=n_4=0$, 
and the other fields $n_1$ and $n_2$ appear with a constraint
$(n_1)^2+(n_2)^2=1$, giving a modulus ${\cal M} \simeq U(1)$ 
to the vortex.
For a fractional instanton, this $U(1)$ modulus is twisted half 
along the vortex string extending to the compactified direction, 
as described below.

An instanton (Skyrmion) can be represented by 
(a global analog of) a vorton \cite{Davis:1988jq}, 
that is, a vortex ring along which a $U(1)$ modulus is twisted.
This fact was first found in the context of 
Bose-Einstein condensates (BEC) 
\cite{Ruostekoski:2001fc,Nitta:2012hy} 
(see also \cite{Metlitski:2003gj}), 
and stable solutions in a Skyrme model were also 
constructed in Refs.~\cite{Gudnason:2014hsa,Gudnason:2014gla,Gudnason:2014jga}.
Configurations of Skyrmions as vortons are shown
 in Fig.~\ref{fig:fractional-O4-2}. 
The decomposition of an instanton into fractional 
instantons can be understood as 
higher dimensional analog  
of a domain wall ring in the $O(3)$ model 
with the twisted boundary condition $(-,-,+)$.
When the size of a vortex ring is the same with 
that of the compactification scale $R$, 
the top and bottom parts of the vortex ring touch each other 
through the compact $x^3$ direction with 
the twisted boundary condition.
Then, a reconnection of two fractions of the ring can occur 
(see \cite{Eto:2006db} for a reconnection of strings with moduli), 
the ring can be split 
into two vortex strings stretched along the compact direction,  
and subsequently 
they are separated into the $x^1$-$x^2$ plane 
as shown in Fig.~\ref{fig:decay-ring-O4}.
The $U(1)$ modulus is twisted half along 
each string, resulting in a fractional (anti-)instanton. 
These twisted vortices in the Skyrme model were numerically constructed in Ref.~\cite{Harland:2008eu}.
By considering all possibilities of twisted vortex rings, 
we find four kinds of fractional (anti-)instantons, 
as summarized in Fig.~\ref{fig:O(4)}  (2a)--(2d). 

The ansatz for fractional (anti-)instanton configurations 
with the boundary condition $(-,-,+,+)$ 
is given as
\beq
&& n_3 + i n_4  = \sin g(r) e^{i\theta}, \qquad
n_1 + i n_2 = \cos g(r) e^{i\zeta(z)},\\
&& g(0) = 0, \qquad
g(\infty) = \pm \frac{\pi}{2},  \\
&& \zeta(z=R) = \zeta(z=0) \pm \pi,
\eeq
where $(r,\theta,z)$ are cylindrical coordinates.
The topological instanton charge (Skyrmion charge or baryon number) 
can be calculated as
\begin{align}
Q_3 = \frac{1}{16\pi^2} \int d^3x \; \frac{1}{r}\sin(g) g_r \zeta_z 
= \1{2\pi} [\zeta]^{z=R}_{z=0} = \pm \1{2}  .
\end{align}
General formula of the instanton (Skyrme) charge 
for a vortex string with the winding number $Q$, 
along which  the $U(1)$ modulus is twisted $P$ 
times,   
was calculated to be $PQ$ in Ref.~\cite{Kobayashi:2013aza} 
in the context of Hopfions
and in Ref.~\cite{Gudnason:2014hsa} for Skyrmions.
The topological charges of fractional (anti-)instantons 
with the boundary condition $(-,-,+,+)$ 
are summarized in 
Table \ref{table:homotopy-O4-2}.
%%%%%%%%%%%%%%%%%%%%%%%%%
\begin{table}[h]
\begin{tabular}{c|c|c|c} 
     & $\pi_1$ & $\pi_1$ & $\pi_3$ \\ \hline
Fig.~\ref{fig:O(4)} (2a) &$ +1$     & $+1/2$ & $+1/2$ \\
Fig.~\ref{fig:O(4)} (2b) &$ -1$     & $-1/2$ & $+1/2$ \\
Fig.~\ref{fig:O(4)} (2c) &$ -1$     & $+1/2$ & $-1/2$ \\
Fig.~\ref{fig:O(4)} (2d) &$ +1$     & $-1/2$ & $-1/2$  
\end{tabular}
\caption{Homotopy groups of fractional instantons in the 
$O(4)$ model with the boundary condition $(-,-,+,+)$.
The columns represent the homotopy groups  
of a host soliton $\pi_1$, a daughter soliton $\pi_1$, 
and the total instanton $\pi_3$ from left to right. 
\label{table:homotopy-O4-2}}
\end{table}
%%%%%%%%%%%%%%%%%%%%%%

Interestingly,
we do not need higher derivative (Skyrme) term 
even though fractional instantons are Skyrmions. 
Indeed, stable configurations 
of (half) Skyrmions inside a vortex string 
was constructed without the Skyrme term 
in Ref.~\cite{Gudnason:2014hsa} 
on ${\mathbb R}^3$ 
without twisted boundary condition.

Fractional instantons with the boundary condition $(-,-,+,+)$ 
are global vortices in the $x^1$-$x^2$ plane
so that the interaction between them 
is $E_{\rm int} \sim \pm \log r$ with distance $r$ for large separation 
(the force is $F \sim \pm 1/r$),
where positive sign is for a pair of (anti-)vortices 
and negative sign is for a pair of a vortex and anti-vortex.

Bions can be constructed by combining 
configurations in (2a) and (2c) in Fig.~\ref{fig:O(4)},
or  (2b) and (2d) in Fig.~\ref{fig:O(4)}.
In the both cases, instanton charges are canceled out.
The interaction between fractional instantons 
constituting a bion is 
$E_{\rm int} \sim - \log r$ with distance $r$ for large separation 
and $F \sim - 1/r$, because they are a pair of a global vortex 
and global anti-vortex.

As in Eq.~(\ref{eq:SS-dim-red})
for the $O(3)$ model with the boundary condition 
$(-,-,+)$, the Scherk-Schwarz dimension reduction 
to two dimensions 
induces a potential term 
\beq 
  V = m^2 (\hat n_1^2 + \hat n_2^2) 
     = m^2 (1- \hat n_3^2 - \hat n_4^2).
\eeq

If one gauges the $U(1)$ symmetry acting on $n_3 + i n_4$, 
vortices become local vortices 
 of the ANO type, having finite energy.

%%%%%%%%%%%%%%%%%
\subsection{$(-,-,-,+)$: a half lump inside a domain wall}

%%%%%%%%%%%%%%%%%%%%%
\begin{figure}
\begin{center}
\includegraphics[width=0.1\linewidth,keepaspectratio]{3d-frame}
\begin{tabular}{cc}
\includegraphics[width=0.30\linewidth,keepaspectratio]{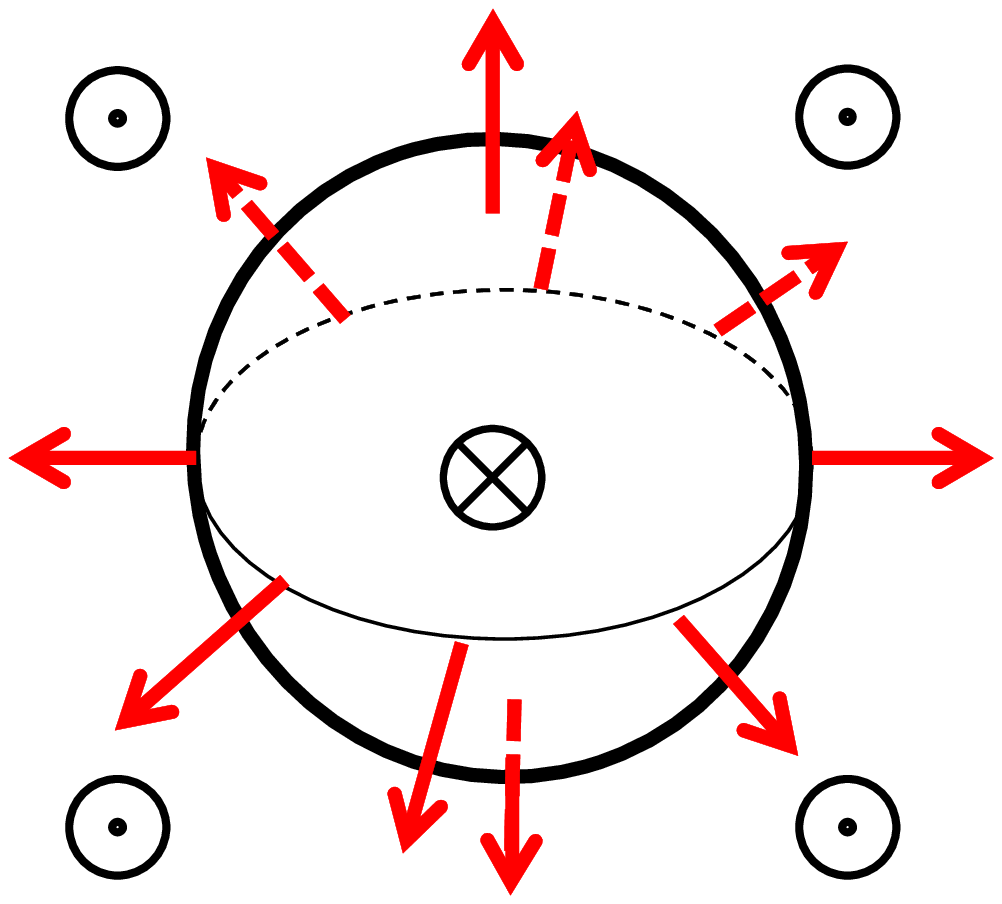} &
\includegraphics[width=0.20\linewidth,keepaspectratio]{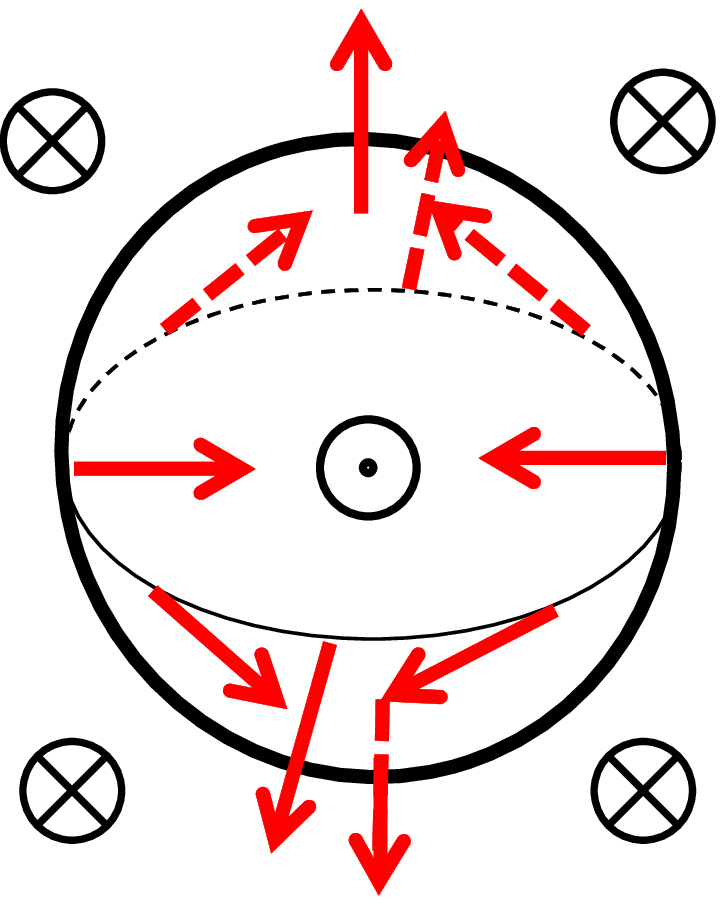} \\
(a) $(+1,+\1{2},+\1{2})+(-1,-\1{2},+\1{2})$ & 
(b) $(+1,-\1{2},-\1{2})+(-1,+\1{2},-\1{2})$ \\
\includegraphics[width=0.20\linewidth,keepaspectratio]{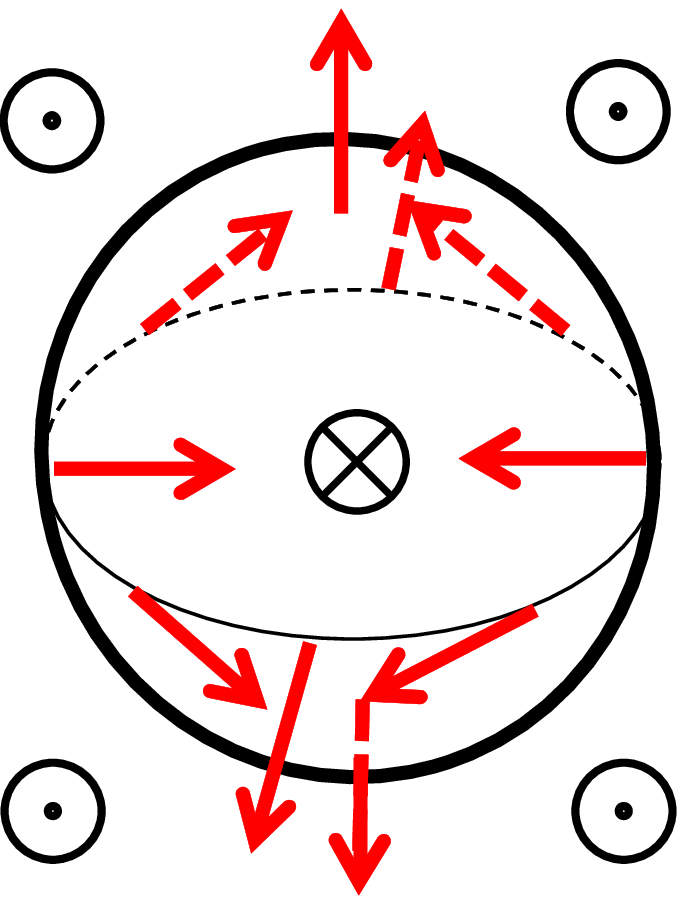} &
\includegraphics[width=0.30\linewidth,keepaspectratio]{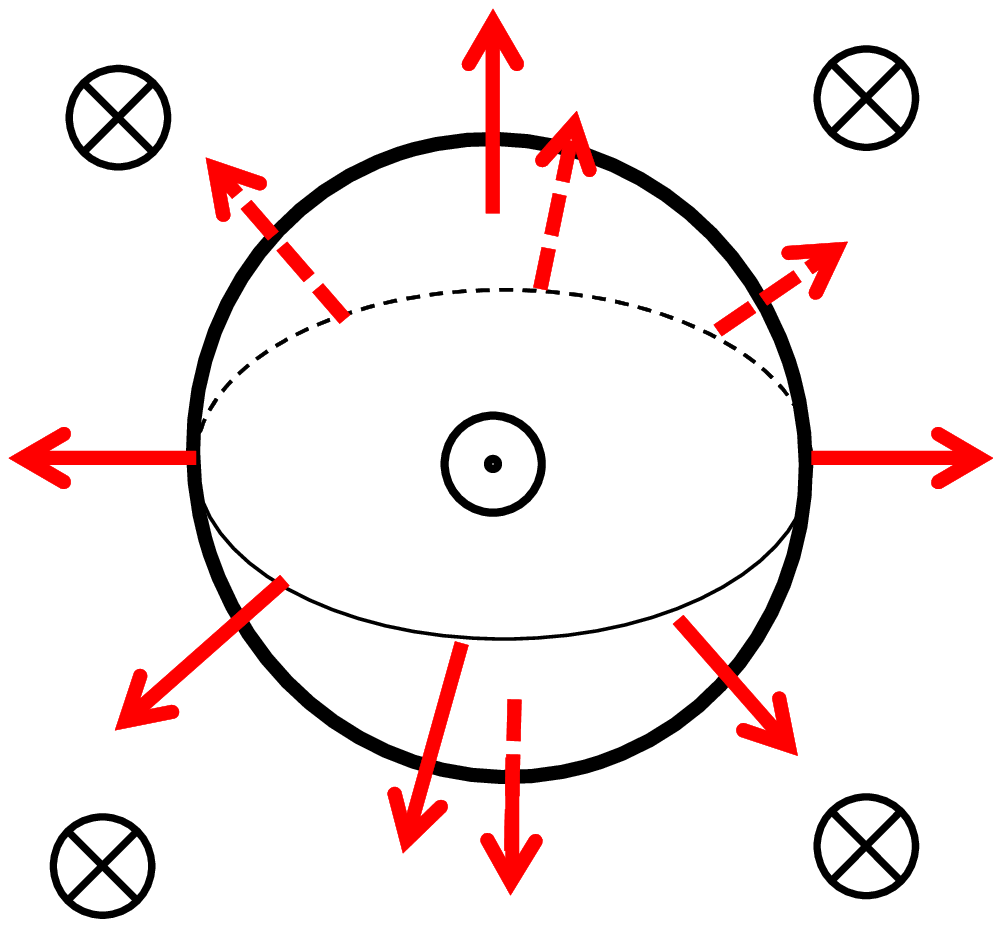} \\
(c) $(-1,-\1{2},+\1{2})+(+1,+\1{2},+\1{2})$ & 
(d) $(-1,+\1{2},-\1{2})+(+1,-\1{2},-\1{2})$ \end{tabular}
\end{center}
\caption{Fractional instantons from instantons in the $O(4)$ model 
with the boundary condition $(-,-,-,+)$. 
$\otimes$ and $\odot$ denote $n_4 = +1$ and $n_4 = -1$, 
representing the vacua ${\cal N}=\{\pm 1\}$. 
Red arrows represent $(n_1,n_2,n_3)$ 
parameterizing the moduli space of a domain wall 
${\cal M}\simeq S^2$. 
An instanton can be represented as a twisted domain wall,
that is, a spherical domain wall around which $S^2$ moduli are 
wound once. 
Brackets  $(*,*,*)$ denote
topological charges for a host domain wall characterized by $\pi_0$, 
that for a lump characterized by $\pi_2$, 
and that for an instanton characterized by $\pi_3$. 
(a) An instanton 
can be split into two fractional instantons 
$(+1,+\1{2},+\1{2})$ and $(-1,-\1{2},+\1{2})$.
(b) An anti-instanton 
can be split into two fractional anti-instantons 
$(+1,-\1{2},-\1{2})$ and $(-1,+\1{2},-\1{2})$.  
(c) and (d) are isomorphic to (a) and (b), respectively, 
by a $2\pi$ rotation along an axis parallel to the $x^1$ axis.
\label{fig:fractional-O4-3}}
\end{figure}
%%%%%%%%%%%%%%%%%%%%%%%%
%%%%%%%%%%%%%%%%%%%%%
\begin{figure}
\begin{center}
\includegraphics[width=1.0\linewidth,keepaspectratio]{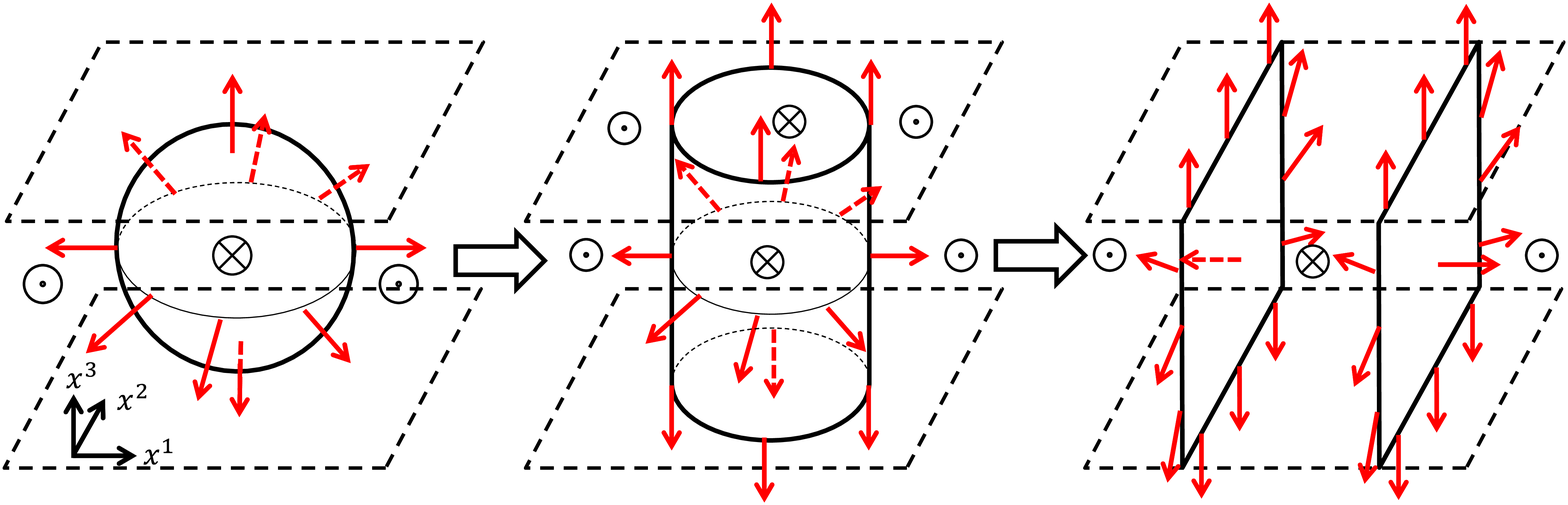} 
\end{center}
\caption{
Deformation of a twisted spherical domain wall 
into  two fractional instantons 
in the $O(4)$ model with the boundary condition $(-,+,+,+)$. 
The notations are the same with Fig.~\ref{fig:fractional-O4-3}.
The dotted planes denote the boundary at $x^3=0$ and $x^3 =R$.
A twisted spherical domain wall turns to a twisted domain wall 
tube 
when the top and bottom of the sphere touch each other through 
the compact direction $x^3$ with 
the twisted boundary condition.
If it is further stretched into the infinities in the $x^2$ direction, 
it can be deformed to two fractional instantons. 
\label{fig:decay-sphere-O4}}
\end{figure}
%%%%%%%%%%%%%%%%%%%%%%%

The fixed manifold is characterized by 
$n_1=n_2=n_3=0$ which is 
two discrete points characterized by $n_4=\pm 1$. 
The moduli space of vacua is ${\cal N} \simeq \{\pm 1\}$. 
It has a nontrivial homotopy 
$\pi_0(\pm 1) \simeq {\mathbb Z}_2$, allowing a domain wall.
In the domain wall core, 
the field making a domain wall vanishes $n_4=0$ 
and the other fields $n_1, n_2$ and $n_3$ appear with a constraint
$(n_1)^2+(n_2)^2+(n_3)^2=1$, giving  the moduli 
${\cal M} \simeq S^2$ to a domain wall. 
For a fractional instanton, these $S^2$ moduli are wound half 
in the wall world volume with the compact direction, 
as described below.

An instanton (Skyrmion) can be represented as 
a twisted spherical domain wall,
that is, a spherical domain wall around which $S^2$ moduli are 
wound \cite{Gudnason:2013qba}.
Configurations of Skyrmions as twisted spherical domain walls are shown
 in Fig.~\ref{fig:fractional-O4-3}.  
Deformation of a twisted spherical domain wall 
into  two fractional instantons  can be explained as follows. 
First, when the size of a sphere is the same with that of 
the compactification radius $R$, 
the top and bottom of 
the sphere touch and join each other 
through the compact direction $x^3$ with 
the twisted boundary condition,
and the sphere turns to 
a twisted domain wall tube  
as in left to middle in Fig.~\ref{fig:decay-sphere-O4}.  
Second, if the tube is further stretched  to the infinities in the $x^2$ direction, 
it can be deformed to 
 two surfaces (domain walls), separated into 
the $x^1$ direction as middle to right in Fig.~\ref{fig:decay-sphere-O4},  
where the domain wall world volume extend
to the $x^2$ and $x^3$ coordinates. 
The $S^2$ moduli are twisted half inside 
the domain wall world volumes, 
giving rise to fractional (anti-)instantons. 
While the first process can occur energetically, 
the second process cannot occur energetically because it 
needs infinite world volumes unless the $x^2$ direction is compactified.
Still the final configurations themselves are possible.
We thus find four possibilities of fractional (anti-)instantons 
as shown in Fig.~\ref{fig:O(4)} (3a)--(3d). 

The ansatz for fractional (anti-)instanton 
with the boundary condition $(-,-,-,+)$  
is given as
\beq
&& \mathbf{n} = (b_1(x^1,x^2)\sin f(x^3),b_2(x^1,x^2)\sin f(x^3),b_3(x^1,x^2)\sin f(x^3),\cos f(x^3)),\\
&& f(0) = 0, \qquad
f(\infty) = \pi.
\eeq
The fields ${\bf b} =(b_1,b_2,b_3)(x^1,x^2)$ are 
induced on a domain wall, 
satisfying the same boundary condition 
with $(-,-,-)$ of the $O(3)$ model.
Then we can consider a half lump given in Eq.~(\ref{eq:O(3)---}).
The topological instanton charge (Skyrmion charge or baryon number) 
can be calculated as \cite{Gudnason:2014nba} 
\beq
&& Q_3 = \frac{1}{\pi} \int d^3x \; \mathcal{Q} f_x = 
 \int d^2x \mathcal{Q} \equiv Q_2,\\
&& \mathcal{Q} = \frac{1}{8\pi} \epsilon^{ij} 
  \mathbf{b}\cdot\del_i\mathbf{b}\times\del_j\mathbf{b}.
\eeq
The topological charges of fractional (anti-)instantons 
with the boundary condition $(-,-,-,+)$ 
are summarized in 
Table \ref{table:homotopy-O4-3}.
%%%%%%%%%%%%%%%%%%%%%%%%%
\begin{table}[h]
\begin{tabular}{c|c|c|c} 
     & $\pi_0$ & $\pi_2$ & $\pi_3$ \\ \hline
Fig.~\ref{fig:O(4)} (3a) &$ +1$     & $+1/2$ & $+1/2$ \\
Fig.~\ref{fig:O(4)} (3b) &$ -1$     & $-1/2$ & $+1/2$ \\
Fig.~\ref{fig:O(4)} (3c) &$ -1$     & $+1/2$ & $-1/2$ \\
Fig.~\ref{fig:O(4)} (3d) &$ +1$     & $-1/2$ & $-1/2$  
\end{tabular}
\caption{Homotopy groups of fractional instantons in the 
$O(4)$ model with the boundary condition $(-,-,-,+)$.
The columns represent the homotopy groups  
of a host soliton $\pi_0$, a daughter soliton $\pi_2$, 
and the total instanton $\pi_3$ from left to right. 
\label{table:homotopy-O4-3}}
\end{table}
%%%%%%%%%%%%%%%%%%%%%%

Interestingly, 
we do not need higher derivative (Skyrme) term 
even though fractional instantons are Skyrmions, 
as in the case of the boundary condition $(-,-,+,+)$.
Indeed, stable configurations 
of a unit (not half) Skyrmion inside a domain wall was constructed 
in ${\mathbb R}^3$  
without twisted boundary condition
\cite{Nitta:2012wi,Gudnason:2014nba}.
We expect that the same holds for half instantons (Skyrmions).

Fractional instantons 
 with the boundary condition $(-,-,-,+)$ 
are domain walls perpendicular to the $x^1$ coordinate
so that the interaction between them 
is $E_{\rm int}  \sim -e^{-mr}$ with distance $r$ for large separation. 
The energy of domain walls are linearly divergent in the $x^2$ direction.

Bions can be constructed by combining 
configurations in (3a) and (3c) in Fig.~\ref{fig:O(4)},
or  (3b) and (3d) in Fig.~\ref{fig:O(4)},  
where the instanton charge is canceled out.
The interaction between fractional instantons 
constituting a bion is attractive and 
exponentially suppressed $E_{\rm int}  \sim -e^{-mr}$.

%%%%%%%%%%%%%%%%%
\subsection{$(-,-,-,-)$}

There are no fixed points for 
the boundary condition $(-,-,-,-)$ 
as the case with the boundary condition $(-,-,-)$
in the $O(3)$ model.
We do not have localized solitons wrapping 
around a fixed manifold. 
Again, we regard that there is a space-filling soliton (brane) 
with the moduli ${\cal M} \simeq  S^3$.

One (anti-)instanton is separated into two 
fractional (anti-)instantons with the boundary condition 
$(-,-,-,-)$,
as summarized in Fig.~\ref{fig:O(4)} (4a)--(4b). 
Each fractional instanton wraps a half sphere of the target space 
$S^3$. 

We need higher derivative (Skyrme) term 
for the stability of fractional instantons (Skyrmions).

%%%%%%%%%%%%%%%%%%%%%%%%%
\begin{table}[h]
\begin{tabular}{c|c|c|c} 
     & $\pi_{-1}$ & $\pi_3$ & $\pi_3$ \\ \hline
Fig.~\ref{fig:O(4)} (4a) &$ +1$     & $+1/2$ & $+1/2$ \\
Fig.~\ref{fig:O(4)} (4b) &$ -1$     & $-1/2$ & $+1/2$ \\
Fig.~\ref{fig:O(4)} (4c) &$ -1$     & $+1/2$ & $-1/2$ \\
Fig.~\ref{fig:O(4)} (4d) &$ +1$     & $-1/2$ & $-1/2$  
\end{tabular}
\caption{Homotopy groups of fractional instantons in the 
$O(4)$ model with the boundary condition $(-,-,-,-)$.
The columns represent the homotopy groups  
of a host soliton $\pi_{-1}$, a daughter soliton $\pi_3$, 
and the total instanton $\pi_3$ from left to right. 
Here, $\pi_{-1}$ is merely formal. 
\label{table:homotopy-O4-4}}
\end{table}
%%%%%%%%%%%%%%%%%%%%%%

The Scherk-Schwarz dimension reduction 
can be discussed 
as in Eq.~(\ref{eq:SS-dim-red}) for $(-,-,+,+)$.
However,  
by assuming the dependence of the fields 
on the compact direction $x^2$ as 
\beq 
(n_1,n_2,n_3,n_4) 
= \left(
\hat n_1(x^1)\cos {\pi \over R} x^2, 
\hat n_2(x^1) \sin {\pi \over R} x^2,
\hat n_3(x^1)\cos {\pi \over R} x^2, 
\hat n_4(x^1) \sin {\pi \over R} x^2
\right)
\eeq
in the presence of the twisted boundary condition 
$(-,-,-,-)$, 
we see that it does not give a nontrivial potential:
\beq 
  V = m^2 (\hat n_1^2 + \hat n_2^2 + \hat n_3^2 + \hat n_4^2) 
= m^2 ,
\eeq
with $m$ in Eq.~(\ref{eq:SS-dim-red}).

%%%%%%%%%%%%%%%%%%%%%
\section{Summary and Discussion \label{sec:summary} }

We have found that 
a fractional instanton in the $O(3)$ model 
is a global vortex with an Ising spin for $(-,+,+)$, 
a half sine-Gordon kink on a domain wall for $(-,-,+)$,
or a half lump on a ``space-filling brane" for $(-,-,-)$,
and that 
a fractional instanton in the $O(4)$ model 
is
a global monopole with an Ising spin for $(-,+,+,+)$, 
a half sine-Gordon kink on a global vortex for $(-,-,+,+)$,
a half lump on a domain wall for $(-,-,-,+)$,
or a half Skyrmion on a ``space-filling brane" for $(-,-,-,-)$. 
As from general argument in Sec.~\ref{sec:general}, 
the above classification holds for the $O(N)$ model 
with arbitrary $N$. 
We have also constructed neutral bions  the $O(3)$ and $O(4)$ models  
but have found that charged bions are not possible.
We have seen that 
when the number of minus signs in the boundary condition 
is even, 
a small compactification limit gives 
the Scherk-Schwarz dimensional reduction 
which induces a potential term as in Eq.~(\ref{eq:SS-dim-red}). 

If the interaction energy of two fractional instantons 
is exponentially suppressed $E_{\rm int} \sim e^{-mr}$ 
when they are well separated at distance $r$, 
the total energy of well separated 
fractional instantons is just of the sum 
of those of individual fractional instantons.
In this case, 
they would play a  role in resurgence of quantum field theory. 
This is indeed the case of the $O(3)$ model 
with the boundary condition $(-,-,+)$ 
\cite{Dunne:2012ae,Dunne:2012zk,Misumi:2014jua}  
in which fractional (anti-)instantons are (anti-)BPS 
so that there exists no interaction between 
fractional BPS  instantons, or between 
fractional anti-BPS  instantons,
and exponentially suppressed interaction 
$E_{\rm int} \sim e^{-mr}$ 
between a BPS and an anti-BPS fractional instantons.

Fractional instantons are not local or BPS in the other cases 
discussed in this paper as they are.
However,
with suitable modifications, some of them may become 
BPS or local as summarized as follows:

\begin{enumerate}
\item
The $O(3)$ model with $(-,+,+)$. 
If the $U(1)$ symmetry acting on $n_2 + i n_3$ is gauged,
half lump-vortices become local vortices 
having finite energy,
in which case the interaction between them  
would be exponentially suppressed. 
This is because 
the $U(1)$ gauged $O(3)$ model
with a potential $V=m^2 n_1^2$ on ${\mathbb R}^2$ 
allows local vortices 
\cite{Schroers:1995he,Schroers:1996zy,Baptista:2004rk,Nitta:2011um,
Alonso-Izquierdo:2014cza}.
If we further choose the gauge coupling to be  
$e^2 = m^2$, fractional (anti-)instantons become (anti-)BPS 
and the theory can be made supersymmetric \cite{Nitta:2011um}.  
In our case, the twisted boundary condition would 
play a role of the potential, so the gauge coupling 
should be correlated to the compactification radius 
for vortices to be BPS.

\item 
The $O(3)$ model with only four derivative term.
An $O(3)$ model consists of only four derivative 
(Skyrme) term and a suitable potential term 
is known as a BPS baby Skyrme model,  
admitting BPS instantons (lumps, baby Skyrmions) 
on ${\mathbb R}^2$ \cite{Adam:2010jr}.
This model can be made supersymmetric 
\cite{Adam:2011hj,Nitta:2014pwa,Bolognesi:2014ova}.
A generalization of the model to ${\mathbb R}^1 \times S^1$ 
with twisted boundary condition is expected to 
admit BPS fractional instantons.

\item
The $O(4)$ model with $(-,+,+,+)$. 
If one gauges the $SO(3)$ symmetry action on $(n_2,n_3,n_4)$, 
a half-Skyrmion monopole  becomes local, that is, of `t Hooft-Polyakov type 
having finite energy, 
in which case the interaction between them 
is exponentially suppressed.
This is because 
an $SO(3)$ gauged Skyrme model
with a potential $V=m^2 n_1^2$ on ${\mathbb R}^3$ 
allows a local `t Hooft-Polyakov type monopole with finite energy 
\cite{Brihaye:1998}.
A BPS limit is not known in this case.
Again in our case, the twisted boundary condition would 
play a role of the potential.

\item
The $O(4)$ model with $(-,-,+,+)$. 
If one gauges the $U(1)$ symmetry acting on 
$n_3 + i n_4$, 
vortices as half instantons become local vortices having finite energy, 
in which case the interaction between them 
would be exponentially suppressed.

\item
The $O(4)$ model on  $S^2 \times S^1$.
If we consider a geometry $S^2 \times S^1$ instead of 
${\mathbb R}^2 \times S^1$, 
instantons (Skyrmions) are BPS 
for untwisted boundary condition \cite{Canfora:2014aia}.
An extension to a twisted boundary condition should be possible, 
in which case
fractional (anti-)instantons may be also (anti-)BPS. 

\item
The $O(4)$ model with only a six derivative term.
If we consider Lagrangian containing 
only a six derivative term, 
which is baryon charge density squared,  
and a suitable potential term,
instantons (Skyrmions) are BPS, which 
is indeed the case of ${\mathbb R}^3$
\cite{Adam:2010fg}. 
It may be generalized to the case of
${\mathbb R}^2 \times S^1$ 
with a twisted boundary condition, 
in which case
fractional (anti-)instantons may be also (anti-)BPS.

\end{enumerate}
In these cases, fractional instantons 
will play a role in resurgence, 
which is indeed the case of 
the $O(3)$ model 
with the boundary condition $(-,-,+)$ 
\cite{Dunne:2012ae,Dunne:2012zk,Misumi:2014jua} 
as denoted above.

When we compactify more than one directions, 
we can consider more general twisted boundary conditions.
For instance, 
we may consider the $O(3)$ model on ${\mathbb R}^n \times (S^1)^2$
with a twisted boundary condition 
$(-,-,+)$ for one direction 
and $(+,-,-)$ for the other direction.
A complete classification of these more general cases remain
as an interesting problem.

A lattice of half Skyrmion appear in finite baryon density \cite{Ma:2013ooa}.
There may be certain relation with our half Skyrmions 
in the presence of a compact direction with twisted boundary conditions.

Hopfions are knot like solitons supported by 
the Hopf charge $\pi_3(S^2) \simeq {\mathbb Z}$
in the $O(3)$ model with four derivative (Faddeev-Skyrme) term 
\cite{Faddeev:1975}.
Since Hopfions on ${\mathbb R}^3$ are closed lump strings 
along which $U(1)$ moduli are twisted 
(see, e.g.~Ref.~\cite{Kobayashi:2013xoa}), 
those on ${\mathbb R}^2 \times S^1$
with an untwisted boundary condition 
can be twisted closed lump strings wrapping around 
$S^1$
\cite{Jaykka:2009ry,Kobayashi:2013aza}.
If we impose twisted boundary conditions, 
we will be able to obtain a fractional Hopfion  
as a half-twisted lump string wrapping around $S^1$.

By applying our method to non-Abelian gauge theories,
classification of fractional Yang-Mills instantons 
may be possible, 
which would be important toward the resurgence 
of gauge theories.
To this end, realizations of Yang-Mills instantons 
as various composite solitons summarized in Ref.~\cite{Nitta:2013vaa} 
will be useful, as has been demonstrated for Skyrmions in this paper. 
Yang-Mills instantons are 
Skyrmions inside a domain wall \cite{Eto:2005cc}, 
lumps inside a vortex \cite{Hanany:2004ea,Eto:2004rz,Eto:2005sw,Fujimori:2008ee}, 
or sine-Gordon kinks on a monopole string 
\cite{Nitta:2013cn}.
Investigating boundary conditions realizing these 
would be an important first step 
toward the resurgence 
of gauge theories.

Finally, let us make a comment on duality. 
As seen in this paper, a ${\mathbb C}P^1$ instanton 
with the boundary condition $(-,-,+)$ 
is decomposed into a set of two fractional instantons 
which are half twisted domain walls, 
as seen in Fig.~\ref{fig:decay-ring-O3}, 
and one of them becomes a domain wall 
in a small compactification radius limit in which 
the other is removed to infinity \cite{Eto:2004rz}. 
The same relation holds between 
a Yang-Mills instanton and a BPS monopole, 
which can be also understood as a T-duality 
acting on D-branes in type-II string theory 
\cite{Giveon:1998sr}.
In Ref.~\cite{Eto:2004rz}, 
 ${\mathbb C}P^{N-1}$ fractional instantons 
were realized as  fractional Yang-Mills instantons 
trapped inside a vortex in a $U(N)$ gauge theory, 
which explains a relation between the above mentioned 
two T-dualities. 
Here, in this paper, we have added one more example, 
that is, 
a T-duality between a Skyrmion and a vortex.
In the $O(4)$ model with the boundary condition $(-,-,+,+)$, 
equivalently Eq.~(\ref{eq:tbcO4--++}),
a Skyrmion is decomposed into a set of 
two fractional instantons which are 
half twisted vortex strings 
as seen in Fig.~\ref{fig:decay-ring-O4}.  
One of them becomes  a vortex 
in a small compactification radius limit, 
in which the other is removed to infinity.
We think that a further T-duality 
maps this configuration to 
a domain wall through a domain wall Skyrmion.

%%%%%%%%%%%%%%
\section*{Acknowledgments}

The author thanks Tatsuhiro Misumi and Norisuke Sakai 
for a discussion on bions 
and Sven Bjarke Gudnason 
for a discussion on Skyrmions.  
This work is supported in part by Grant-in-Aid for Scientific Research 
No.~25400268
and by the ``Topological Quantum Phenomena'' 
Grant-in-Aid for Scientific Research 
on Innovative Areas (No.~25103720)  
from the Ministry of Education, Culture, Sports, Science and Technology 
(MEXT) of Japan.

%%%%%%%%%% References %%%%%%%%%%%%%%%%%%%%%%%%%
\newcommand{\J}[4]{{\sl #1} {\bf #2} (#3) #4}
\newcommand{\andJ}[3]{{\bf #1} (#2) #3}
\newcommand{\AP}{Ann.\ Phys.\ (N.Y.)}
\newcommand{\MPL}{Mod.\ Phys.\ Lett.}
\newcommand{\NP}{Nucl.\ Phys.}
\newcommand{\PL}{Phys.\ Lett.}
\newcommand{\PR}{ Phys.\ Rev.}
\newcommand{\PRL}{Phys.\ Rev.\ Lett.}
\newcommand{\PTP}{Prog.\ Theor.\ Phys.}
\newcommand{\hep}[1]{{\tt hep-th/{#1}}}
%%%%%%%%%%%%%%%%%%%%%%%%%%%%%%%%%%%%%%%%%%%%%%%

\end{document}